\documentclass[a4paper,fleqn,usenatbib]{mnras}

\usepackage{newtxtext,newtxmath}

\usepackage[T1]{fontenc}
\usepackage{ae,aecompl}

\usepackage{graphicx}	% Including figure files
\usepackage{amsmath}	% Advanced maths commands
\usepackage{amssymb}	% Extra maths symbols

\usepackage{cancel}
\usepackage{amsfonts}
\usepackage{algorithm}
\usepackage[noend]{algpseudocode}

\makeatletter
\def\BState{\State\hskip-\ALG@thistlm}
\makeatother

\usepackage{listings}
\usepackage{enumitem}
\usepackage{calc}
\usepackage{bm}
\usepackage{times}
\usepackage{float}

\usepackage{color}

\usepackage[english]{babel}
\graphicspath{{./images/}}
\newcommand{\modif}[1]{\textcolor{black}{#1}}
\newcommand{\pows}{\ensuremath{\mathcal{P}}}
\newcommand{\bis}{\ensuremath{\mathcal{B}}}

\newtheorem{theorem}{Theorem}

\title[A regularized tri-linear approach for optical interferometric imaging]{A regularized tri-linear approach for optical interferometric imaging}

\author[J. Birdi et al.]{
Jasleen Birdi\thanks{E-mail: jb36@hw.ac.uk},
Audrey Repetti,
and Yves Wiaux
\\
\\
Institute of Sensors, Signals and Systems, Heriot-Watt University, Edinburgh EH14 4AS, UK
}

\date{Accepted XXX. Received YYY; in original form ZZZ}

\pubyear{2016}

\begin{document}
\label{firstpage}
\pagerange{\pageref{firstpage}--\pageref{lastpage}}
\maketitle

% Abstract of the paper
\begin{abstract}
In the context of optical interferometry, only undersampled power spectrum and bispectrum data are accessible. It poses an ill-posed inverse problem for image recovery.
Recently, a tri-linear model was proposed for monochromatic imaging, leading to an alternated minimization problem. In that work, only a positivity constraint was considered, and the problem was solved by an approximated Gauss-Seidel method.
In this paper, we propose to improve the approach on three fundamental aspects. 
First, we define the estimated image as a solution of a regularized minimization problem, promoting sparsity in a fixed dictionary using either an $\ell_1$ or a (re)weighted-$\ell_1$ regularization term. 
Secondly, we solve the resultant non-convex minimization problem using a block-coordinate forward-backward algorithm. This algorithm is able to deal both with smooth and non-smooth functions, and benefits from convergence guarantees even in a non-convex context. 
Finally, we generalize our model and algorithm to the hyperspectral case, promoting a joint sparsity prior through an $\ell_{2,1}$ regularization term.
We present simulation results, both for monochromatic and hyperspectral cases, to validate the proposed approach.

\end{abstract}


%%%%%%%%%%%%%%%%%%%%%%%%%%%%%%%%%%%%%%%%%%%%%%%%%%

%%%%%%%%%%%%%%%%% BODY OF PAPER %%%%%%%%%%%%%%%%%%

\section{Introduction}
With the advent of astronomical interferometers, it has become possible to image the sky at very high angular resolution.
An interferometer basically consists of an array of telescopes such that each pair of telescopes probes a spatial frequency in the Fourier plane (denoted by $u-v$ plane) of the image of interest. Given the limited number of telescopes, incomplete sampling of the $u-v$ plane is obtained. 
In particular, for radio interferometry, measurements consist of complex visibilities, related to Fourier coefficients of the intensity image of interest \citep{Thompson2001}. In this context, the incomplete Fourier sampling leads to a linear ill-posed inverse problem for image reconstruction, and iterative algorithms need to be designed to solve this problem. 
Classical reconstruction methods for radio interferometry are mainly based on iterative deconvolution (\textsc{clean}; \cite{Hogbom1974}), and on {maximum entropy methods} (MEM) to impose smoothness on the sought image by maximizing the entropy of the image \citep{Cornwell1985}. More recently, imaging techniques within the framework of compressive sensing have been proposed  \citep{Wiaux2009}. 
These methods rely on finding an image that is sparse in a given dictionary, using convex optimization algorithms \citep{Boyd2004, Combettes2010}. 

As compared to the radio interferometers, optical interferometers involve a less number of telescopes, which in turn provides a sparser $u-v$ coverage. Moreover, atmospheric turbulence at optical wavelengths causes random phase fluctuations leading to cancellation of the visibility values. Indeed the measurements consist of phase insensitive observables: power spectrum and bispectrum, resulting into loss of partial phase information \citep{Thiebaut2010}. 
This induces non-linearity in the inverse problem for image reconstruction in optical interferometry. Thus, the image recovery methods used in radio interferometry cannot be directly applied, and new methods need to be developed.

Research in this direction has led to the development of various algorithms, based on different approaches. In \citet{Thiebaut2008}, the so-called MIRA method has been developed, using a {maximum a posteriori} (MAP) approach to recover the image, where different types of quadratic regularization can be considered. 
The author proposed to solve the resulting minimization problem using
a limited variable metric algorithm which accounts for parameter bounds (namely, the VMLMB algorithm \citep{Thiebaut2002}).  
Another technique, proposed by \citet{Meimon2005}, namely WISARD, makes use of a self-calibration approach to solve for missing phase information, using smooth regularizations. 
The so-called BSMEM method, proposed in \citet{Buscher1994}, consists of using MEM to impose smoothness on the estimated image. Recently, \citet{Hofmann2014} proposed the IRBis method (image reconstruction software using the bispectrum), which solves the minimization problem from a MAP approach, considering smooth regularization terms, and employing a non-linear optimization algorithm based on conjugate gradients \citep{Hager2005,Hager2006a}. 
However, due to the non-linearity of the considered inverse problem, the minimization problems solved by the above methods perform only local optimization. 
For global minimum search, different approaches have been proposed these last years. In particular, techniques based on a {Markov Chain Monte Carlo} (MCMC) method \citep{Gamerman1997} have been adopted in MACIM \citep{Ireland2006a} and SQUEEZE \citep{Baron2012}, while in \citet{Auria2013}, a tensor approach has been proposed. 
In the latter, following the idea of phase-lift methods for phase retrieval problems \citep{Cand2011, Waldspurger2013}, the data model is lifted from a vector to a super-symmetric rank-1 order-3 tensor formed by the tensor product of the vector representing the sought image with itself. This yields a linear inverse problem, and a convex minimization problem can be deduced from a MAP approach. In \citet{Auria2014}, the tensor approach has been extended to account for the signal sparsity and thereby improving the reconstruction quality. However, solving for order-3 tensor  instead of an image (i.e. a vector) increases the dimensionality of the problem drastically and makes this approach computationally very expensive. 
Thus, \citet{Auria2013} proposed another method which involves solving linear and convex sub-problems alternately and iteratively for 3 images. 
Although the global minimization problem remains non-convex and dependent on the initial guess, in practice it has been shown that it provides much better reconstruction quality and accelerates the convergence speed as compared to the tensor approach. 
Moreover, contrary to the state-of-the-art-methods, it brings convexity to the sub-problems. 
However, \citet{Auria2013} proposed to solve the tri-linear problem using a Gauss-Seidel method {(\citet{Zangwill1969}, \citet[Chap~7]{Ortega1970}, \citet[Chap.2]{Bertsekas1999})}, which does not have any convergence guarantees in this context. Additionally, only positivity constraints have been considered, without imposing any other \emph{a priori} information on the underlying image.

All of the above mentioned methods are designed to reconstruct monochromatic images. 
However, electromagnetic radiations at different wavelengths can be emitted from an astrophysical source, corresponding to its spectrum.
In order to exploit the spectrum of the source, modern optical interferometers are paving the way for multi-wavelength imaging. Instruments such as AMBER \citep{Petrov2001}, GRAVITY \citep{Eisenhauer2008} and MATISSE \citep{Lopez2009}, can take measurements at multiple wavelength channels. 
This necessitates the progression of imaging techniques from monochromatic to hyperspectral case. 
Lately, initial work are done in the direction of hyperspectral imaging for optical interferometry. 
In particular, the method proposed by \citet{Kluska2014}, namely SPARCO, is a semi-parametric approach for image reconstruction of chromatic objects, whereas the method proposed by \citet{Thiebaut2013} deals with a sparsity regularized approach considering the observed scene to be a collection of point-like sources. Recently the use of differential phases for hyperspectral imaging has been proposed in PAINTER \citep{Schutz2014}. The methods proposed by \citet{Thiebaut2013} and \citet{Schutz2014} use the {alternating direction method of multipliers} (ADMM) algorithm \citep{Boyd2010} to solve the considered minimization problem. 

In this article, we propose an image reconstruction algorithm which can be applied both for monochromatic and hyperspectral cases in optical interferometry. 
More precisely, in the monochromatic case, we propose to improve the method based on the tri-linear data model proposed by \citet{Auria2013}.
First, we propose to impose sparsity as a regularization term, by means of an $\ell_1$-norm, either in the image domain or in a given basis \citep{Wiaux2009, Carrillo2012}, leveraging the recent compressive sensing theory \citep{Donoho2006}. 
In addition, we have developed an algorithm, based on the block-coordinate forward-backward algorithm recently proposed, e.g., by \citet{Bolte2014,Frankel2015,Chouzenoux2016}, which allows to deal with non-necessarily smooth regularization terms such as the $\ell_1$ norm. Moreover, this algorithm benefits from the convergence guarantees even for the non-convex global minimization problems. 
Finally, we generalize the proposed method to the hyperspectral case. It translates to a new approach for hyperspectral imaging in optical interferometry. We exploit the joint sparsity of the image cube through an $\ell_{2,1}$ norm \citep{Thiebaut2013}.

The rest of the article is organized as follows. Section~\ref{sec:optical_imaging} describes the observation model, whereas the corresponding regularized minimization problem is detailed in Section~\ref{sec:min_prob}. In Section~\ref{sec:proposed_algo}, the proposed algorithm to solve the resultant minimization problem is presented along with the implementation details, incorporating various regularization terms. The simulations performed and the results obtained thereby for monochromatic case are discussed in Section~\ref{sec:simulations}.  Section~\ref{sec:hyperspectral} is devoted to the hyperspectral case. Starting with the problem statement, the optimization details and the simulations performed are then presented with the results obtained. Finally, the conclusion is provided in Section~\ref{sec:Conclusion}.

%---------------------------------------------------------

\section{Optical Interferometric Observation Model} \label{sec:optical_imaging}

Consider the intensity image of interest be represented by real and positive vector ${\overline{\bm{x}}} = (\overline{x}_n)_{1 \leq n \leq N} \in\mathbb{R}_+^{N}$.
Its discrete Fourier transform is denoted by ${\widehat{\bm{x}}} = (\widehat{x}_n)_{1 \leq n \leq N} \in \mathbb{C}^N$. An interferometer probes discrete spatial frequencies in the $u-v$ plane of the image of interest. Each spatial frequency sampled by a pair of telescopes, separated by a distance $d$, is given by $({d}/{\lambda})$, with $\lambda$ being the observation wavelength \citep{Thiebaut2010}. 
Note that the total flux is assumed to be measured independently and the zero frequency Fourier coefficient, denoted by $\widehat{x}_{c}$, is normalized to be equal to 1. 

In optical interferometry, the measurements are composed by $M_{\pows}$ power spectrum measurements, corresponding to the squared modulus of the complex visibilities, and by $M_{\bis}$ bispectrum measurements, corresponding to a triple product of three different complex visibilities. 
Thus, each measurement can be represented by the triple product of Fourier coefficients of the image of interest, i.e. $\widehat{x}_i \widehat{x}_j \widehat{x}_k$, where $i$, $j$ and $k$ belong to $\{1, \ldots, N\}$. 
Considering the Hermitian symmetry, we denote by $\widehat{x}_{i^*}$ the Fourier coefficient at the opposite spatial frequency to that related with $\widehat{x}_i$. Following this notation, the power spectrum measurements are obtained by choosing indices $j=i^*$ and $k = c$, thus giving triple product of the form $\widehat{x}_i\widehat{x}_{i^*} \widehat{x}_c = |\widehat{x}_i|^2$.	 Similarly, for the bispectrum measurements, phase closure should be satisfied so that the spatial frequencies corresponding to $\widehat{x}_i , \widehat{x}_j$ and $ \widehat{x} _k$ sum to zero \citep{Monnier2007}. As a result, the bispectrum measurements are given by $\widehat{x}_i\widehat{x}_{j} \widehat{x}_{{(i+j)}^{*}}$.

It is to be mentioned here that in general, for a fixed number $A$ of telescopes in an interferometer, the independent spatial frequencies sampled, each probed by a pair of telescopes, are equal to ${A \choose 2} = A(A-1)/2$, and the number of possible closing triangles (i.e. phase closures) is ${A \choose 3} = A(A-1)(A-2)/(3 \times 2)$. 
However, out of these only ${A-1 \choose 2} = (A-1)(A-2) / 2$ number of phase closures are independent \citep{Monnier2007}. As a result, most of the Fourier phase information is missing. Combined with the sparseness in the $u-v$ coverage, this poses a highly under-determined inverse problem. 

In view of the description provided above, the inverse problem can be written as follows:
\begin{equation}	\label{eq:meas_pb}
\bm{y} = 
\big[ (\bm{\mathsf{T}}_1 {\overline{\bm{x}}})\cdot (\bm{\mathsf{T}}_2 {\overline{\bm{x}}})\cdot (\bm{\mathsf{T}}_3 {\overline{\bm{x}}}) \big] 
+ \bm{\eta} ,
\end{equation}
where $\cdot$ denotes the Hadamard product, 
$\bm{y} = (y_m)_{1 \le m \le M} \in \mathbb{C}^M$, with $M = M_{\pows} + M_{\bis}$, $\bm{\eta} \in \mathbb{C}^M $ is a realization of an additive i.i.d. Gaussian noise, and $\bm{\mathsf{T}}_1$, $\bm{\mathsf{T}}_2$, $\bm{\mathsf{T}}_3$ are linear operators from $\mathbb{R}^N$ to $\mathbb{C}^M$. 
More precisely, for every $p \in \{1,2,3\}$, $\bm{\mathsf{T}}_p$ performs a discrete 2D Fourier transform $\bm{\mathsf{F}} \in \mathbb{C}^{N\times N}$, followed by selection operators, denoted by $\bm{\mathsf{S}}  \in \mathbb{R}^{M_{\pows}\times N}$ and $\bm{\mathsf{L}}_p \in \mathbb{R}^{M\times M_{\pows}}$, i.e.
\begin{equation} \label{eq:mask}
\bm{\mathsf{T}}_p = \bm{\mathsf{L}}_p \bm{\mathsf{S F}} .
\end{equation}
Firstly, the operator $\bm{\mathsf{S}}$  selects $M_{\pows}$ Fourier coefficients corresponding to the spatial frequencies given by the telescopes' position. Note that due to Hermitian symmetry, only half of the Fourier plane is sampled. Then, the operators $\bm{\mathsf{L}}_1$, $\bm{\mathsf{L}}_2$ and $\bm{\mathsf{L}}_3$ select the different coefficients from $\bm{\mathsf{S F}} {\overline{\bm{x}}}$, in order to construct the triple products corresponding to the power spectrum and bispectrum measurements.
This makes these three operators different from each other.

%-----------------------------------------------------------

\section{Proposed regularized minimization problem}
\label{sec:min_prob}

\subsection{Problem formulation}
\label{ssec:MAP}

The data model in equation~\eqref{eq:meas_pb} being non-linear, 
applying directly a MAP approach would lead to a non-convex minimization problem. 
To bring linearity in \eqref{eq:meas_pb}, following the model proposed by \citet{Auria2013}, we introduce $( \overline{\bm{u}}_1, \overline{\bm{u}}_2, \overline{\bm{u}}_3 )$ $\in (\mathbb{R}_+^N)^3$ such that 
\begin{equation}	\label{ass:eq_eq_u}
\overline{\bm{u}}_1 = \overline{\bm{u}}_2 = \overline{\bm{u}}_3 = \overline{\bm{x}}. 
\end{equation}
Then, the data model \eqref{eq:meas_pb} is equivalent to
\begin{equation} \label{eq:data_mod}
\bm{y} = \big[ (\bm{\mathsf{T}}_1 \overline{\bm{u}}_1 ) \cdot (\bm{\mathsf{T}}_2 \overline{\bm{u}}_2 ) \cdot (\bm{\mathsf{T}}_3 \overline{\bm{u}}_3 ) \big] + \bm{\eta},
\end{equation}
where $\overline{\bm{u}}_1$, $\overline{\bm{u}}_2$ and $\overline{\bm{u}}_3$ correspond to the unknown image which is to be estimated. 
The new model described in \eqref{eq:data_mod} is tri-linear, i.e., it is linear in each of the variables $\overline{\bm{u}}_1$, $\overline{\bm{u}}_2$, and $\overline{\bm{u}}_3$. Thus, the problem can be solved separately for each of these variables, keeping other two fixed.

We propose to use a MAP approach to find an estimation of the original image $\overline{\bm{x}}$. More precisely, we propose to define the estimation of $( \overline{\bm{u}}_1, \overline{\bm{u}}_2, \overline{\bm{u}}_3 )$  as a solution to 
\begin{equation} \label{eq:overall_min_}
\underset{(\bm{u}_1, \bm{u}_2, \bm{u}_3)\in (\mathbb{R}^N)^3}{\operatorname{minimize}} \, 
f(\bm{u}_1, \bm{u}_2, \bm{u}_3) 
+ \sum_{p=1}^3 r(\bm{u}_p)\, ,
\end{equation}
where $f \colon \mathbb{R}^N \to ]-\infty, +\infty[$
is the data fidelity term ensuring consistency of the solution with the measurements, and $r \colon \mathbb{R}^N \to ]-\infty, +\infty]$ is a regularization term incorporating \emph{a priori} information on the target image $\overline{\bm{x}}$. Here, due to equality~\eqref{ass:eq_eq_u}, we propose to choose the same regularization for ${\bm{u}}_1$, ${\bm{u}}_2$ and ${\bm{u}}_3$.

Since $\bm{\eta}$ in (\ref{eq:data_mod}) is assumed to be a realization of an i.i.d. Gaussian noise, the usual least-squares criterion can be used for the data fidelity term:
\begin{equation} \label{eq:data_fid}
f(\bm{u}_1, \bm{u}_2, \bm{u}_3) 
= \dfrac{1}{2} \big\|{\bm{y} - (\bm{\mathsf{T}}_1 \bm{u}_1 ) \cdot (\bm{\mathsf{T}}_2 \bm{u}_2 ) \cdot (\bm{\mathsf{T}}_3 \bm{u}_3 )} \big\|_2^2  .
\end{equation}
\modif{Note that here we have assumed that the noise variance is the same for both the power spectrum and bispectrum measurements. However, in practice, the bispectrum measurements are degraded by a noise with greater variance than that of the noise associated to the power spectrum \citep{Pauls2005}.
In such scenario, one can use a weighted least-squares data fidelity term in order to incorporate information from the noise covariance matrix \citep{Hofmann2014}.}

In order to ensure a good reconstruction quality, 
we propose to use a hybrid regularization term: 
\begin{equation}	\label{eq:reg_term1}
(\forall \bm{x} \in \mathbb{R}^N)\quad
r(\bm{x}) = \iota_{\mathbb{R}_+^{N}}(\bm{x}) + \mu g(\bm{x}),
\end{equation}
where $\iota_{\mathbb{R}_+^{N}}(\bm{x}) $ denotes the indicator function equal to $0$ if $\bm{x} \in \mathbb{R}_+^N$, and $+\infty$ otherwise, $\mu \in ]0, +\infty[$ is a regularization parameter, and $g \colon \mathbb{R}^N \to ]-\infty, +\infty]$ is a convex non-necessarily smooth function. 
Thus, the proposed formulation can be seen as a generalization of the model proposed in \citet{Auria2013}. Indeed, \citet{Auria2013} proposed to solve \eqref{eq:overall_min_} using $f$ defined in \eqref{eq:data_fid}, and $r$ given by \eqref{eq:reg_term1} when $\mu \equiv 0$.

\subsection{Symmetrized data fidelity term}
\label{ssec:AM_symm}

Problem \eqref{eq:overall_min_} can be solved by alternating sequentially between the estimation of each variable $\bm{u}_1$, $\bm{u}_2$ and $ \bm{u}_3$ while keeping the other two fixed.
Since the vectors are solved separately in each sub-problem, the 3 estimated vectors can converge to different estimations. 
One method to avoid this issue is to add the information \eqref{ass:eq_eq_u} in the regularization term, e.g. to consider quadratic terms controlling the distance between the variables $\bm{u}_1$, $\bm{u}_2$ and $ \bm{u}_3$. 
However, introducing such regularization terms involve additional regularization parameters to be tuned.
Thus, to ensure convergence of the 3 vectors to similar estimations, while avoiding to complicate the minimization problem with additional regularization parameters, we propose to consider a symmetric data fidelity term 
for $\bm{u}_1$, $\bm{u}_2$ and $\bm{u}_3$, instead of considering the usual least-squares criterion \eqref{eq:data_fid}. 
More precisely, in order to take into account the symmetry between $\overline{\bm{u}}_1$, $\overline{\bm{u}}_2$ and $\overline{\bm{u}}_3$, we propose to consider the following data fidelity term :
\begin{align}
\widetilde{f}(\bm{u}_1, \bm{u}_2, \bm{u}_3) 
=	
&	\dfrac{1}{6} \Big( f(\bm{u}_1, \bm{u}_2, \bm{u}_3) + f(\bm{u}_1, \bm{u}_3, \bm{u}_2) 	\nonumber	\\
&	+ f(\bm{u}_2, \bm{u}_1, \bm{u}_3) + f(\bm{u}_2, \bm{u}_3, \bm{u}_1) 	\nonumber	\\
&	+ f(\bm{u}_3, \bm{u}_1, \bm{u}_2) + f(\bm{u}_3, \bm{u}_2, \bm{u}_1)  \Big) ,	\label{eq:data_fid_sym}
\end{align}
where $f$ is given by \eqref{eq:data_fid}. In this case, it can be noticed that ${\bm{u}}_1$, ${\bm{u}}_2$ and ${\bm{u}}_3$ are commutative in \eqref{eq:data_fid_sym}, i.e. we have
\begin{align}
\widetilde{f}(\bm{u}_1, \bm{u}_2, \bm{u}_3) 
&	=	\widetilde{f}(\bm{u}_1, \bm{u}_3, \bm{u}_2) 
	=	\widetilde{f}(\bm{u}_2, \bm{u}_1, \bm{u}_3)  
	=	\widetilde{f}(\bm{u}_2, \bm{u}_3, \bm{u}_1)  \nonumber		\\
&	=	\widetilde{f}(\bm{u}_3, \bm{u}_1, \bm{u}_2)  
	=	\widetilde{f}(\bm{u}_3, \bm{u}_2, \bm{u}_1) .
\end{align}

The symmetrization of the data fidelity term can be explained as follows. 
Due to equality~\eqref{ass:eq_eq_u}, images $\overline{\bm{u}}_1$, $\overline{\bm{u}}_2$ and $\overline{\bm{u}}_3$ correspond to the sought image $\overline{\bm{x}}$. 
Let $\widehat{\bm{u}}_p = \big( \widehat{u}_{p,i} \big)_{1 \le i \le N}$ denote the Fourier transform of $\overline{\bm{u}}_p$, for $p \in \{1,2,3\}$. Then, for a given frequency index $i$, we have
$\widehat{u}_{1,i} = \widehat{u}_{2,i} = \widehat{u}_{3,i} $.
This implies that each measurement $y_{ijk}$, %given by the triple product $\widehat{x}_i \widehat{x}_j \widehat{x}_k$, 
where $(i,j,k)$ is a triplet of frequency indices, can be given by $\widehat{u}_{p,i} \widehat{u}_{q,j} \widehat{u}_{s,k}$, for all the possible permutations of $(p,q,s) \in (\{1,2,3\})^3$, with $p \neq q \neq s$.

Thus, following this symmetrized approach, we propose to 
\begin{equation} \label{eq:overall_min}
\underset{(\bm{u}_1, \bm{u}_2, \bm{u}_3)\in (\mathbb{R}^N)^3}{\operatorname{minimize}} \, 
\Big\{ h(\bm{u}_1, \bm{u}_2, \bm{u}_3)  = \widetilde{f}(\bm{u}_1, \bm{u}_2, \bm{u}_3) 
+ \sum_{p=1}^3 r(\bm{u}_p) \Big\} ,
\end{equation}
where $\widetilde{f}$ is defined by \eqref{eq:data_fid_sym}, and $r$ is given by \eqref{eq:reg_term1}. Note that, since the data fidelity term is symmetrized and the same regularization term is used for $\bm{u}_1, \bm{u}_2, \bm{u}_3$,  the global cost function $h$ is symmetric as well with respect to $\bm{u}_1, \bm{u}_2, \bm{u}_3$.
Furthermore, the minimization problem is solved using identical initialization for the unknown vectors $\bm{u}_1$, $\bm{u}_2$, and $\bm{u}_3$, and the final estimation $\bm{x}^{\star}$ of $\overline{\bm{x}}$ is taken to be the mean of the 3 estimated vectors. 

We will demonstrate in Section~\ref{sec:simulations}, through simulation results, that the recovered estimations of $\overline{\bm{u}}_1$, $\overline{\bm{u}}_2$ and $\overline{\bm{u}}_3$ are very close.

\subsection{Alternated minimization} \label{ssec:Alter_min}

As discussed earlier, problem~\eqref{eq:overall_min} can be solved sequentially, alternating between the estimations of $\overline{\bm{u}}_1$, $\overline{\bm{u}}_2$, and $\overline{\bm{u}}_3$. 
To describe the three corresponding sub-problems, additional notations are introduced. 

In particular, according to Section  \ref{ssec:AM_symm}, let us rewrite the considered symmetrized data fidelity term \eqref{eq:data_fid_sym} as follows
\begin{equation}
\widetilde{f}(\bm{u}_1, \bm{u}_2, \bm{u}_3) = \dfrac{1}{2} \big\| \widetilde{\bm{y}} - (\widetilde{\bm{\mathsf{T}}}_1 \bm{u}_1)  \cdot (\widetilde{\bm{\mathsf{T}}}_2 \bm{u}_2)  \cdot (\widetilde{\bm{\mathsf{T}}}_3 \bm{u}_3)   \|_2^2,
\end{equation}
where $\widetilde{\bm{\mathsf{T}}}_1$, $\widetilde{\bm{\mathsf{T}}}_2$, and $\widetilde{\bm{\mathsf{T}}}_3$ are linear operators defined to be the concatenations of the permutations of the operators $ \big( \bm{\mathsf{T}}_p \big)_{1 \le p \le 3}$:
\begin{equation}
\widetilde{\bm{\mathsf{T}}}_1 = \frac{1}{6^{1/6}} \,
\left[ \begin{matrix}
{\bm{\mathsf{T}}}_1 \\
{\bm{\mathsf{T}}}_1 \\
{\bm{\mathsf{T}}}_2 \\
{\bm{\mathsf{T}}}_2 \\
{\bm{\mathsf{T}}}_3 \\
{\bm{\mathsf{T}}}_3 
\end{matrix} \right] , \quad
\widetilde{\bm{\mathsf{T}}}_2 = \frac{1}{6^{1/6}} \;
\left[ \begin{matrix}
{\bm{\mathsf{T}}}_2 \\
{\bm{\mathsf{T}}}_3 \\
{\bm{\mathsf{T}}}_1 \\
{\bm{\mathsf{T}}}_3 \\
{\bm{\mathsf{T}}}_1 \\
{\bm{\mathsf{T}}}_2 
\end{matrix} \right], \;  \text{and} \;
\widetilde{\bm{\mathsf{T}}}_3 = \frac{1}{6^{1/6}} \,
\left[ \begin{matrix}
{\bm{\mathsf{T}}}_3 \\
{\bm{\mathsf{T}}}_2 \\
{\bm{\mathsf{T}}}_3 \\
{\bm{\mathsf{T}}}_1 \\
{\bm{\mathsf{T}}}_2 \\
{\bm{\mathsf{T}}}_1 
\end{matrix} \right] ,
\end{equation}
and $\widetilde{\bm{y}} \in \mathbb{C}^{(6M)}$ is the concatenation of the corresponding 6 permutations of the observation vector $\bm{y}$, divided by $6^{1/2}$.
Let $(p,q,s) \in \{1,2,3\}$. Fix $\bm{u}_q \in [0, +\infty[^N$ and $\bm{u}_s \in [0, +\infty[^N$ such that $p \neq q \neq s$, and consider the operator $\widetilde{\bm{\mathsf{T}}}_{(\bm{u}_q, \bm{u}_s)} \colon \mathbb{R}^N \to \mathbb{C}^M$ defined by
\begin{equation} \label{eq:operator_T}
\widetilde{\bm{\mathsf{T}}}_{(\bm{u}_q, \bm{u}_s)}\bm{u}_p = \big[ (\widetilde{\bm{\mathsf{T}}}_1\bm{u}_q) \cdot (\widetilde{\bm{\mathsf{T}}}_2\bm{u}_s) \cdot (\widetilde{\bm{\mathsf{T}}}_3 \bm{u}_p) \big] .
\end{equation}
Then, the minimization of $h$ with respect to $\bm{u}_p$ (while $\bm{u}_q$ and $\bm{u}_s$ are fixed) can be rewritten as follows
\begin{equation}	\label{eq:convex}
\underset{\bm{u}_p \in \mathbb{R}^N}{\operatorname{minimize}}\quad \widetilde{f}_p (\bm{u}_p \, | \, \bm{u}_q, \bm{u}_s) + r(\bm{u_}p) ,
\end{equation}
where $r$ is given by \eqref{eq:reg_term1} and
\begin{equation} \label{eq:f_p}
\widetilde{f}_p(\bm{u}_p \, | \, \bm{u}_q, \bm{u}_s) = \frac12 \|\widetilde{\bm{y}} - \widetilde{\bm{\mathsf{T}}}_{(\bm{u}_q, \bm{u}_s)} \bm{u}_p \|_2^2  .
\end{equation}
Note that the data fidelity term $\widetilde{f}_p( \cdot  \, | \, \bm{u}_q, \bm{u}_s )$ defined by \eqref{eq:f_p} is a convex differentiable function, with its gradient given by
\begin{equation}
\nabla \widetilde{f}_p (\bm{u}_p \, | \, \bm{u}_q, \bm{u}_s) = \widetilde{\bm{\mathsf{T}}}_{(\bm{u}_q,\bm{u}_s)}^{\,\dagger} \big(\widetilde{\bm{\mathsf{T}}}_{(\bm{u}_q,\bm{u}_s)} \bm{u}_p - \widetilde{\bm{y}}\big).
\end{equation}
Moreover, $\nabla \widetilde{f}_p$ is $\kappa(\bm{u}_q, \bm{u}_s)$-Lipschitzian \citep[definition 1.46]{Bauschke2011}
with
\begin{equation} \label{eq:Lipschitz_const}
 \kappa(\bm{u}_q, \bm{u}_s)= \big\| \widetilde{\bm{\mathsf{T}}}_{(\bm{u}_q, \bm{u}_s)}^{\,\dagger} \widetilde{\bm{\mathsf{T}}}_{(\bm{u}_q, \bm{u}_s)} \big\|_S  ,
\end{equation}
$\| \cdot \|_S$ denoting the spectral norm.

\subsection{Choice of the regularization term} \label{ssec:choose_reg}

Concerning the choice of $g$ in \eqref{eq:reg_term1}, it is important to emphasize that astronomical images are usually sparse, otherwise they can have sparse representation \citep{Starck2010}. 
Mathematically, this means that the original image can be expressed as 
\begin{equation}
\overline{\bm{x}} = \bm{{\Psi}} \bm{\alpha}, 
\end{equation}
where $\bm{\Psi}\in \mathbb{R}^{N\times J}$ is a given dictionary such that $\overline{\bm{x}} $ is represented by a sparse vector $\bm{\alpha} \in \mathbb{R}^J$  in this dictionary.  
\modif{For instance, in the particular case when the image $\overline{\bm{x}}$ itself is assumed to be sparse (such as the point sources image), one can choose $\bm{\Psi}$ to be the identity matrix (i.e.  the Dirac basis).
%\bm{\Psi}$ is nothing but the identity matrix (known as the Dirac basis). 
More generally, for the sparse representation of continuous extended structures, $ \bm{\Psi}$ can be considered as wavelet basis \citep{Mallat2009}, a possibly redundant or a concatenation of non-redundant wavelet basis \citep{Carrillo2012}.} 

\modif{In this context,} the theory of compressive sensing has proven its worth in numerous cases to obtain a unique solution from a highly under-determined problem, relying on the sparsity of the underlying signal \citep{Wiaux2009,Duarte2011}. This drives us to use the regularization function $g$ in \eqref{eq:reg_term1} to promote sparsity in our minimization problem.
 
A natural way to find the sparsest solution is by considering regularization term of the form 
\begin{equation}
g(x) = \|\bm{\Psi}^\dagger \bm{x}\|_0,
\end{equation} 
where $\| \cdot \|_0$ denotes the $\ell_0$ pseudo norm counting the non-zero entries of its argument \citep{Donoho2006}. Note that in practice this function is difficult to manage due to its non-convexity and non-differentiability. 
Thus, non-convexity is often relaxed by the use of the $\ell_1$ norm 
\citep{Chen2001}, so that the sparsity prior is taken to be 
\begin{equation}	\label{eq:l1_reg}
g(x) = \|{\bm{\Psi} ^\dagger \bm{x}}\|_1.
\end{equation}  

However, unlike the $\ell_0$ pseudo norm, the $\ell_1$ norm is dependent on the magnitude of the coefficients of the signal. Thus, these last years, several approximations of the $\ell_0$ pseudo norm have been proposed \citep{Candes2008,Chouzenoux2013,Repetti2015}. 
In particular, as proposed in \citet{Candes2008}, $\ell_0$ minimization behaviour can be nicely approximated by reweighted-$\ell_1$ minimization. 
The authors have shown through several experiments that in many sparse signal recovery problems, reweighted-$\ell_1$ minimization can outperform $\ell_1$ minimization. In the context of radio interferometry, this has been demonstrated numerically by \citet{Carrillo2012}. Thus, in our approach we will consider both $\ell_1$ and reweighted-$\ell_1$ to promote sparsity. 
In the reweighted-$\ell_1$ method, a sequence of weighted-$\ell_1$ minimization problems is considered, i.e. problem \eqref{eq:overall_min} with
\begin{equation} \label{eq:reg_term}
g(x) =  \| \bm{\mathsf{W}} \bm{\Psi}^\dagger \bm{x} \|_1 \, ,
\end{equation}
where the weights ${\bm{\mathsf{W}}} = \operatorname{Diag}(w_1, \ldots, w_J)$, with $(w_j)_{1\le j \le J} \in ]0,+\infty[^J$, are computed from the current estimation of $\overline{\bm{x}}$. Note that in the case when $\bm{\mathsf{W}}$ is the identity matrix, the usual $\ell_1$ regularization \eqref{eq:l1_reg} is recovered. The calculation of weights will be discussed more in detail in Section~\ref{ssec:re_weight}.

%---------------------------------------------------------

\section{Proposed Algorithm} \label{sec:proposed_algo}
\subsection{Algorithm formulation} \label{ssec:algo_form}

In this section, we will describe more in detail the proposed alternating minimization algorithm to solve problem~\eqref{eq:overall_min}. 
We exploit the variable block structure described in Section~\ref{ssec:Alter_min} using a block-coordinate forward-backward algorithm \citep{Bolte2014,Frankel2015,Chouzenoux2016}. 
In this method, $\bm{u}_1$, $\bm{u}_2$ and $\bm{u}_3$ are updated sequentially, by solving \eqref{eq:convex}, as described in Algorithm~\ref{algo_blockfb}. 
More precisely, this algorithm consists in computing, at each iteration $k\in \mathbb{N}$, 
\begin{enumerate}[labelindent=0pt,labelwidth=\widthof{\ref{algo_desc:iii}},label=(\roman*),itemindent=1em,leftmargin=!]
\item \label{algo_desc:i}
$ \bm{u}_1^{(k+1)}$ while $ \big( \bm{u}_2^{(k)}, \bm{u}_3^{(k)}\big)$ are fixed, $\vspace*{0.1cm}$
\item \label{algo_desc:ii}
$ \bm{u}_2^{(k+1)}$ while $ \big( \bm{u}_1^{(k+1)}, \bm{u}_3^{(k)}\big)$ are fixed, $\vspace*{0.1cm}$
\item \label{algo_desc:iii}
$ \bm{u}_3^{(k+1)}$ while $ \big( \bm{u}_1^{(k+1)}, \bm{u}_2^{(k+1)} \big)$ are fixed. 
\end{enumerate}
The update of each variable $ \big(\bm{u}_p^{(k+1)}\big)_{1 \le p \le 3}$ is computed with the forward-backward iterations described in steps~\ref{algo:init}-\ref{algo:up_update} of Algorithm~\ref{algo_blockfb}. Each iteration involves alternating between 
\begin{itemize}[labelindent=0pt, labelwidth=\widthof{(\ref{algo:prox_step})}, itemindent=1em, leftmargin=!]
\item
Step~\ref{algo:grad_step}: 
gradient step (or forward step) on the corresponding differentiable function, i.e., 
$\widetilde{f}_1( \cdot \,|\, \bm{u}_2^{(k)}, \bm{u}_3^{(k)})$ for $\bm{u}_1$, 
$\widetilde{f}_2( \cdot \,|\, \bm{u}_1^{(k+1)}, \bm{u}_3^{(k)})$ for $\bm{u}_2$ 
and $\widetilde{f}_3( \cdot \,|\, \bm{u}_1^{(k+1)}, \bm{u}_2^{(k+1)})$ for $\bm{u}_3$, 
\vspace*{0.1cm}
\item
Step~\ref{algo:prox_step}:
proximity step (or backward step) on the non-necessarily smooth function $r$. 
\end{itemize}
The proximity operator of $r$ at $\bm{x} \in \mathbb{R}^N$ is defined as 
\begin{equation}
\operatorname{prox}_{r} (\bm{x}) 
= \underset{\bm{u}\in \mathbb{R}^N}{\operatorname{argmin}} \,\, r(\bm{u}) + \frac{1}{2}\|\bm{u}-\bm{x}\|_2^2 .
\label{eq:prox}
\end{equation}  
\modif{This operator has been introduced by \cite{Moreau1965} and extensively used in signal and image processing to deal with non-smooth functions \citep{Combettes2011}. An interesting fact concerning this operator is that it admits explicit formulae for a wide class of functions such as $\ell_p$ norms, for $p>0$. Finally, it can be seen as a generalization of the projection operator onto a non-empty closed convex set $\mathcal{C}$ when $r$ is chosen to be the indicator function of $\mathcal{C}$:
\begin{align}
P_\mathcal{C}(x)
& = \underset{\bm{u}\in \mathbb{R}^N}{\operatorname{argmin}} \,\, \iota_\mathcal{C}(\bm{u}) + \frac{1}{2}\|\bm{u}-\bm{x}\|_2^2, \nonumber\\
& = \underset{\bm{u}\in \mathcal{C}} {\operatorname{argmin}}\,\, \frac{1}{2}\|\bm{u}-\bm{x}\|_2^2\, ,
\label{eq:projection}
\end{align}
thus finding the closest point to $\bm{x}$ belonging to the set $\mathcal{C}$.}

\modif{Intuitively, the forward-backward iterations can be understood as follows: For each of the variables $\bm{u}_p$, consider the minimization problem~\eqref{eq:convex}. Here the objective function is a sum of smooth and non-smooth functions.
Firstly, a gradient step is performed on the differentiable function $\widetilde{f}_p (\cdot | \bm{u}_q, \bm{u}_s)$, giving $\bm{z}^{(t)}$ (step~\ref{algo:grad_step} in Algorithm~\ref{algo_blockfb}). 
Then a proximity step (step~\ref{algo:prox_step}) is applied to the non-smooth function $r$. Here in the computation of the proximity operator of $r$, the quadratic term controls the distance between the solution of this step and $\bm{z}^{(t)}$.
Finally, as a result of these forward-backward iterations, the solution obtained is basically the minimizer of the global objective function in~\eqref{eq:convex}.}

Note that, in Algorithm~\ref{algo_blockfb}, for every $k \in \mathbb{N}$, the gradient of $\widetilde{f}_1( \cdot \,|\, \bm{u}_2^{(k)}, \bm{u}_3^{(k)})$ (resp. $\widetilde{f}_2( \cdot \,|\, \bm{u}_1^{(k+1)}, \bm{u}_3^{(k)})$ and $\widetilde{f}_3( \cdot \,|\, \bm{u}_1^{(k+1)}, \bm{u}_2^{(k+1)})$) depends on the current iterates $(\bm{u}_2^{(k)}, \bm{u}_3^{(k)})$ (resp. $(\bm{u}_1^{(k+1)}, \bm{u}_3^{(k)})$ and $(\bm{u}_1^{(k+1)}, \bm{u}_2^{(k+1)})$). Thus, the linear operator $\widetilde{\bm{\mathsf{T}}}_{(\bm{u}_2^{(k)}, \bm{u}_3^{(k)})}$ (resp. $\widetilde{\bm{\mathsf{T}}}_{(\bm{u}_1^{(k+1)}, \bm{u}_3^{(k)})}$ and $\widetilde{\bm{\mathsf{T}}}_{(\bm{u}_1^{(k+1)}, \bm{u}_2^{(k+1)})}$) needs to be updated at each iteration $k \in \mathbb{N}$.

\begin{algorithm}
\caption{Block coordinate Forward-Backward algorithm}\label{algo_blockfb}
\begin{algorithmic}[1]
% -------------------------------------------------------------
\State \textbf{Initialization:} 
%$k=0$, 
Let $\bm{u}_1^{(0)} = \bm{u}_2^{(0)} = \bm{u}_3^{(0)} \in \mathbb{R}_+^{N}$,
$t_{\max} \in \mathbb{N}^*$, 
and, for every $k \in \mathbb{N}$, let $(\delta_1^{(k,t)})_{0 \le t \le t_{\max}-1}$, $(\delta_2^{(k,t)})_{0 \le t \le t_{\max}-1}$ and $(\delta_3^{(k,t)})_{0 \le t \le t_{\max}-1}$ be positive sequences.

\vspace*{0.3cm}

\State 
\textbf{For} $k = 0, 1, \ldots$

\State	\label{algo:start_for_subprob}
\quad \textbf{for} $p = 1, 2, 3$

\vspace*{0.1cm}

\State \label{algo:T1}
\quad\quad \textbf{if} $p = 1$ ; $\bm{\mathsf{T}} = \widetilde{\bm{\mathsf{T}}}_{(\bm{u}_2^{(k)},\bm{u}_3^{(k)})}$ ; \textbf{end if}

\State \label{algo:T2}
\quad\quad \textbf{if} $p = 2$ ; $\bm{\mathsf{T}} = \widetilde{\bm{\mathsf{T}}}_{(\bm{u}_1^{(k+1)},\bm{u}_3^{(k)})}$ ; \textbf{end if}

\State \label{algo:T3}
\quad\quad \textbf{if} $p = 3$ ; $\bm{\mathsf{T}} = \widetilde{\bm{\mathsf{T}}}_{(\bm{u}_1^{(k+1)},\bm{u}_2^{(k+1)})}$ ; \textbf{end if}

\State \label{algo:init}
\quad\quad $\widetilde{\bm{u}}^{(0)} = \bm{u}_p^{(k)}$

\vspace*{0.1cm}

\State \label{algo:fb_start}
\quad\quad \textbf{for} $t = 0, \ldots, t_{\max}-1$

\State	\label{algo:grad_step}
\quad\quad\quad 
$\bm{z}^{(t)} = \widetilde{\bm{u}}^{(t)} - \delta_p^{(k,t)} \bm{\mathsf{T}}^\dagger \big(\bm{\mathsf{T}} \widetilde{\bm{u}}^{(t)} - \bm{y}\big)$

\State	\label{algo:prox_step}
\quad\quad\quad 
$\widetilde{\bm{u}}^{(t+1)} = \operatorname{prox}_{\delta_p^{(k,t)} \, r} \big( \bm{z}^{(t)} \big)$

\State
\quad\quad \textbf{end for} \label{algo:fb_end}

\vspace*{0.1cm}

\State \label{algo:up_update}
\quad\quad $\bm{u}_p^{(k+1)} = \widetilde{\bm{u}}^{(t_{\max})}$

\vspace*{0.1cm}

\State	\label{algo:end_for_subprob}
\quad \textbf{end for}

\State
\textbf{end for}

\vspace*{0.3cm}

\State \textbf{Return:} $\bm{x}^{\star} = \big(\bm{u}_1^{\star} + \bm{u}_2^{\star} + \bm{u}_3^{\star} \big)/3$, where $\bm{u}_1^{\star} = \lim_k \bm{u}_1^{(k)}$, $\bm{u}_2^{\star} = \lim_k \bm{u}_2^{(k)}$, $\bm{u}_3^{\star} = \lim_k \bm{u}_3^{(k)}$.

\vspace{0.2cm}

\end{algorithmic}
\end{algorithm}

\subsection{Convergence Results} \label{ssec:conv}

The key point of the proposed Algorithm~\ref{algo_blockfb} is that its convergence can be derived from \citet{Bolte2014,Chouzenoux2016}. We present the convergence results in the following theorem:
\begin{theorem}
Let $(\bm{u}_1^{(k)})_{k \in \mathbb{N}}$,  $(\bm{u}_2^{(k)})_{k \in \mathbb{N}}$ and $(\bm{u}_3^{(k)})_{k \in \mathbb{N}}$ be sequences generated by Algorithm~\ref{algo_blockfb}. 
Assume that, for every $k\in \mathbb{N}$ and $t \in \{0, \ldots,t_{\max}-1 \}$,
\begin{equation}
\begin{cases}
\displaystyle \delta_1^{(k,t)} \in \Big] 0, {2}/{\kappa \big(\bm{u}_2^{(k)}, \bm{u}_3^{(k)}\big)} \Big[ \, ,	\vspace*{0.15cm} \\
\displaystyle \delta_2^{(k,t)} \in \Big] 0, {2}/{\kappa \big(\bm{u}_1^{(k+1)}, \bm{u}_3^{(k)}\big)} \Big[ \, , \vspace*{0.15cm} \\
\displaystyle \delta_3^{(k,t)} \in \Big] 0, {2}/{\kappa \big(\bm{u}_1^{(k+1)}, \bm{u}_2^{(k+1)}\big)} \Big[ \, , 
\end{cases}
\end{equation}
where $\kappa(\cdot,\cdot)$ is defined by \eqref{eq:Lipschitz_const}.
If $g$ is a semi-algebraic function\footnote{A function is semi-algebraic if its graph is a finite union of sets defined by a finite number of polynomial inequalities. Semi-algebraicity property is satisfied by a wide class of functions. In particular, it is satisfied by the different functions $g$ described in Section~\ref{ssec:choose_reg}.}, then $(\bm{u}_1^{(k)}, \bm{u}_2^{(k)}, \bm{u}_3^{(k)})_{k \in \mathbb{N}}$ converges to a critical point $(\bm{u}_1^{\star}, \bm{u}_2^{\star}, \bm{u}_3^{\star})$ of $h $, and $\big( h(\bm{u}_1^{(k)}, \bm{u}_2^{(k)}, \bm{u}_3^{(k)}) \big)_{k \in \mathbb{N}}$ is a non-increasing function converging to $h(\bm{u}_1^{\star}, \bm{u}_2^{\star}, \bm{u}_3^{\star})$. 
\end{theorem}

Note that, according to \citet{Chouzenoux2016}, to ensure the convergence of Algorithm~\ref{algo_blockfb}, $t_{\max}$ needs to be finite (and equal to $1$ in \citet{Bolte2014}).
In the limit case that $t_{\max} \to + \infty$, Algorithm~\ref{algo_blockfb} can be viewed as an approximated Gauss-Seidel algorithm (\citet{Zangwill1969}, \citet[Chap~7]{Ortega1970}, \citet[Chap.2]{Bertsekas1999}).
However, up to the best of our knowledge, the most general convergence results for the Gauss-Seidel method are presented in \citet{Tseng2001}, 
and require technical assumptions on $\widetilde{f}_p+r$ that are not necessarily satisfied in our minimization problem, due to the selection operators involved in \eqref{eq:data_mod}\footnote{In particular, convexity of sub-problems $\widetilde{f}_p+r$, $p\in \{1,2,3\}$ is not enough to ensure the convergence of the Gauss-Seidel algorithm \citep{Powell1973}.
}. 
Thus, it is important to note that our method is in contrast with the algorithm proposed by \citet{Auria2013}, where an approximated Gauss-Seidel method is adopted.

\subsection{Implementation details}
\label{ssec:implement}
As mentioned in Section~\ref{ssec:algo_form}, each sub-problem~\eqref{eq:convex} is solved using the forward-backward iterations.
At each sub-iteration $t \in \{0,...,t_{\max}-1\}$, for every $p \in \{1,2,3\}$, step~\ref{algo:prox_step} performs the proximity operator of $r$, computed as follows:
\begin{align}
%(\forall p \in \{1,2,3\})\quad
\widetilde{\bm{u}}^{(t+1)} 
&= \operatorname{prox}_{\delta_p^{(k,t)} \, r} (\bm{z}^{(t)}) 	\nonumber\\
%&= \underset{\bm{u}\in \mathbb{R}^N}{\operatorname{argmin}} \, \,r(\bm{u}) + \frac{1}{2} \|\bm{u}-\bm{z}^{(t)}\|_{2}^{2} \, , \nonumber
%\\
&=\underset{\bm{u}\in \mathbb{R}^N}{\operatorname{argmin}} \, \,\iota_{\mathbb{R}_+^{N}}(\bm{u}) + \zeta_p^{(k,t)} g(\bm{u})+ \frac{1}{2} \|\bm{u}-\bm{z}^{(t)}\|_{2}^{2} \,	,	 \label{eq:u_prox}
\end{align}
where $\zeta_p^{(k,t)} = \delta_p^{(k,t)}\mu$. The computation of the proximity operator in~\eqref{eq:u_prox} depends on the choice of $g$. It can either have an explicit formulation or need to be computed using sub-iterations. In the following, we briefly describe the proximity steps obtained for the different regularization terms $g$ discussed in Section~\ref{ssec:choose_reg}.

\subsubsection{Positivity and reality}
\label{sssec:impl:pos}
In \citet{Auria2013}, only positivity and reality constraints have been considered. Thus, the regularization term~\eqref{eq:reg_term1} corresponds to the case when $\mu = 0$. In this case, the proximity step~\ref{algo:prox_step} boils down to the projection of the current iterate onto the real positive set $\mathbb{R}^N_+$, and is given by
\begin{equation}	\label{eq:proj_real_pos}
\widetilde{\bm{u}}^{(t+1)} 
= \operatorname{Proj}_{\mathbb{R}^N_+} \big( \bm{z}^{(t)} \big)
=  \Big( \max \big\{ \operatorname{Re} ( z_n^{(t)} ), 0 \big\} \Big)_{1 \leq n \leq N} \, ,
\end{equation}
where $\operatorname{Re}(\cdot)$ denotes the real part operator.

\subsubsection{Positivity, reality and sparsity in the image space}
\label{sssec:impl:spars_im}
In the case when the original image is known to be sparse, function $g$ can be used to promote sparsity directly in the image space. This corresponds to regularization~\eqref{eq:l1_reg} (resp. \eqref{eq:reg_term}) with $\bm{\Psi}$ chosen equal to the identity matrix. Then, according to \citet[Table 10.2(ix)]{Combettes2010}, the proximity step~\ref{algo:prox_step} has an explicit formulation, given by
\begin{equation}
\widetilde{\bm{u}}^{(t+1)} 
= \big(p_n^{(t+1)}\big)_{1 \leq n \leq N},
\end{equation}
where, for every $n \in \{1 \leq n \leq N\}$,
\begin{equation}	
p_n^{(t+1)} = 
\begin{cases}
 \operatorname{Re} (z_n^{(t)}) - \omega_n	&	\text{if }  \operatorname{Re} (z_n^{(t)}) \ge  \omega_n	, \\
 0		&	\text{otherwise},
 \end{cases}
\end{equation}
with $\omega_n=\zeta_p^{(k,t)}$. 
%(resp. $\omega_n = w_n \zeta_p^{(k,t)}$).
\modif{This operator is called the \emph{positive soft- thresholding operator}. It involves setting all the components of $z_n^{(t)}$ smaller than the soft-thresholding parameter $\omega_n$ to zero, while the other components are reduced by $\omega_n$. Thus, this operator sparsifies the vector $z_n^{(t)}$, while imposing positivity and reality. }

\subsubsection{Positivity, reality and sparsity in a given dictionary} 
\label{sssec:impl:spars_dic}

As discussed in Section~\ref{ssec:choose_reg}, if an astronomical image is not sparse, it can have a sparse representation in a given dictionary $\bm{\Psi}$. In this case, regularization~\eqref{eq:l1_reg} or \eqref{eq:reg_term} can be used, where $\bm{\Psi}^\dagger$ is a general dictionary. However, the proximity operator~\eqref{eq:u_prox} does not have a closed form solution. Its computation in step~\ref{algo:prox_step} involves sub-iterations, which we propose to perform using the so-called \emph{dual forward-backward algorithm} \citep{Combettes2010,Combettes2011}, described in Algorithm~\ref{algo_dualfb}.

\begin{algorithm}
\caption{Dual Forward-Backward algorithm to compute \eqref{eq:u_prox}}
\label{algo_dualfb}
\begin{algorithmic}[1]

\State \textbf{Initialization:} Let $\widetilde{\bm{p}}^{\text{(0)}}\in\mathbb{R}^N$, $\epsilon\in\, ]0, \text{min}\{1, 1/\|\mathsf{W\Psi}^\dagger\|^2\}[$, and ${\gamma \in [\epsilon , 2/\|\mathsf{W\Psi}^\dagger\|^2 - \epsilon]}$.

\vspace{0.1cm}

\State 
\textbf{For} $\ell = 0, 1, \ldots$

\vspace{0.1cm}

\State \label{algoDFB:proj}
\quad $\bm{v}^{(\ell)} = \text{Proj}_{\mathbb{R}_+^{N}} \big(\bm{z}^{(t)} - \bm{\Psi} \bm{\mathsf{W}}^\dagger \widetilde{\bm{p}}^{(\ell)}\big)$

\vspace{0.1cm}

\State \label{algoDFB:change_dom}
\quad $\bm{s}^{(\ell)} = \widetilde{\bm{p}}^{(\ell)} + \gamma \bm{\mathsf{W}}\bm{\Psi}^\dagger \bm{v}^{(\ell)} $

\vspace{0.1cm}

\State \label{algoDFB:prox}
\quad $\widetilde{\bm{p}}^{(\ell+1)} = \bm{s}^{(\ell)} - \gamma \, \operatorname{prox}_{\gamma^{-1} \zeta_p^{(k,t)} g}\big(\gamma^{-1} \bm{s}^{(\ell)}\big)$

\vspace{0.1cm}

\State 
\textbf{end for}

\vspace{0.1cm}

\State 
\textbf{Return:} $ \widetilde{\bm{u}}^{(t+1)} =  \lim_\ell \bm{v}^{(\ell)}$.

\end{algorithmic}
\end{algorithm}

In the Algorithm~\ref{algo_dualfb}, $\bm{\mathsf{W}}$ is the identity matrix if the $\ell_1$ regularization \eqref{eq:l1_reg} is used, or $\bm{\mathsf{W}}$ corresponds to a diagonal matrix with positive weights $(w_1, \ldots, w_J)$ if weighted-$\ell_1$ regularization \eqref{eq:reg_term} is chosen. Moreover, step~\ref{algoDFB:proj} is computed using definition~\eqref{eq:proj_real_pos} in the image space, while the proximity operator in step~\ref{algoDFB:prox} corresponds to the \emph{soft-thresholding operator} \citep{Chaux2007} computed in the dictionary space. It is given by
\begin{equation}
\operatorname{prox}_{\gamma^{-1} \zeta_p^{(k,t)} g}\big(\gamma^{-1} \bm{s}^{(\ell)}\big)
= \big( p_j^{(\ell)} \big)_{1 \le j \le J},
\end{equation}
where, for every $j \in \{1,\ldots,J\}$, $p_j^{(\ell)}$ is defined by 
\begin{equation}	
p_j^{(\ell)} = 
\begin{cases}
\gamma^{-1} \big(s_j^{(\ell)} + \omega_j\big)
 &	\text{if } s_j^{(\ell)} < -\omega_j	, \\
 0		
&	\text{if } -\omega_j \le s_j^{(\ell)} \le \omega_j,	\\
\gamma^{-1} \big(s_j^{(\ell)} - \omega_j\big)
&	\text{otherwise},
 \end{cases}
\end{equation}
where $\omega_j = \zeta_p^{(k,t)}$ if regularization \eqref{eq:l1_reg} or \eqref{eq:reg_term} is considered. \modif{The \emph{soft-thresholding operator} sparsifies the vector $\gamma^{-1} \bm{s}^{(\ell)}$, by setting all its components satisfying $\big| \gamma^{-1} s_j^{(\ell)}\big| \leq \gamma^{-1}\omega_j $ to zero. Note that unlike the \emph{positive soft-thresholding operator} described in Section~\ref{sssec:impl:spars_im}, it does not impose positivity. }

\vspace*{-0.2cm}
\subsection{Reweighting approach} \label{ssec:re_weight}

As discussed in Section~\ref{ssec:choose_reg}, we propose to use a reweighted-$\ell_1$ regularization term to promote sparsity. 
In particular, we propose to compute the weights $\bm{\mathsf{W}}$ in \eqref{eq:reg_term} according to the weighting procedure developed in \citet{Candes2008}.
More precisely, let $\bm{x}^{\star}$ be a critical point obtained from Algorithm~\ref{algo_blockfb}, 
where the function $r$ is defined by \eqref{eq:reg_term1} with, either $\mu=0$, or $g$ given by an $\ell_1$ regularization function \eqref{eq:l1_reg}. 
Then, $\bm{x}^{\star}$ is used to compute the weights for the first weighting procedure, essentially computed from  the inverse of the values of $\bm{\Psi}^\dagger \bm{x}^{\star}$:
\begin{equation} \label{eq:weights}
(\forall j \in \{1,\ldots,J\})\quad
w_{j} = \dfrac{1}{\epsilon + |[\bm{\Psi}^\dagger \bm{x}^{\star}]_j|} \, ,
\end{equation}
where $\epsilon>0$, and $[\bm{\Psi}^\dagger \bm{x}^{\star}]_j$ denotes the $j$-th component of $\bm{\Psi}^\dagger \bm{x}^{\star}$ (if $\bm{\Psi}$ is chosen to be identity, or if $\mu=0$, then $J=N$ and $w_n = \frac{\epsilon}{\epsilon + |x^{\star}_n|}$). 
Note that $\epsilon$ in \eqref{eq:weights} can be viewed as a stabilization parameter (see \citet[Sect. 2]{Candes2008}). 
In particular, choosing $\epsilon \to 0$ leads to an approximation of the $\ell_0$ pseudo norm, limiting the dependence of the weighted-$\ell_1$ regularization on the magnitude of the signal coefficients.

Finally, Algorithm~\ref{algo_blockfb} is used again to solve the new minimization problem, taking into account the weighted-$\ell_1$ regularization \eqref{eq:reg_term} with weights computed by \eqref{eq:weights}. 
The new solution $\bm{x}^{\star}_1$ obtained from the weighted-$\ell_1$ minimization problem can be used to compute new weights from \eqref{eq:weights}, where $\bm{x}^{\star}$ is replaced by $\bm{x}^{\star}_1$. The reweighted-$\ell_1$ minimization problem obtained can be solved in turn using Algorithm~\ref{algo_blockfb}. This reweighting procedure can be repeated until a stable solution is obtained.

%-----------------------------------------------------------

\section{Simulations and Results} \label{sec:simulations}

In this section, to show the good behaviour of the proposed method, we will present simulation results, obtained by implementing the proposed algorithm in \textsc{matlab} [version R2015a].

\subsection{Simulation setting} \label{ssec:simulation_setting}

\begin{figure}
\centering
\includegraphics[width=3.7cm]{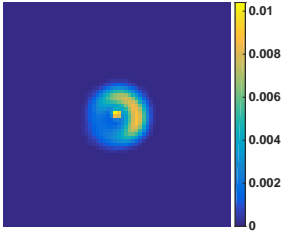}
\caption{Original image \texttt{LkH$\alpha$}, of size $64 \times 64$, used for simulations, taken from the 2004 Imaging Beauty Contest \citep{Lawson2004}.}
\label{fig:true_image}
\end{figure}

\begin{figure}
\centering
\begin{tabular}{c c}
\hspace*{-0.4cm}\includegraphics[width=3.5cm]{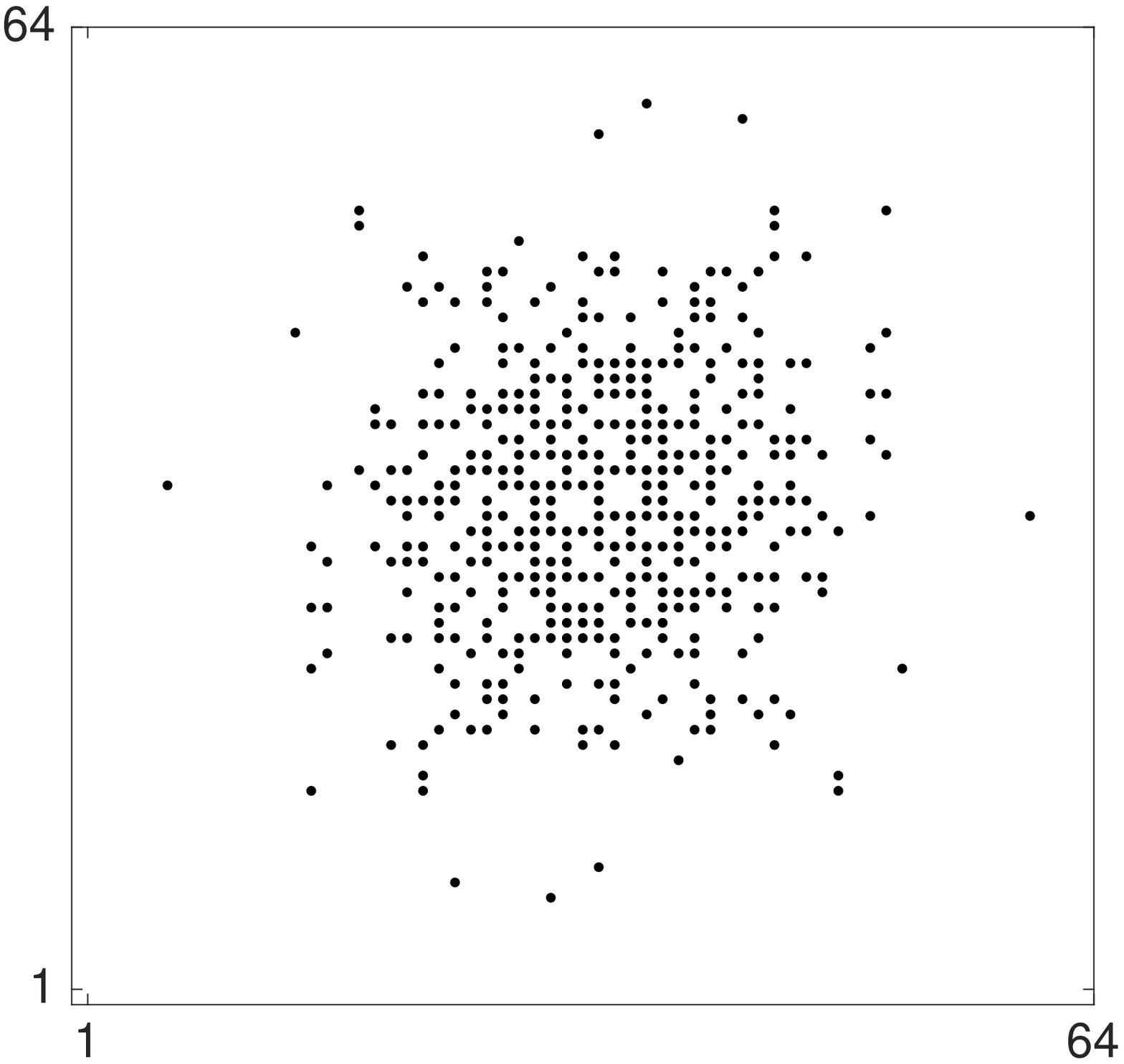} &
\hspace*{-0.2cm}\includegraphics[width=3.5cm]{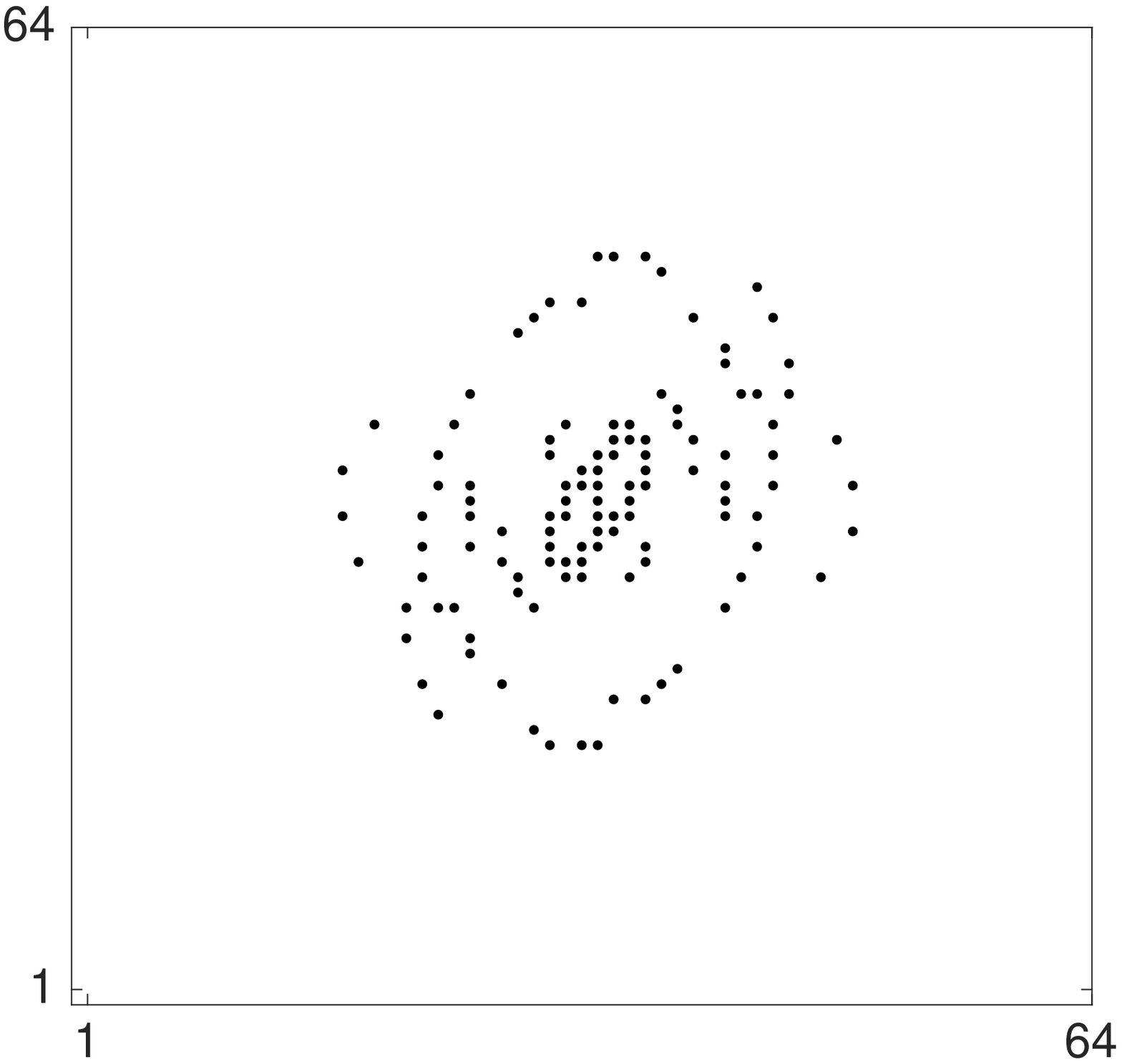} \\
(a) & (b)
\end{tabular}
\caption{Discretized spatial frequencies coverage plans for the image of size $64 \times 64$. (a) Synthetic $u-v$ coverage for $M_{\pows}/N = 0.05$: consists of random variable-density sampling scheme. (b) Realistic $u-v$ coverage: taken from the 2016 Interferometric Imaging Beauty Contest \citep{Sanchez2016}.}
\label{fig:uv_cover}
\end{figure}

All the simulations are performed on the image \texttt{LkH$\alpha$} shown in Figure \ref{fig:true_image}, taken from the 2004 Optical Interferometric Imaging Beauty Contest \citep{Lawson2004}, with $N = 64^2$. 
Two types of $u-v$ coverages are considered:
\begin{itemize}
\item 
Figure~\ref{fig:uv_cover}(a):
Synthetic $u-v$ coverage, which consists of random variable-density sampling scheme in 2D discrete Fourier space. 
In this case, the $u-v$ coverage is generated by random Gaussian sampling such that low frequencies are more likely to be sampled than high frequencies. 
\item 
Figure~\ref{fig:uv_cover}(b): 
Realistic $u-v$ coverage, corresponding to discretized version of 2016 Optical Interferometric Imaging Beauty Contest coverage plan \citep{Sanchez2016}. 
It corresponds to the measurements made by the GRAVITY instrument at the Very Large Telescope Interferometer (VLTI)\footnote{https://www.eso.org/sci/facilities/paranal/telescopes/vlti.html}. The observation wavelength is $1.95 \mu m$. It samples 72 points in the $u-v$ plane resulting into 72 power spectrum measurements. 
\end{itemize}

For both coverages, the bispectrum points are chosen at random, relaxing the phase closure constraint, mainly from the low frequency region. It is taken care that no two bispectrum measurements correspond to the same triple product.

In both the cases, the simulated measurements in \eqref{eq:data_mod} are obtained by taking the input signal-to-noise ratio (iSNR) equal to 30 dB, where
\begin{equation}
\displaystyle \text{iSNR} = 20 \, \log_{10} \left(\dfrac{\| \bm{y} \|_2}{\sqrt{M} \sigma_\eta}\right)  ,
\end{equation}
$\sigma_{\eta}^{2}$ being the variance of the noise.


For quantitative comparison of the reconstructed images, signal-to-noise ratio (SNR) is considered. For a given estimated image $\bm{x}^{\star}$ of an original image $\overline{\bm{x}}$, the SNR is defined as
\begin{equation}
\text{SNR} = 20 \log_{10} \left(\frac{||\overline{\bm{x}}||_2}{\|\bm{x}^{\star} - \overline{\bm{x}}\|_2}\right). 
\end{equation}
In our simulations, results are presented considering a stopping criterion for Algorithm~\ref{algo_blockfb}, given by 
$\underset{p \in \{1,2,3\}}{\max} \big(\|\bm{u}_p^{(k)} - \bm{u}_p^{(k-1)}\|_2 / \|\bm{u}_p^{(k)}\|_2\big) \le 10^{-2}$.

Finally, let us define the power spectrum undersampling ratio as $u_{\pows} = M_{\pows}/N$, and the bispectrum undersampling ratio as $u_{\bis} = M_{\bis}/N$. 
Note that due to the Hermitian symmetry, $M_{\pows}$ power spectrum measurements in fact correspond to $2 M_{\pows}$ sampled spatial frequencies in the Fourier plane. This implies that in the particular case when $u_{\pows} = 0.5$, all the spatial frequencies in the Fourier plane are sampled. 

As discussed in Section~\ref{sec:optical_imaging}, the number of spatial frequencies probed $M_{\pows}$ depends on the number of telescopes $A$. 
Thus, $u_{\pows}$ will change, depending on $A$. 
Also for a given $u_{\pows}$, there can be at most ${A \choose 3}$ possible bispectrum measurements, i.e. $M_{\bis} \le {A \choose 3}$. Keeping this in mind, for a fixed $u_{\pows}$, we have performed simulations by varying the number of bispectrum measurements considered, which results into varying $u_{\bis}$.
Furthermore, for each pair $(u_{\pows}, u_{\bis})$, 10 simulations are performed, varying the noise realization, and, for the synthetic $u-v$ coverage, the sampling pattern as well.

\subsection{Synthetic $u-v$ coverage}
\label{Ssec:synt_uv}

This section presents the simulations performed on the image \texttt{LkH$\alpha$} considering the synthetic $u-v$ coverage given in Figure~\ref{fig:uv_cover}(a). Simulations corresponding to the different regularization terms are described below.

\subsubsection{Positivity and reality constraints} 
\label{Sssec:case1}
We consider the simplest case, described by \citet{Auria2013}, corresponding to the minimization problem~\eqref{eq:overall_min} with only positivity and reality constraints taken into account. 
Details of the implementation of the Algorithm~\ref{algo_blockfb} in this case are described in Section~\ref{sssec:impl:pos}. 

As mentioned in Section~\ref{ssec:conv}, given the non-convexity of the minimization problem~\eqref{eq:overall_min}, 
Algorithm~\ref{algo_blockfb} can only converge to a critical point of $h$. Thus, the reconstructed image depends on the initialization. 
To avoid local minima, we propose to run Algorithm~\ref{algo_blockfb} several times, for $I$ random initializations $\bm{x}_i^{(0)} = \bm{u}_1^{(0)} = \bm{u}_2^{(0)} = \bm{u}_3^{(0)}$, with $i\in \{1, \ldots, I\}$. 
Let $\bm{x}^{\star}_{i}$ be the estimation found with initialization $\bm{x}_i^{(0)}$. 
Then the best estimation $\bm{x}^{\star}$ is selected by taking $\bm{x}^{\star} = \bm{x}^{\star}_{i^{\star}}$, where $i^{\star}$ corresponds to the initialization index with minimum value of the objective function, i.e. for every $ i \in \{1, \ldots, I\} $, $ f(\bm{x}^{\star}_{i^{\star}}) + r(\bm{x}^{\star}_{i^{\star}}) \le f(\bm{x}^{\star}_{i}) + r(\bm{x}^{\star}_{i}) $.

\begin{figure}
\vspace*{-0.2cm}
\centering
\hspace*{-0.1cm}\includegraphics[width=1.05\columnwidth]{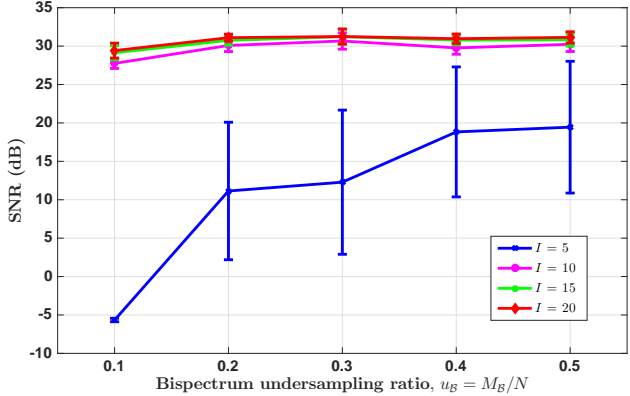}
\vspace*{-0.3cm}
\caption{SNR graph obtained for positivity and reality constrained case with \texttt{LkH$\alpha$} image and synthetic $u-v$ coverage for $u_{\pows} =  0.2$, considering iSNR = 30 dB and varying $u_{\bis}$. The graph shows the comparison of average SNR values (over 10 simulations), and corresponding 1-standard-deviation error bars, for different number of initializations $I$: $I=5$ (blue), $I=10$ (pink), $I=15$ (green), and $I=20$ (red). }
\label{fig:diff_init}
\end{figure}

To choose the number $I$ of random initializations, first tests for different $I$ are performed and presented in Figure~\ref{fig:diff_init}. 
Four curves are depicted, corresponding to the different number of initializations considered, $I \in \{5, 10, 15, 20\}$. Each curve represents the average SNR values over 10 simulations, along with 1-standard-deviation error bars, as a function of the undersampling ratio $u_{\bis}$, for a fixed $u_{\pows} =  0.2$.
It can be seen from the graph that the SNR changes a lot as the number of random initializations increases from 5 to 20. It reflects the sensitivity of the minimization problem to the number of initializations. However, between 15 and 20 initializations, the SNR not only saturates, in fact it exhibits very small standard deviation error bars. 
Thus, in all the subsequent simulations, when only positivity and reality constraints are taken into account, we consider $I=15$ random initializations for each pair $(u_{\pows},u_{\bis})$.

\subsubsection{$\ell_1$ and weighted-$\ell_1$ regularizations} 
\label{Sssec:cases2_3}
In order to solve the minimization problem~\eqref{eq:overall_min} promoting sparsity, 
we consider the regularization function given by~\eqref{eq:reg_term1}, and
we examine both $\ell_1$ and weighted-$\ell_1$  regularizations defined respectively by \eqref{eq:l1_reg} and \eqref{eq:reg_term}, 
using $\bm{\Psi}$ to be Daubechies~8 wavelet basis \citep{Mallat2009}. 
In this case, we use Algorithm~\ref{algo_blockfb} with the implementation details given in Section~\ref{sssec:impl:spars_dic}, and the reweighting process described in Section~\ref{ssec:re_weight}.

%----------------------------------------------------------------------------
\begin{figure*} 
\begin{tabular}{@{}c@{}c@{}}
\includegraphics[width = 1.05\columnwidth]{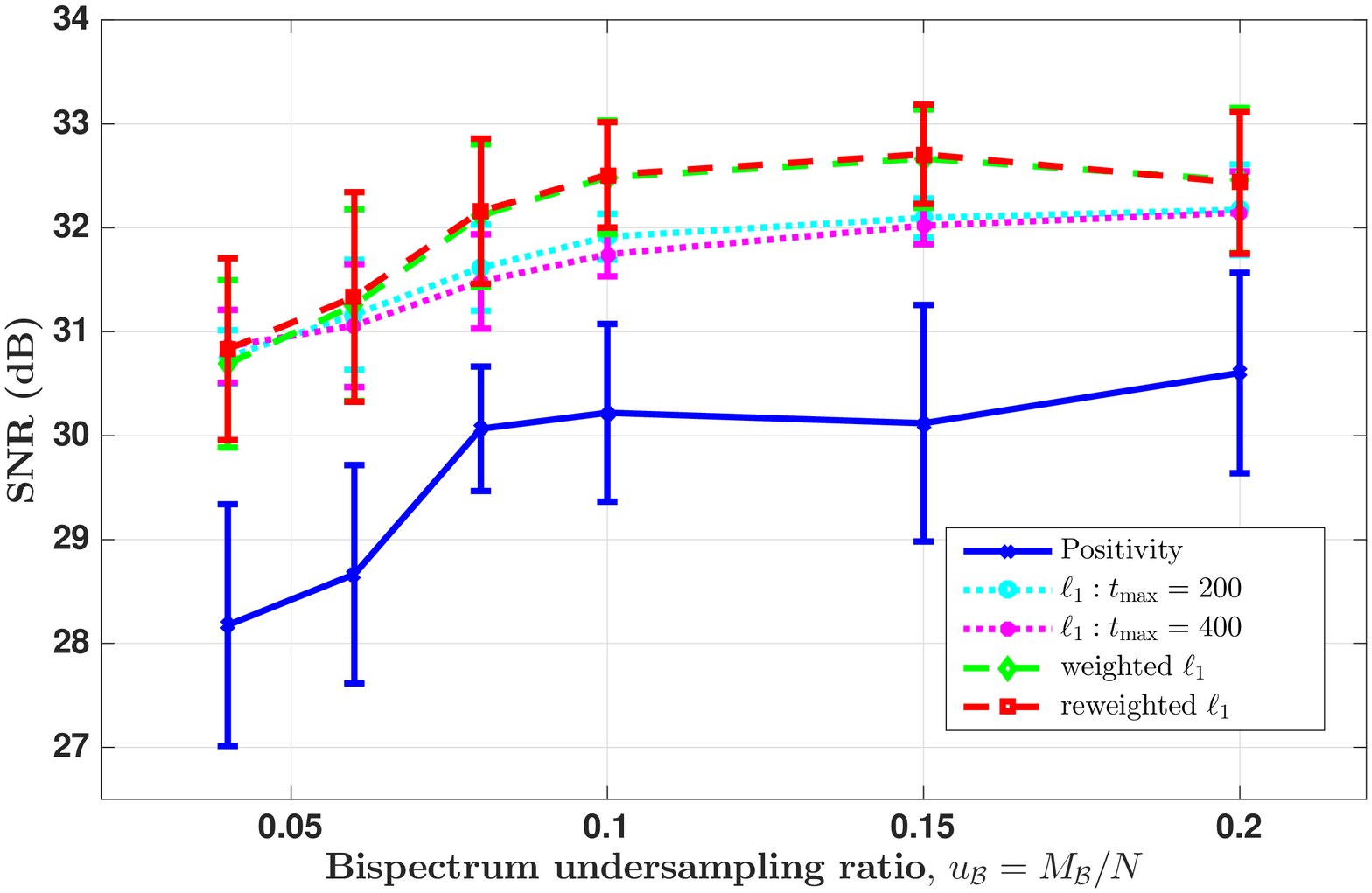}	
& \includegraphics[width = 1.05\columnwidth]{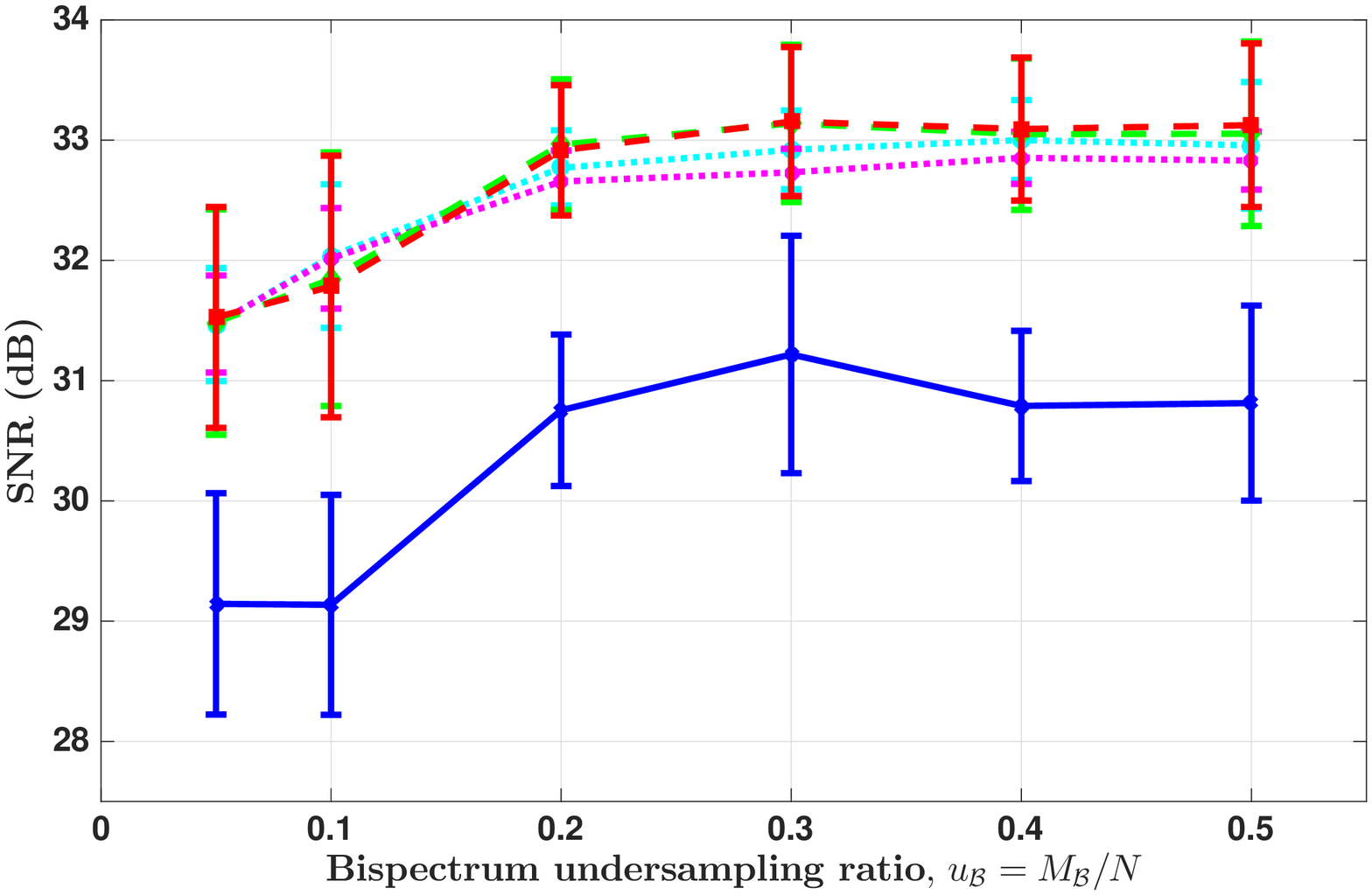} \\
{(a)} 
&	{(b)}
\end{tabular}
\vspace*{-0.1cm}
\caption{ SNR graphs obtained with \texttt{LkH$\alpha$} image and synthetic $u-v$ coverage, considering iSNR = 30 dB, varying $u_{\bis}$ for two different power spectrum undersampling ratios: $\textbf{(a)} \, u_{\pows} = 0.05$ and $\textbf{(b)} \, u_{\pows} = 0.2$. In each graph, comparison of average SNR values (over 10 simulations), and corresponding 1-standard-deviation error bars, for different regularization terms is shown: positivity constraints (solid blue), $\ell_1$ regularization with $t_{\max} = 200$ (dotted cyan) and $t_{\max} = 400$ (dotted pink),  weighted-$\ell_1$ regularization (dashed green) and reweighted-$\ell_1$ regularization (dashed red).} 
\label{fig:synthetic_plots}
\end{figure*}

%----------------------------------------------------------------------------
\begin{figure}
\centering
\begin{tabular}{c c}
\includegraphics[width = 3.7cm]{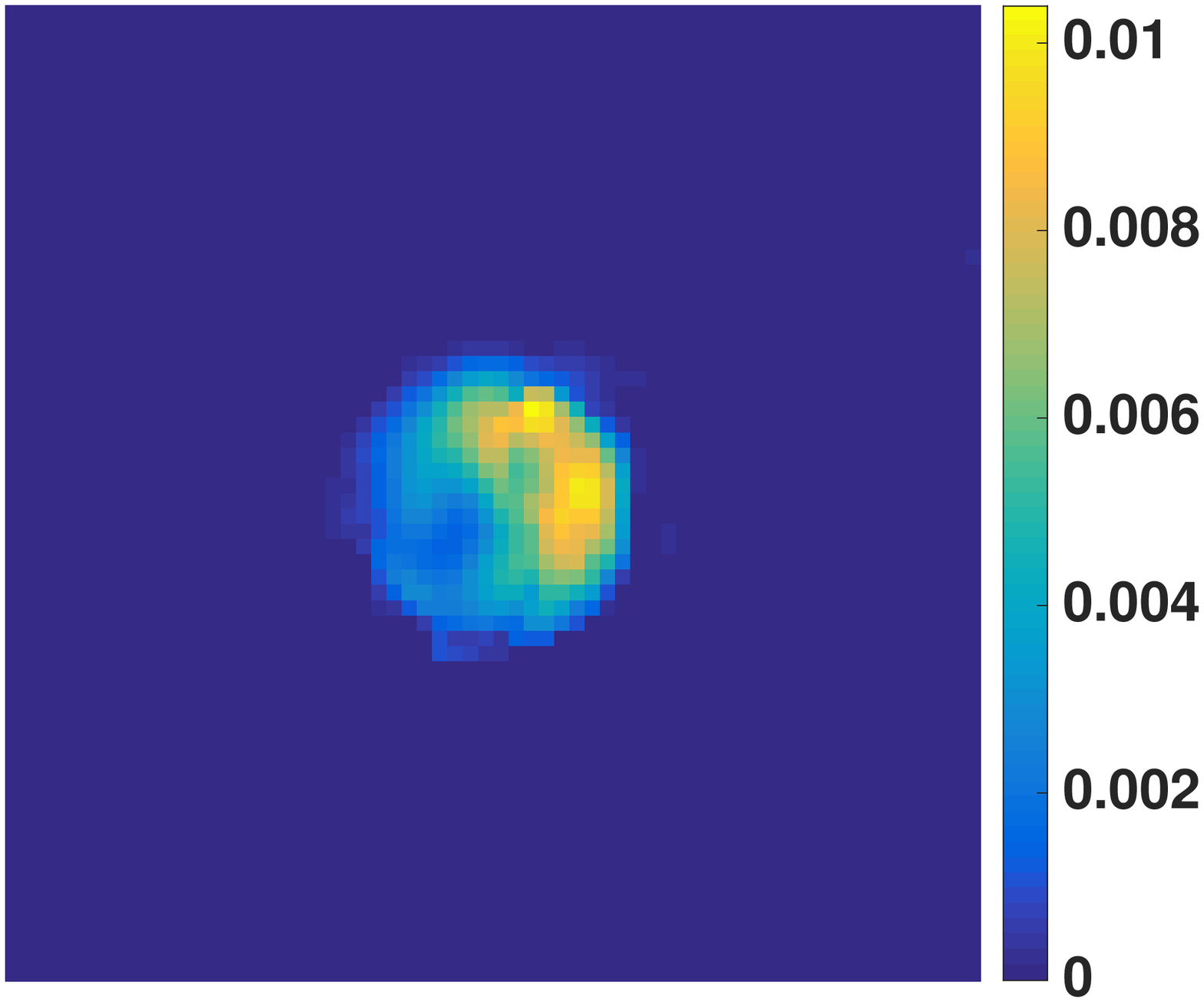} &
\hspace*{-0.2cm}\includegraphics[width = 3.7cm]{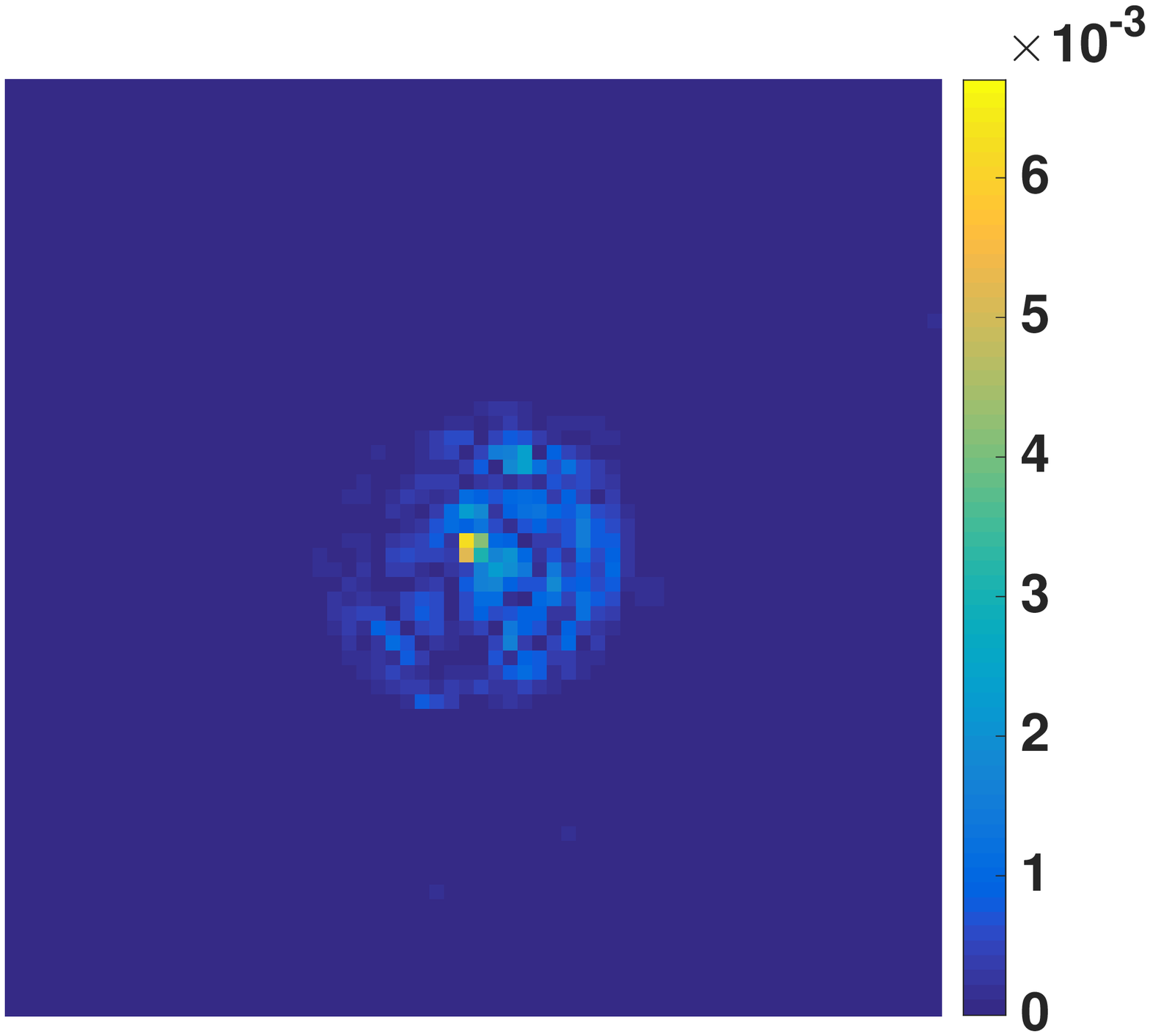} \\ 
\includegraphics[width = 3.7cm]{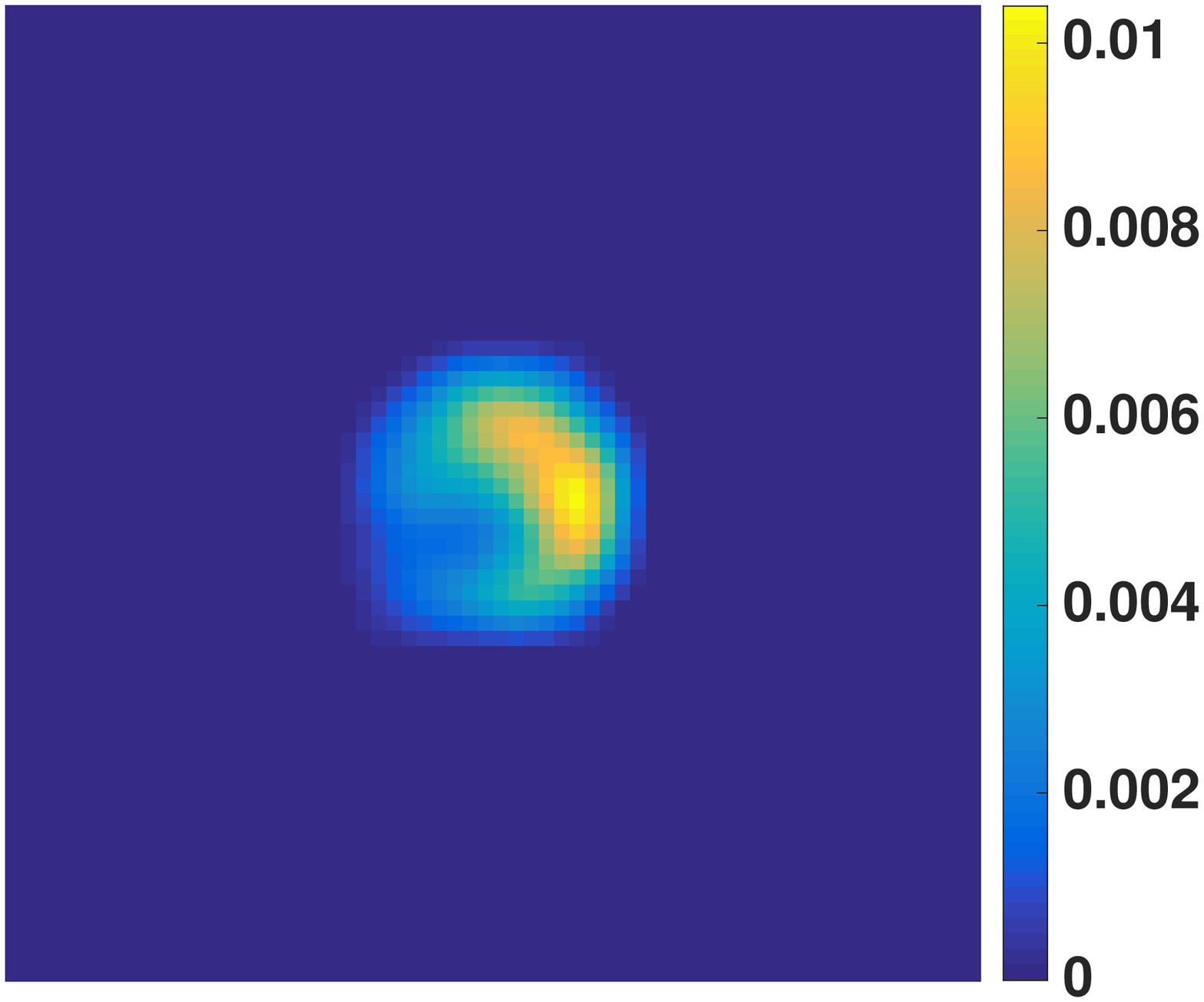} & 
\hspace*{-0.2cm}\includegraphics[width = 3.7cm]{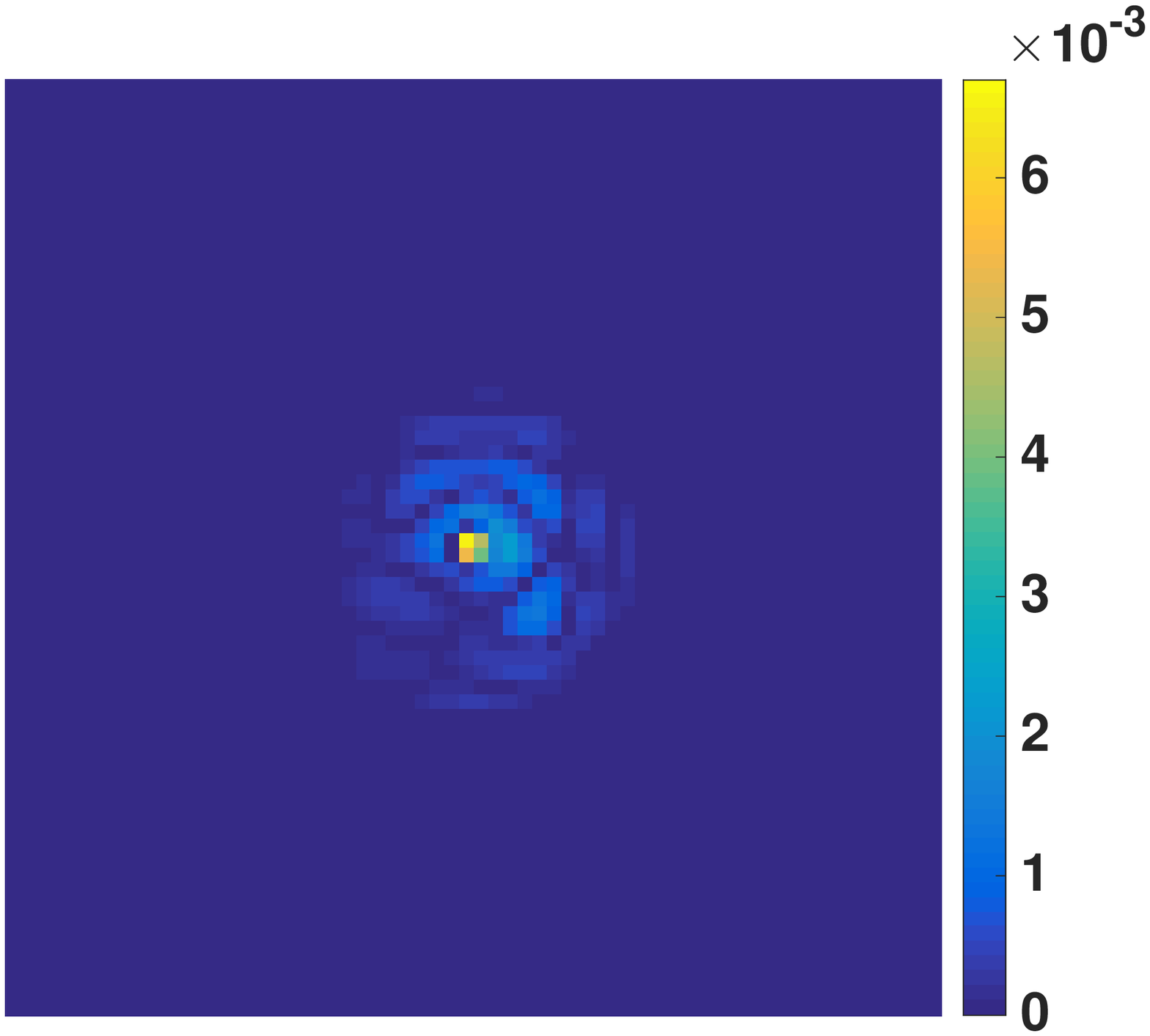} \\
\includegraphics[width = 3.7cm]{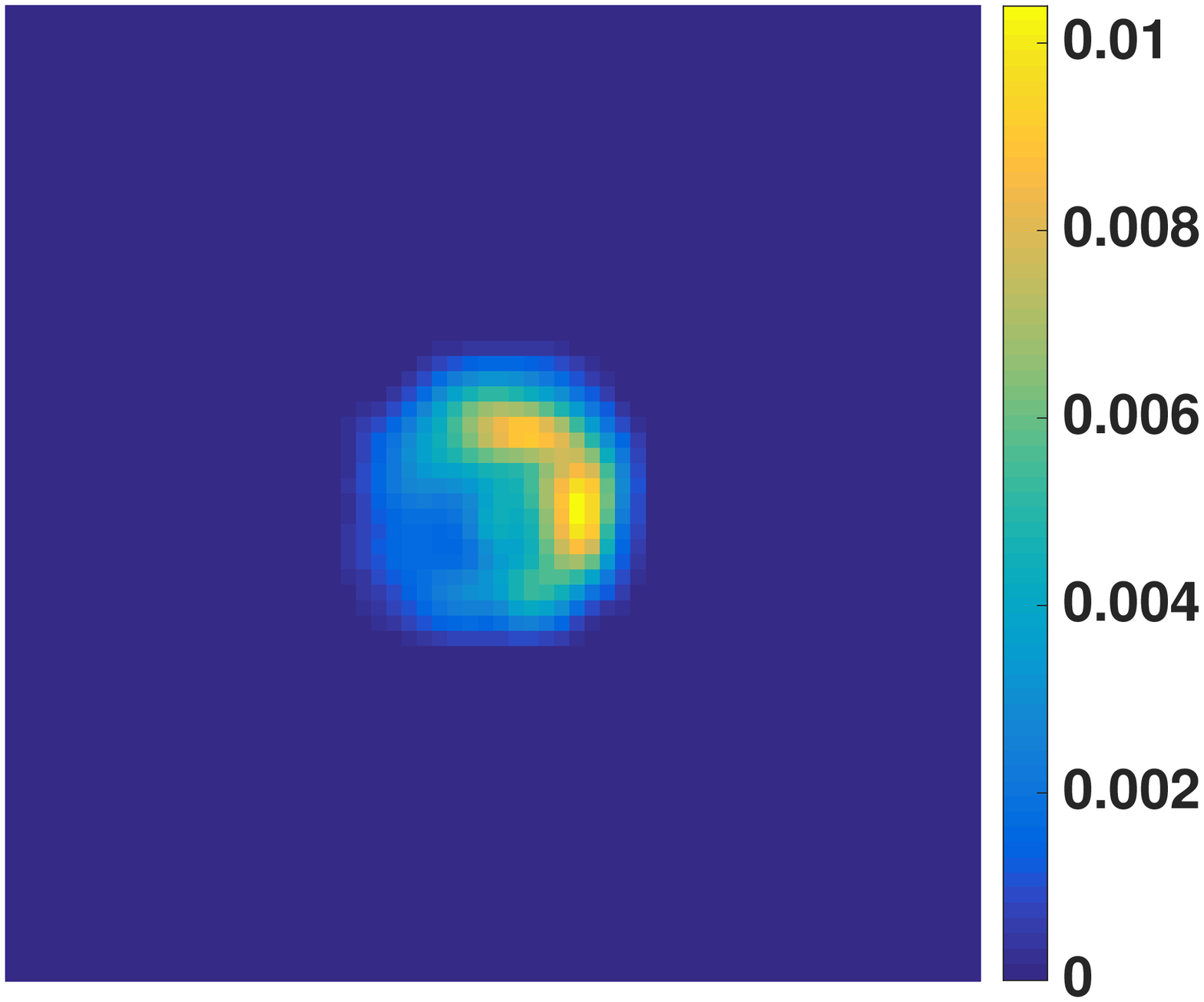} &
\hspace*{-0.2cm}\includegraphics[width = 3.7cm]{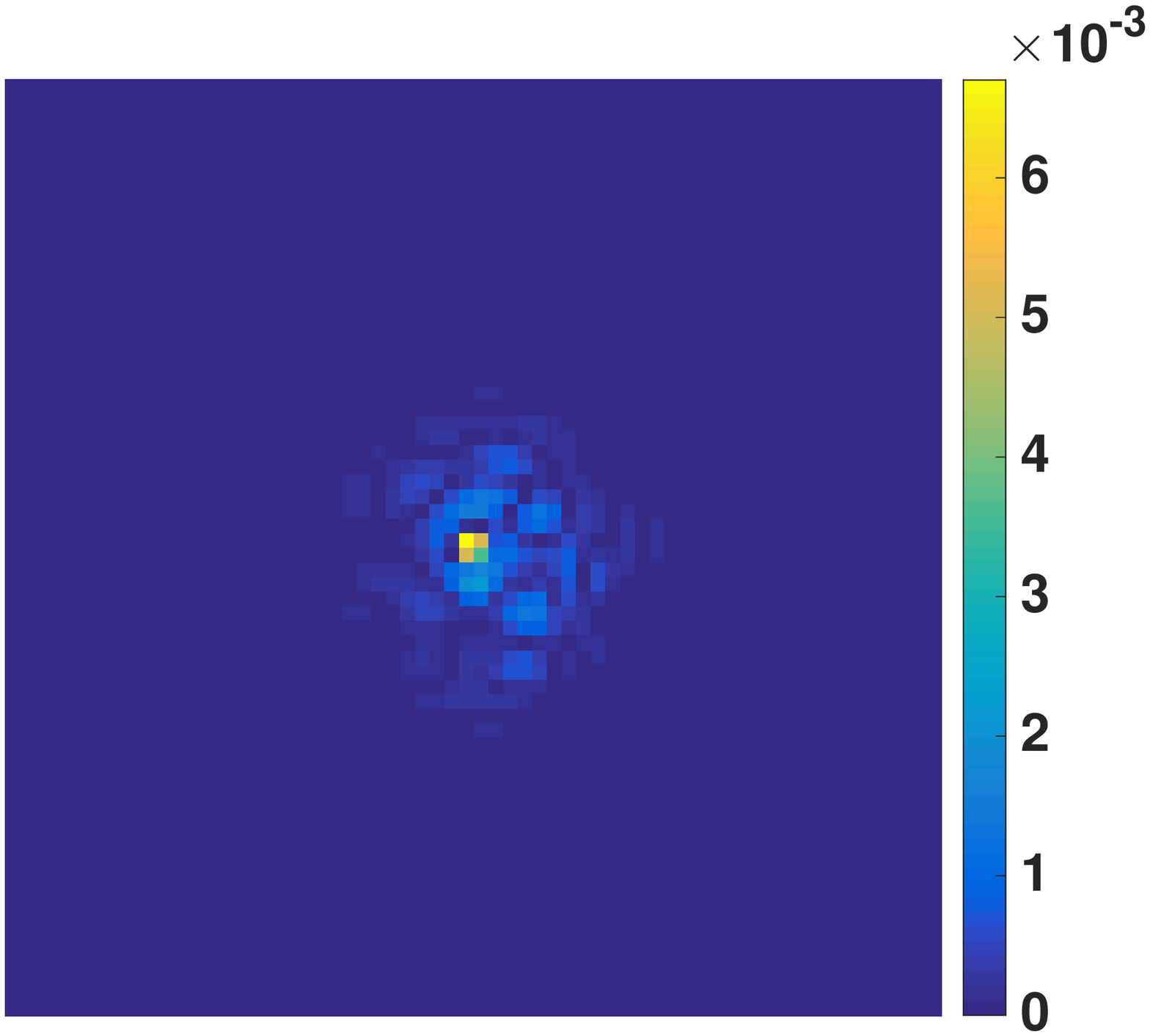}
\end{tabular}
\caption{
Reconstructed images (first column) and error images (second column) obtained by considering the true image \texttt{LkH$\alpha$}, corresponding to median SNR (over 10 simulations), with synthetic $u-v$ coverage for $(u_{\pows},u_{\bis})  = (0.05, 0.1)$. For both the columns, in each row, images corresponding to different regularization terms are shown- First row: positivity constraint, second row: $\ell_1$ regularization with $t_{\max} = 200$, and third row: reweighted-$\ell_1$ regularization.}
\label{fig:images_synthetic_uf1}
\end{figure}

\begin{figure}
\centering
\begin{tabular}{c c}
\includegraphics[width = 3.7cm]{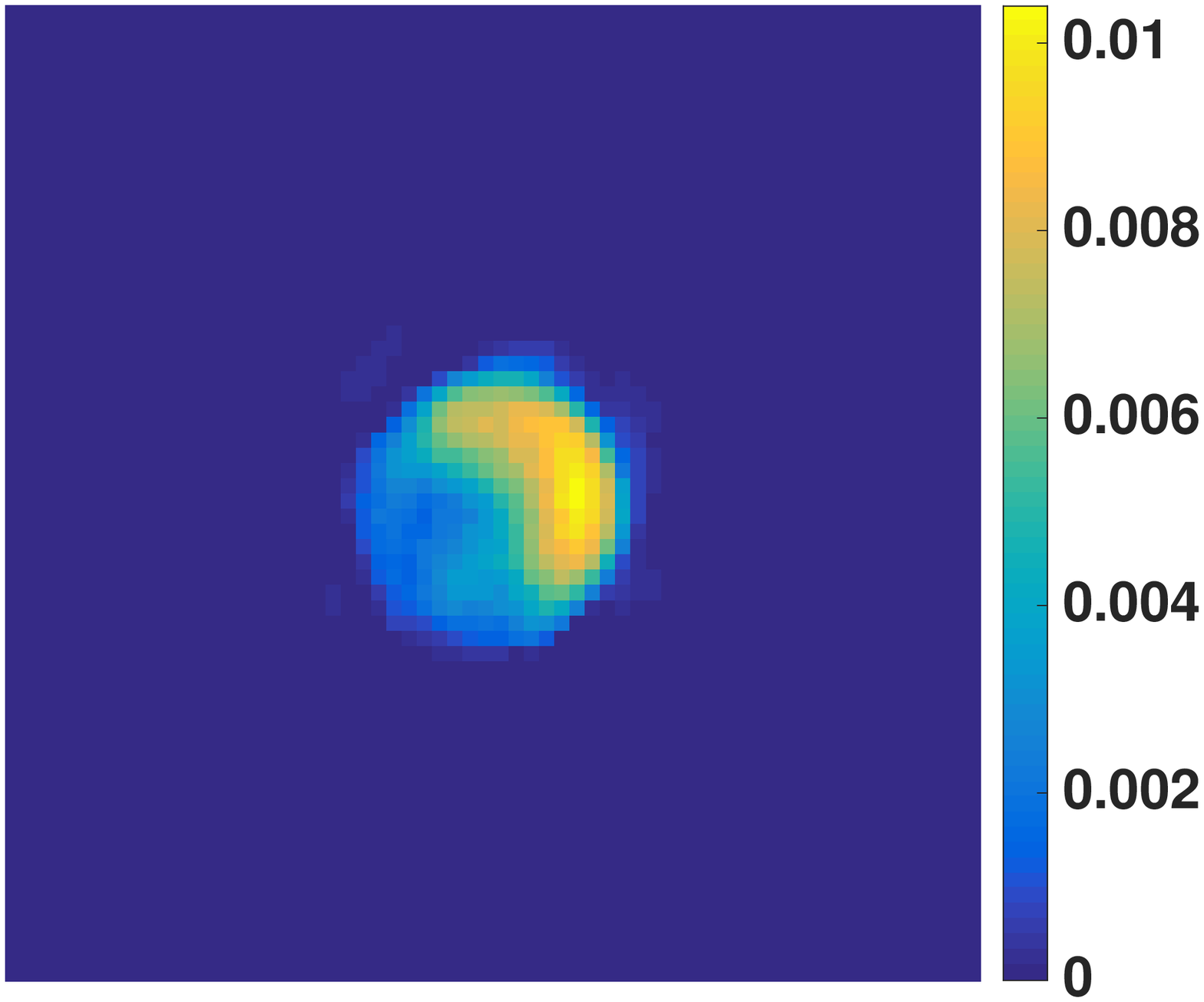} &
\hspace*{-0.2cm}\includegraphics[width = 3.7cm]{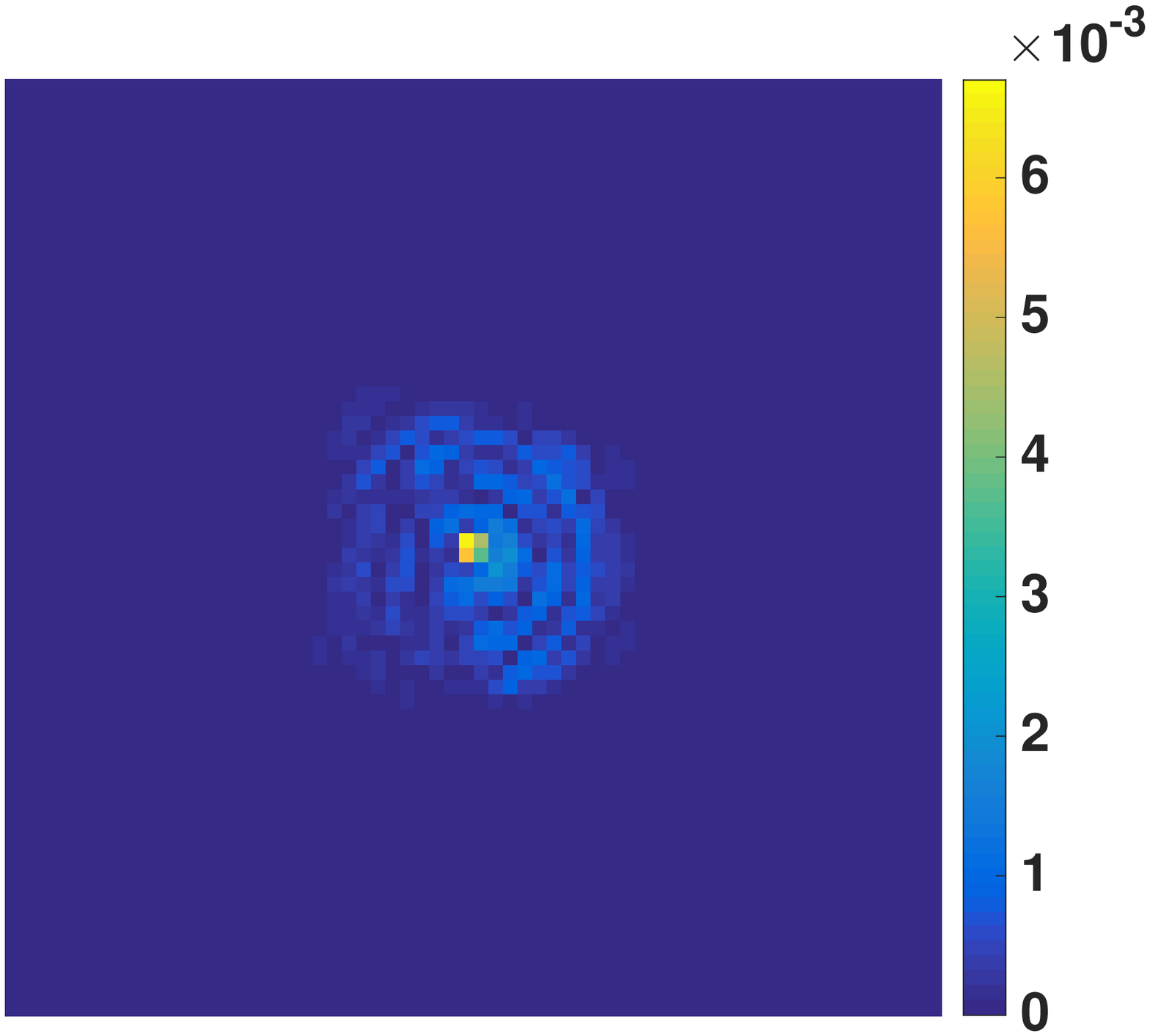} \\ 
\includegraphics[width = 3.7cm]{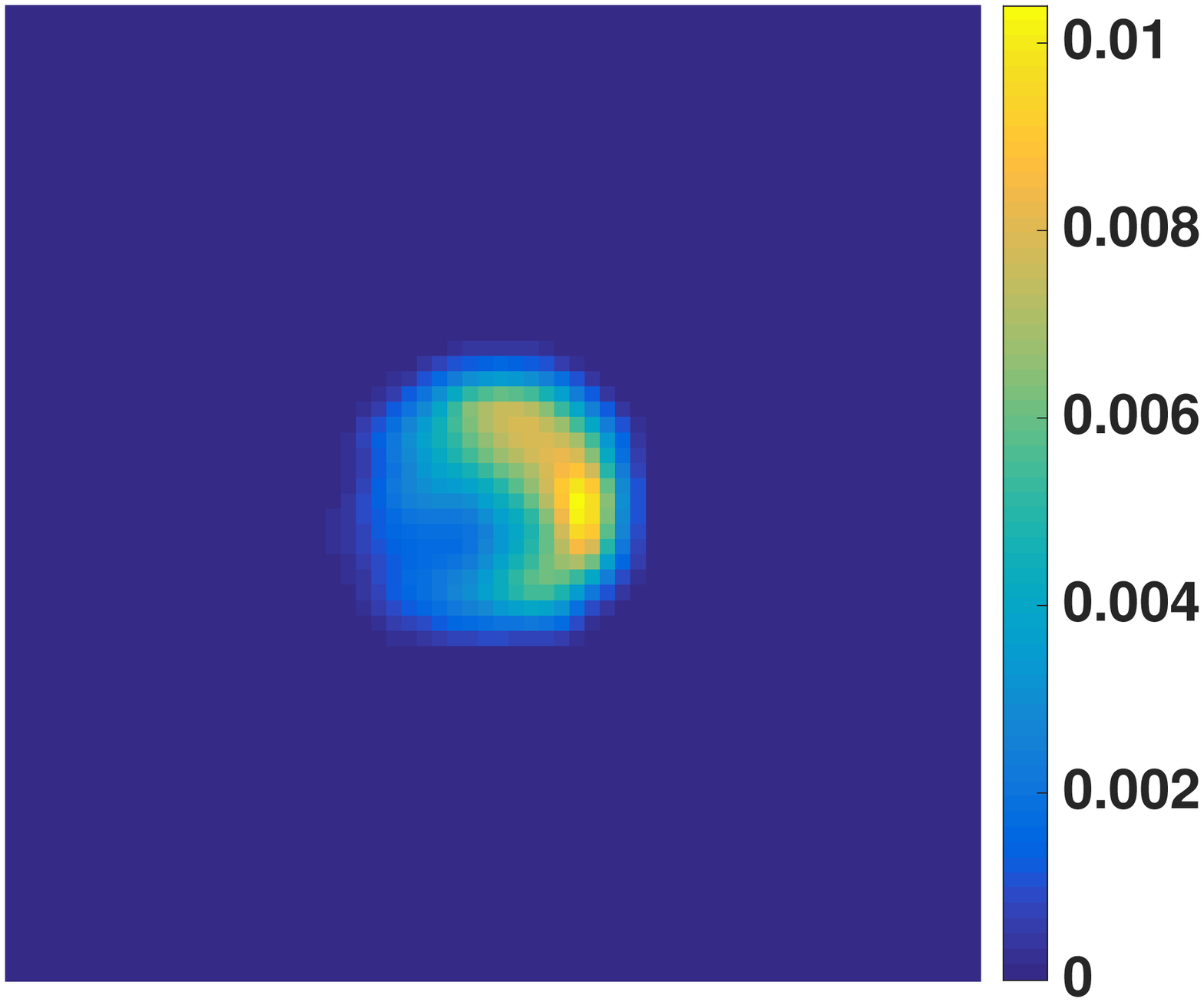} & 
\hspace*{-0.2cm}\includegraphics[width = 3.7cm]{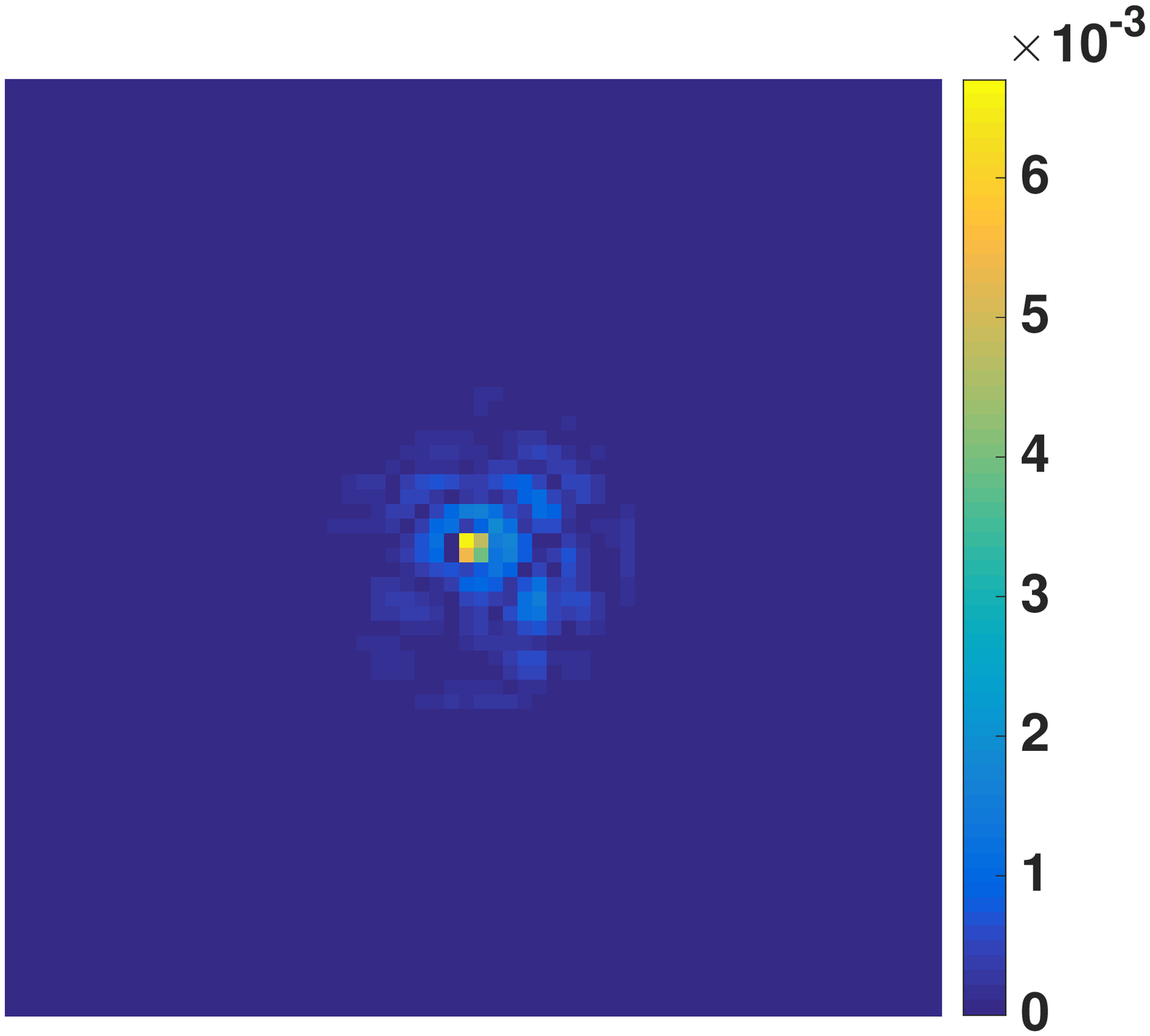} \\
\includegraphics[width = 3.7cm]{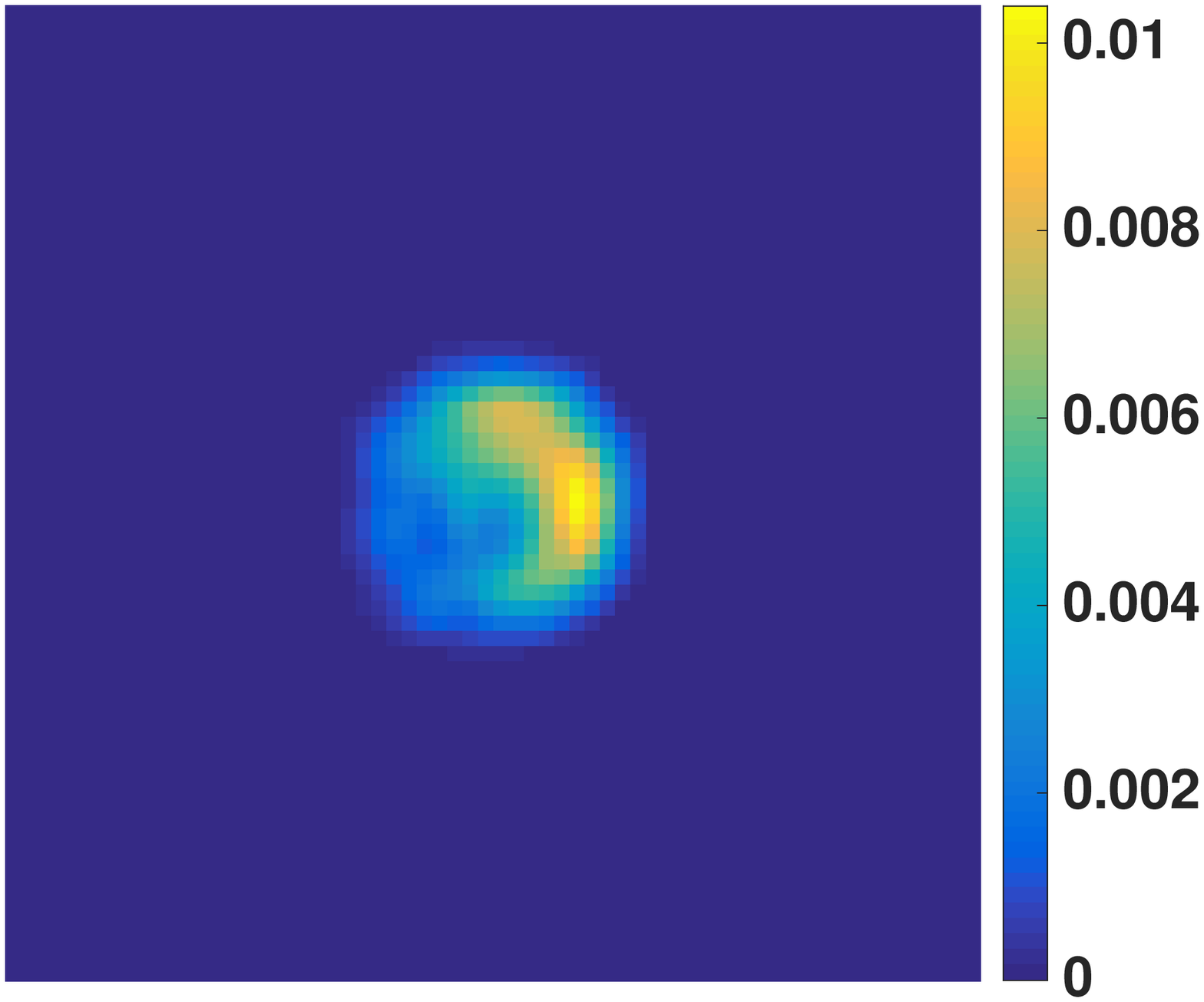} &
\hspace*{-0.2cm}\includegraphics[width = 3.7cm]{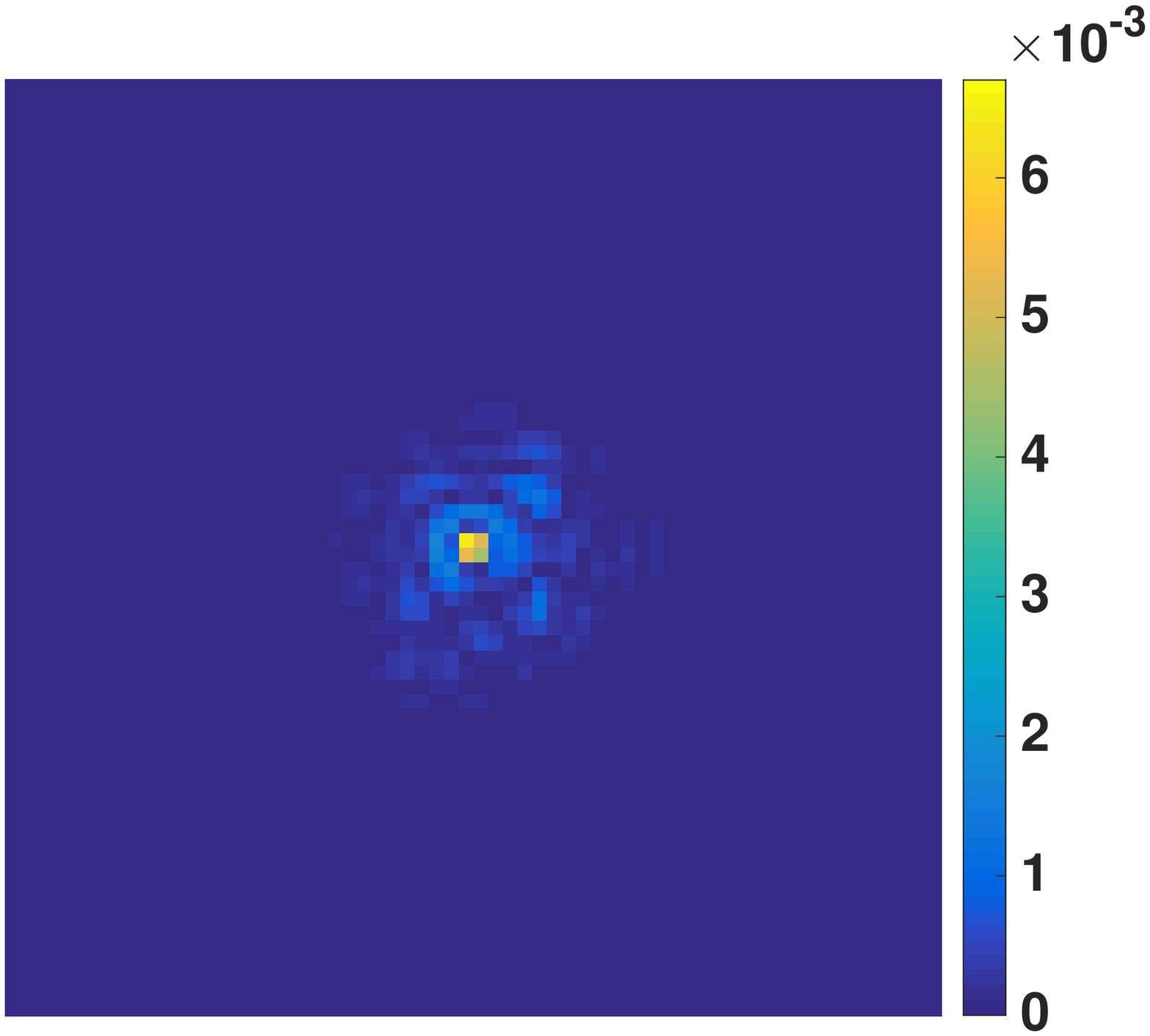}
\end{tabular}
\caption{
Reconstructed images (first column) and error images (second column) obtained by considering the true image \texttt{LkH$\alpha$}, corresponding to median SNR (over 10 simulations), with synthetic $u-v$ coverage for $(u_{\pows},u_{\bis})  = (0.2, 0.3)$. For both the columns, in each row, images corresponding to different regularization terms are shown- First row: positivity constraint, second row: $\ell_1$ regularization with $t_{\max} = 200$, and third row: reweighted-$\ell_1$ regularization.}
\label{fig:images_synthetic_uf4}
\end{figure}   %----------------------------------------------------------------------------

Concerning the initialization, both for $\ell_1$ and weighted-$\ell_1$ minimization problems, two different cases have been tested.
On the one hand, we considered the same initialization strategy as described in Section~\ref{Sssec:case1}, with $I=15$. On the other hand, we used the final estimation obtained from the positivity constrained problem, itself initialized with $I=15$ (Section~\ref{Sssec:case1}). 
Preliminary simulations indicated that the results obtained in the two cases have the similar reconstruction quality. 
However, the computation time was much longer considering several random initializations than using the solution obtained from the positivity constrained problem. 
Thus, for computational efficiency, all further simulations for $\ell_1$ and weighted-$\ell_1$ regularization are performed using the final solution obtained when only positivity constraint is considered, as described in Section~\ref{Sssec:case1}.

To inspect the quality of reconstruction, we consider two sub-cases for $\ell_1$ minimization with different number of forward-backward sub-iterations (corresponding to steps~\ref{algo:grad_step}-\ref{algo:prox_step} in Algorithm~\ref{algo_blockfb}): $t_{\max} = 200$ and $t_{\max} = 400$.
In addition, for the weighting scheme, two sub-cases are considered for different number of weighting iterations: a weighted-$\ell_1$ regularization (with only one weighting computation), and a second weighting iteration (i.e. reweighted-$\ell_1$)\footnote{Note that the simulations were performed with more than 2 weighting iterations. However, preliminary results indicated that after the second weighting iteration, a stable solution was achieved both in terms of the SNR and visual quality.}. 
As discussed in Section~\ref{ssec:re_weight}, the weights are computed using \eqref{eq:weights}, where, for the weighted-$\ell_1$ regularization, we take $\bm{x}^{\star}$ to be the solution obtained from the positivity constrained minimization problem, whereas for the reweighted-$\ell_1$ regularization, $\bm{x}^{\star}$ is the solution obtained from the weighted-$\ell_1$ minimization problem.

Note that during weighted and reweighted-$\ell_1$, $t_{\max}$ is taken to be 200. 
In the simulations performed, regularization parameter $\mu$ in \eqref{eq:reg_term1} is tuned to maximize the SNR: $\mu = 10^{-5}$ (resp. $\mu=1.5 \times 10^{-5}$) for $\ell_1$ (resp. weighted and reweighted-$\ell_1$) minimization problem.

\subsubsection{Simulation results} \label{sssec:simulation_results}

\modif{We have implemented several tests to analyze the performance of the proposed method with respect to the number of measurements made by the interferometer. More precisely, to take into account different undersampling ratios of the $u$-$v$ plane, we have performed simulations by varying $u_{\pows}$ and $u_{\bis}$. Firstly, concerning the choice of $u_{\pows}$, we have considered two cases: $u_{\pows} = 0.05$ corresponding to highly undersampled $u$-$v$ plane, and $u_{\pows} = 0.2$ to simulate a less undersampled data set. Secondly, for each of the considered values of $u_{\pows}$, we have varied number of bispectrum measurements, i.e. $u_{\bis}$. Taking these different values of $u_{\pows}$ and $u_{\bis}$ into account, Figure~\ref{fig:synthetic_plots} shows the SNR graphs corresponding to the reconstructed images, as a function of $u_{\bis}$ for $u_{\pows} = 0.05$ (left)  and $0.2$ (right), respectively. Typically, the range over which $u_{\bis}$ is varied is chosen depending on the value of $u_{\pows}$. As such, we have taken the values of $u_{\bis}$ comparable to and greater than $u_{\pows}$. Consequently, for the smaller value of $u_{\pows}$ = 0.05, we have considered less number of bispectrum measurements with $u_{\bis} \in \{0.04,0.2\}$, whereas for the larger value of $u_{\pows}$ = 0.2, the number of bispectrum measurements considered are also increased, $u_{\bis} \in \{0.05,0.5\}$.}

In each graph of Figure~\ref{fig:synthetic_plots}, comparisons are given for the results obtained using the different regularizations described in Sections~\ref{Sssec:case1} and \ref{Sssec:cases2_3}. For visual assessment, reconstructed images corresponding to median SNR are shown in Figures~\ref{fig:images_synthetic_uf1} and \ref{fig:images_synthetic_uf4}. \modif{The reconstructed images for $\ell_1$ regularization with different $t_{\max}$ are visually very similar. Same is the case for reconstructed images with weighted $\ell_1$ and reweighted $\ell_1$ regularization. Hence, in Figure~\ref{fig:images_synthetic_uf1} and Figure~\ref{fig:images_synthetic_uf4}, we show the images corresponding to positivity constrained case, $\ell_1$ regularization with $t_{\max} = 200$ and reweighted $\ell_1$ regularization. The respective error images are also displayed to show the absolute error $|\bm{x}^{\star} - \bm{\overline{x}}|$ between the reconstructed image $\bm{x}^{\star}$ and the true image $\bm{\overline{x}}$.}

From Figures~\ref{fig:synthetic_plots}, \ref{fig:images_synthetic_uf1} and \ref{fig:images_synthetic_uf4}, we can observe that promoting sparsity, either by $\ell_1$, weighted-$\ell_1$, or reweighted-$\ell_1$ regularization term, \modif{gives better reconstruction quality, and hence lesser residual in the error images,} than the positivity and reality constrained case (SNR improves between $2$ and $3$ dB depending on the considered $(u_{\pows}, u_{\bis})$). 

Moreover, from the results given in Figure~\ref{fig:synthetic_plots}, it can be seen that when $u_{\pows} = 0.2$ (Figure~\ref{fig:synthetic_plots}(b)), the  quality of reconstruction obtained with the $\ell_1$ regularization and  the (re)weighted-$\ell_1$ regularization is almost the same. 
In contrast, when $u_{\pows} = 0.05$ (Figure~\ref{fig:synthetic_plots}(a)), as $u_{\bis}$ is increased, the SNR values obtained with either of the weighted-$\ell_1$ or reweighted-$\ell_1$ regularization terms are greater than the SNR obtained using an $\ell_1$ regularization. 
%This implies that weighting scheme tends to be more beneficial if the number of power spectrum measurements is small.
\modif{This implies that weighting scheme tends to be more beneficial for the case of highly undersampled $u$-$v$ plane.}

\modif{Considering the importance of symmetrization, it is worth mentioning here that the reconstructed images for the final solution $\bm{x}^{\star} = (1/3) \big( \bm{u}_1^{\star} + \bm{u}_2^{\star} + \bm{u}_3^{\star})$ as well as for the solutions of $\bm{u}_1^{\star}$, $\bm{u}_2^{\star}$, $\bm{u}_3^{\star}$ are visually indistinguishable. This observation is supported by the small values of the variations between the solutions : $\|\bm{u}_1^{\star} - \bm{u}_2^{\star}\|_2, \|\bm{u}_2^{\star} - \bm{u}_3^{\star}\|_2$ and $\|\bm{u}_3^{\star} - \bm{u}_1^{\star}\|_2$, which are of the order of $10^{-2}, 10^{-4}$ and $10^{-2},$ respectively. }

% --------------------------------------------
\begin{figure}
\centering
\begin{tabular}{c c}
\hspace*{-0.4cm}\includegraphics[width=3.1cm]{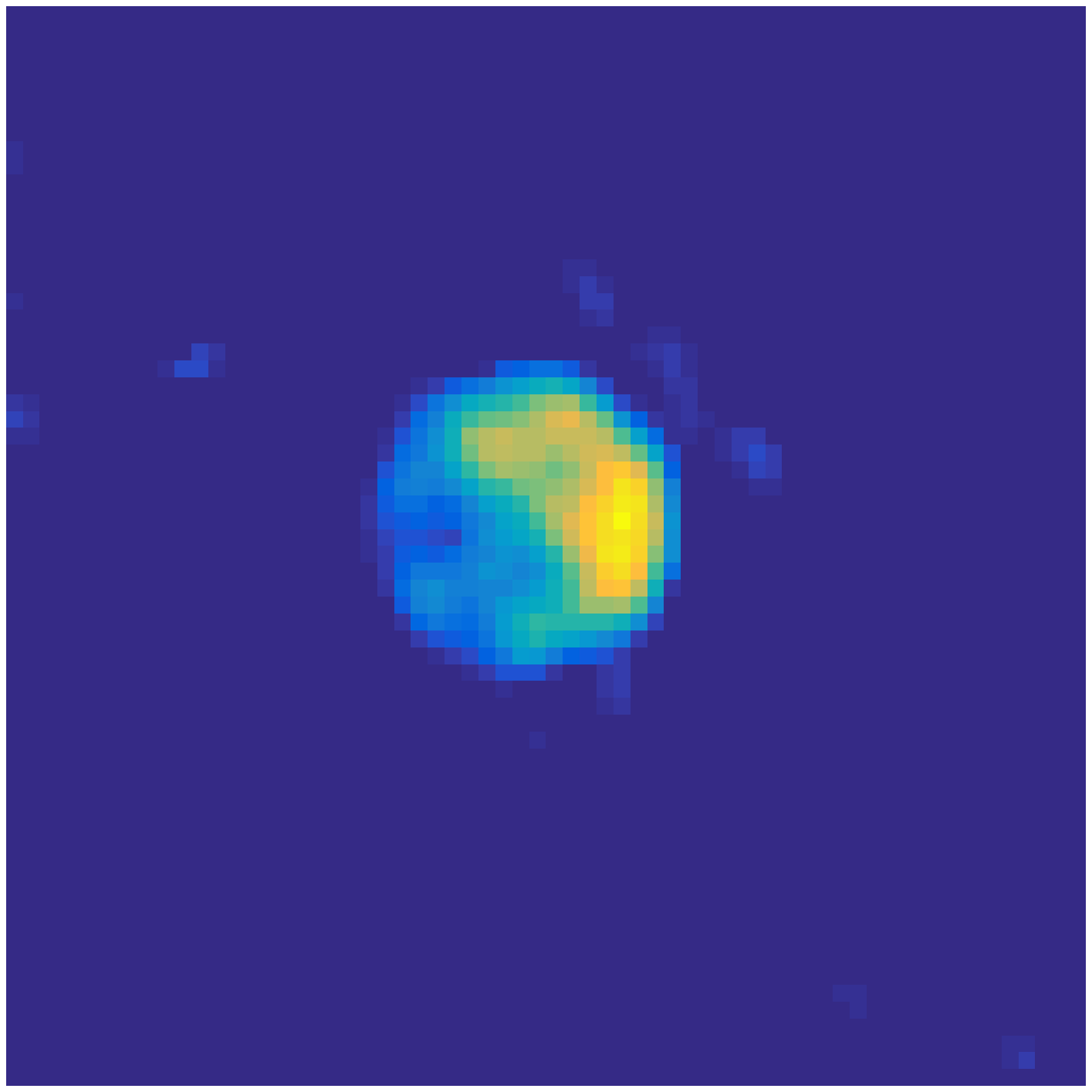} &
\hspace*{-0.05cm} \includegraphics[width=3.1cm]{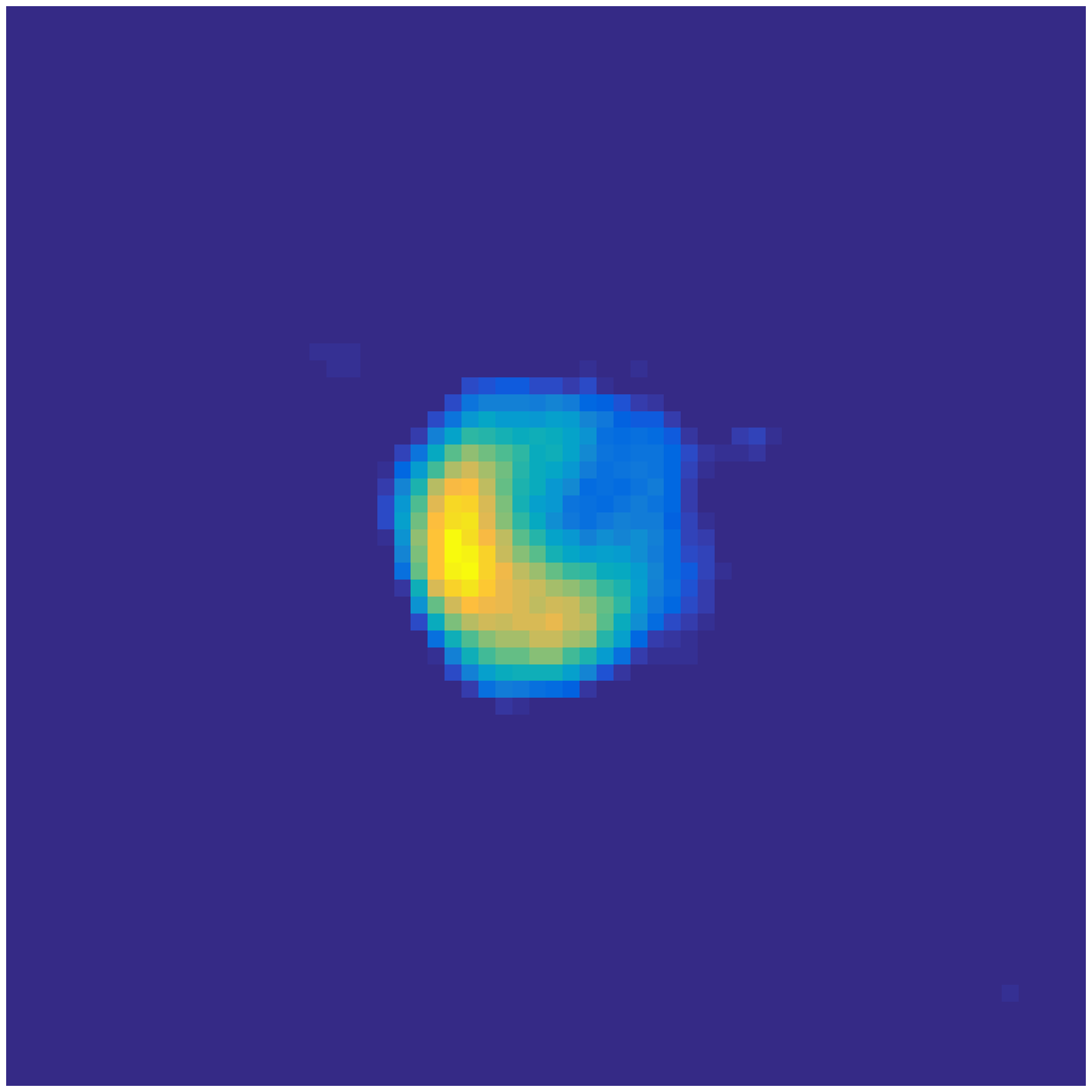} \\
(a) & (b)
\end{tabular}
\caption{ Reconstructed images obtained for the true image \texttt{LkH$\alpha$}, corresponding to two different initializations and the respective median SNR (over 10 simulations) for positivity and reality constrained case, with synthetic $u-v$ coverage for $(u_{\pows}, u_{\bis}) = (0.05, 0)$ (considering only power spectrum measurements).
The figure illustrates the orientation uncertainty when no phase information is taken into account: (a) Reconstructed image with the correct orientation of the true image \texttt{LkH$\alpha$}, (b) Reconstructed image with the opposite orientation.}
\label{fig:only_power}
\end{figure}
%---------------------------------------------

\subsubsection{Image reconstruction without the bispectrum measurements}

In order to emphasize the benefits of using phase information from bispectrum measurements, 
simulations have been performed considering only the power spectrum measurements, i.e. with $u_{\bis} = M_{\bis} = 0$. 
In this case, the Algorithm~\ref{algo_blockfb} has been implemented by considering only positivity and reality constraints, as described in Section~\ref{Sssec:case1}. 
Moreover, as explained in this section, owing to the non-convexity of the minimization problem~\eqref{eq:overall_min}, several simulations are performed with different random initializations.

Considering the synthetic $u-v$ coverage with $u_{\pows} = 0.05$ and $u_{\bis} = 0$ (no bispectrum measurements),  
the reconstructed images obtained from two different random initializations for positivity and reality constrained case are shown in Figure~\ref{fig:only_power}. 
Since the power spectrum measurements do not contain any phase information, it can be observed that the reconstructed images suffer from phase ambiguity. This arises from the space-reversal property of the Fourier transform, i.e., if a signal is inverted in the spatial domain, then in the Fourier domain, this inversion only reverses the sign of the phase of the Fourier coefficients. It implies that with no phase information, the uncertainty related to signal inversion remains.
While the image in Figure~\ref{fig:only_power}(a) is recovered with correct orientation, i.e., the same orientation as that of the original image \texttt{LkH$\alpha$} given in Figure~\ref{fig:true_image}, the image in Figure~\ref{fig:only_power}(b) is recovered with the opposite orientation.
%

% --------------------------------------------------------------
\begin{figure} 
\centering
\vspace*{-0.4cm}
\hspace*{-0.2cm}\includegraphics[width = 1.05\columnwidth]{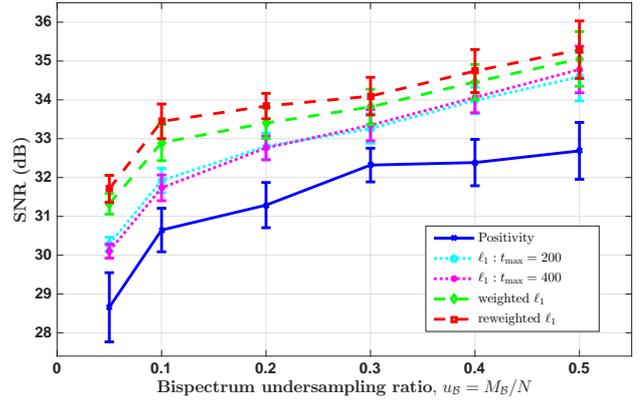}
\vspace*{-0.3cm}
\caption{SNR graph obtained with \texttt{LkH$\alpha$} image and realistic $u-v$ coverage, considering iSNR = 30 dB, varying $u_{\bis}$. In the graph, comparison of average SNR values (over 10 simulations), and corresponding 1-standard-deviation error bars, for different regularization terms is shown: positivity constraints (solid blue), $\ell_1$ regularization with $t_{\max} = 200$ (dotted cyan) and $t_{\max} = 400$ (dotted pink),  weighted-$\ell_1$ regularization (dashed green) and reweighted-$\ell_1$ regularization (dashed red). }
\label{fig:real_plots}
\end{figure}
% --------------------------------------------------------------

On the one hand, this indicates that the proposed Algorithm~\ref{algo_blockfb} is still able to restore images with only power spectrum measurements, i.e., without any phase information, though with the uncertainty in the orientation. 
On the other hand, the results obtained from the case when $u_{\bis} > 0$ highlight that the incorporation of phase information is essential to recover properly oriented images.

\subsection{Realistic $u-v$ coverage}

The performance of the proposed algorithm has been assessed for the realistic $u-v$ coverage given in Figure~\ref{fig:uv_cover}(b). 
We have performed several simulations by varying the number of bispectrum measurements and thus in turn the bispectrum undersampling ratio $u_{\bis}$. Note that, as mentioned in the Section~\ref{ssec:simulation_setting}, for the considered realistic $u-v$ coverage, $M_{\pows} = 72$. With $N = 64^{2}$, this implies that $u_{\pows} \simeq 0.018$. 

\begin{figure}
\vspace{-0.22cm}
\centering
\begin{tabular}{c c}
\includegraphics[width = 3.7cm]{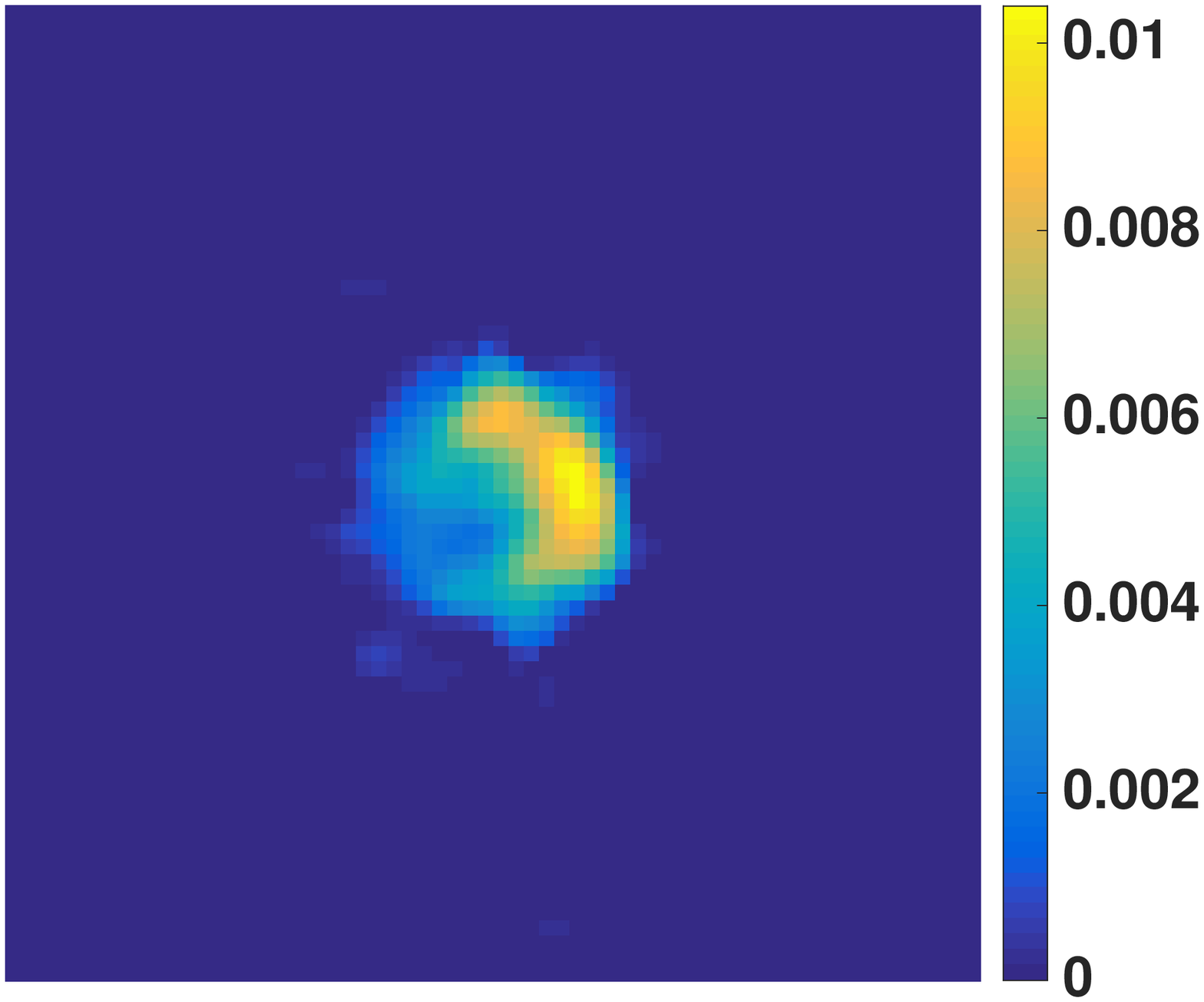} &
\hspace*{-0.2cm}\includegraphics[width = 3.7cm]{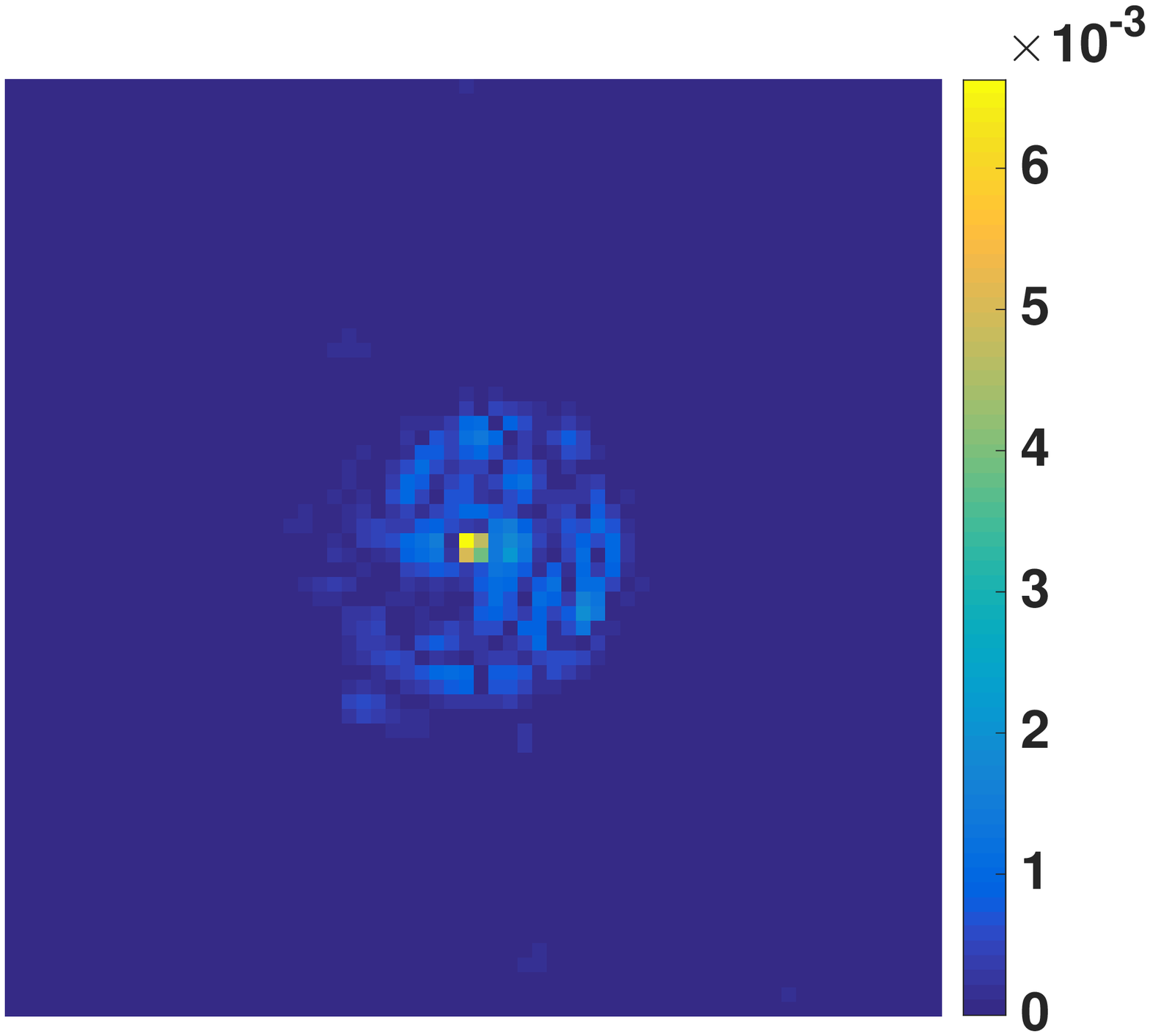} \\
\includegraphics[width = 3.7cm]{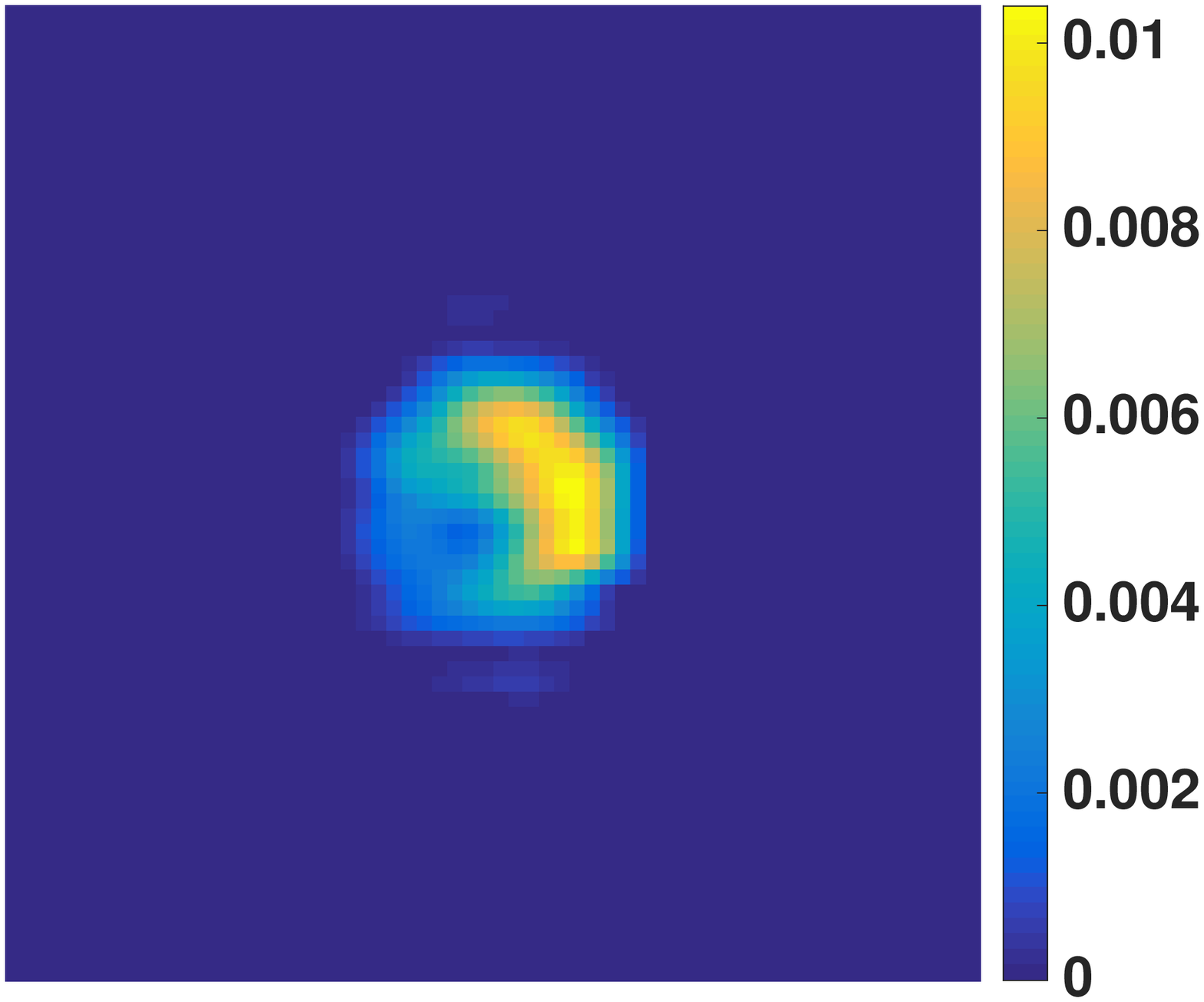} & 
\hspace*{-0.2cm}\includegraphics[width = 3.7cm]{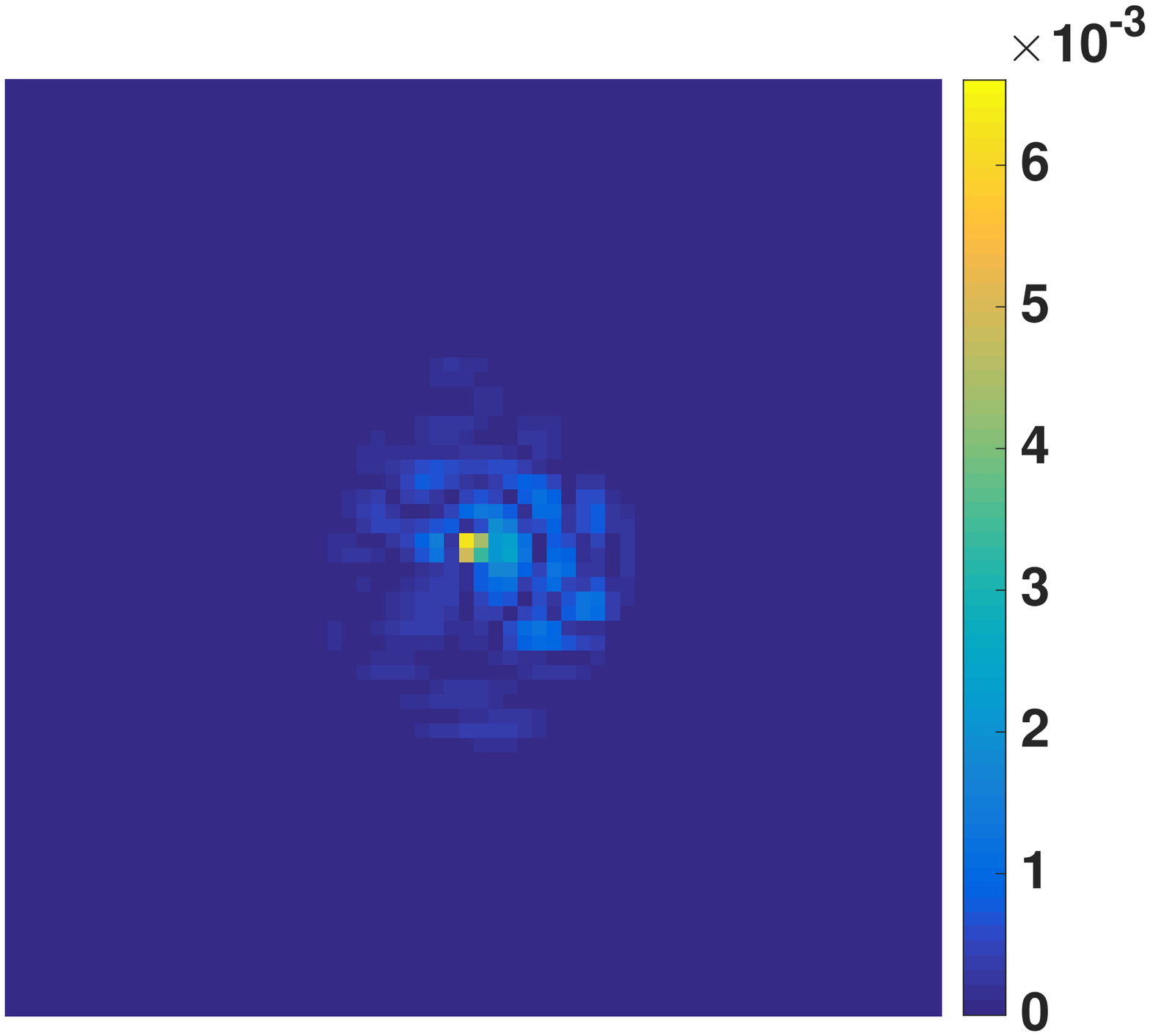} \\
\includegraphics[width = 3.7cm]{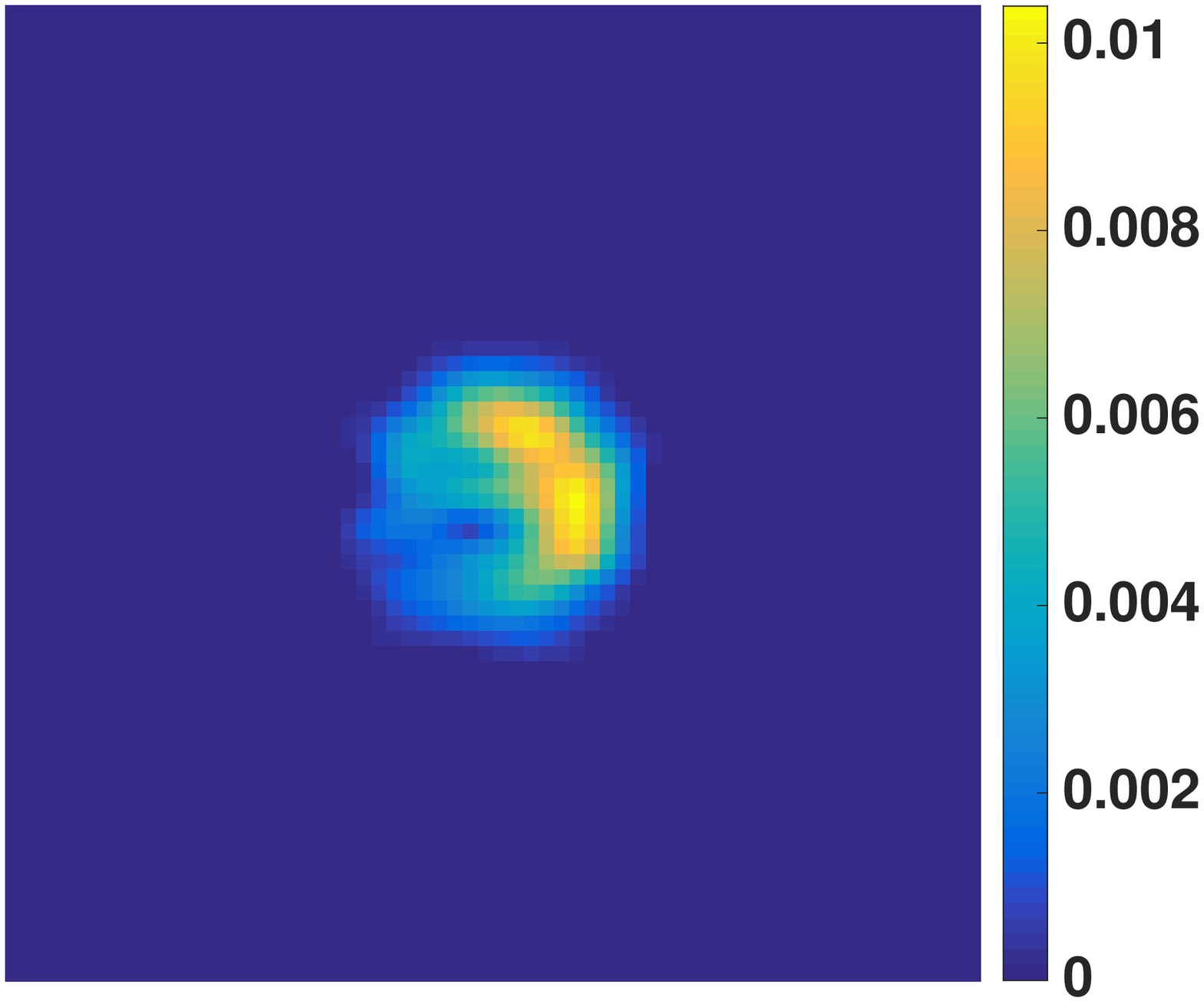}
\vspace*{0.2cm} &
\hspace*{-0.2cm}\includegraphics[width = 3.7cm]{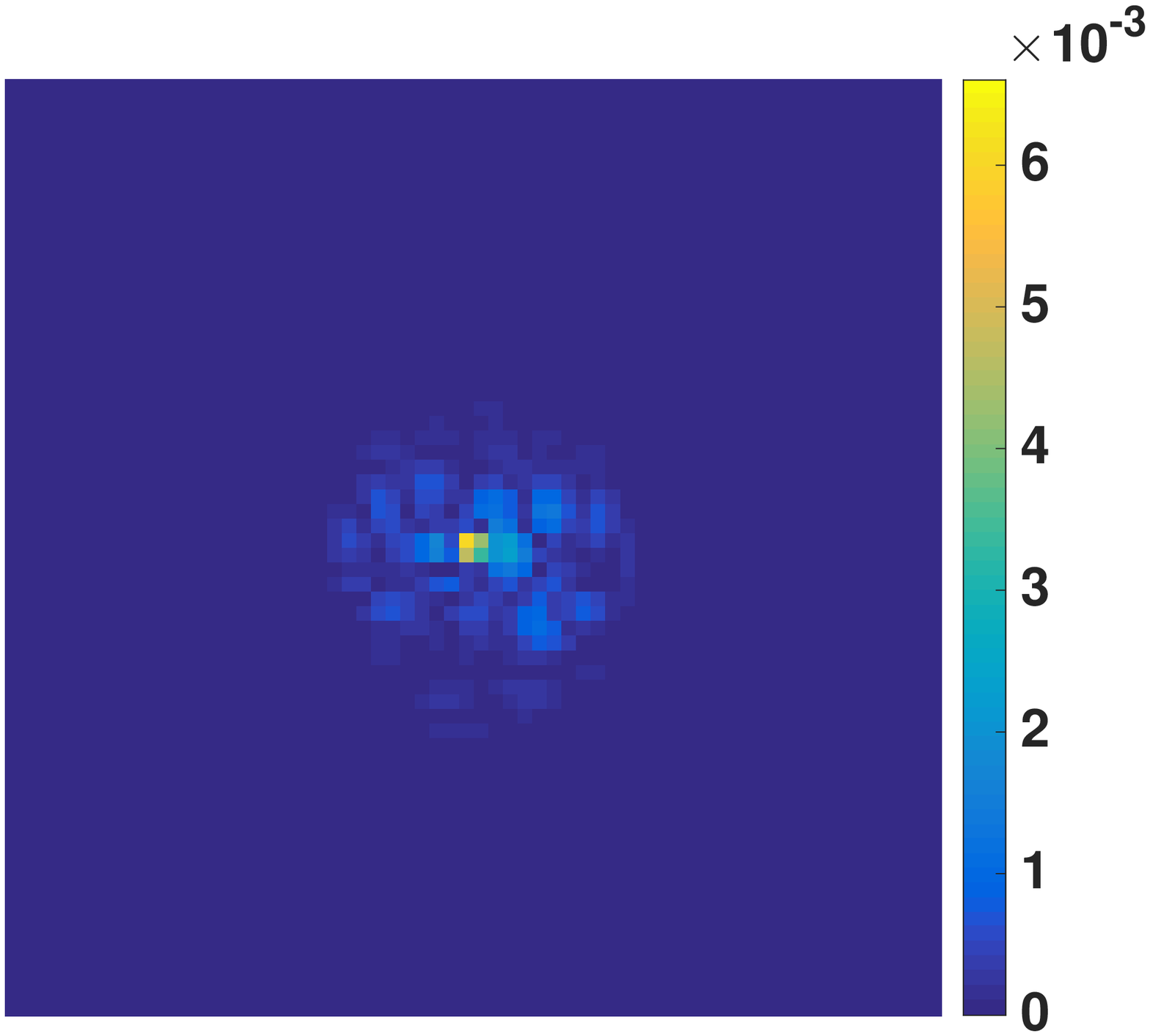}
\end{tabular}
\caption{
Reconstructed images (first column) and error images (second column) obtained by considering the true image \texttt{LkH$\alpha$}, corresponding to median SNR (over 10 simulations), with realistic $u-v$ coverage for $u_{\bis}  = 0.2$. For both the columns, in each row, images corresponding to different regularization terms are shown- First row: positivity constraint, second row: $\ell_1$ regularization with $t_{\max} = 200$, and third row: reweighted-$\ell_1$ regularization. }
\label{fig:images_real}
\end{figure}

Figures~\ref{fig:real_plots} and \ref{fig:images_real} illustrate the results obtained for different regularization terms, as discussed in Section~\ref{Ssec:synt_uv}.
\modif{While Figure~\ref{fig:real_plots} depicts the SNR graph for the reconstructed images as a function of $u_{\bis} \in \{0.05,0.5\}$, the corresponding recovered images and the error images for $u_{\bis} = 0.2$, with median SNR, are shown in Figure~\ref{fig:images_real}. Here again considering the visual similarity between the reconstructed images for $\ell_1$ regularization with different $t_{\max}$, and that between images for weighted and reweighted $\ell_1$ regularization, we only show the images for positivity constraint, $\ell_1$ with $t_{\max}$ = 200 and reweighted $\ell_1$.}

It is to be remarked here that the results obtained 
are in coherence with the observations made for the synthetic $u-v$ coverage in Section~\ref{sssec:simulation_results}. 
More precisely, the results indicate the superiority of promoting sparsity relative to just positivity and reality over the full undersampling range, leading to an improvement of the SNR between $3$ and $4$ dB, depending on the considered value of $u_{\bis}$. 
Moreover, given the small value of $u_{\pows}$, the SNR gets better not only  with increasing $u_{\bis}$, but also by considering the (re)weighted-$\ell_1$ regularization term.

% --------------------------------------------------------------

% -------------------------------------------

%-------------------------------------------------------

\vspace*{-0.4cm}

\section{Hyperspectral imaging} \label{sec:hyperspectral}

\subsection{Problem statement}
\label{Ssec:Pb_hyper}

As described in Section~\ref{sec:optical_imaging}, the sampled spatial frequencies depend on the observation wavelength. Thus, interferometric measurements made at different wavelengths correspond to probing different spatial frequencies in the $u-v$ plane of the image of interest. 
Considering $L$ spectral channels, in accordance with the data model proposed for the monochromatic case \eqref{eq:meas_pb}, the measurement equation at each spectral channel $l \in \{1, \ldots, L\}$, can be written as:
\begin{equation}	\label{eq:hyper_meas}
\bm{y}_l = 
\big[ (\bm{\mathsf{T}}_{1,l} \bm{\overline{x}}_l)\cdot (\bm{\mathsf{T}}_{2,l} \bm{\overline{x}}_l)\cdot (\bm{\mathsf{T}}_{3,l} \bm{\overline{x}}_l) \big] 
+ \bm{\eta}_l  ,
\end{equation}
where $\bm{y}_l \in \mathbb{C}^M$ denotes the measurement vector, 
$\bm{\overline{x}}_l \in \mathbb{R}_+^N$ is the intensity image, 
$\bm{\eta}_l \in \mathbb{C}^M$ is a realization of an additive Gaussian noise, 
and, in analogy with \eqref{eq:mask}, the $l$-th measurement operators are given by $\bm{\mathsf{T}}_{p,l} = \bm{\mathsf{L}}_{p,l} \bm{\mathsf{S}}_{l} \bm{\mathsf{F}}$, for every $p \in \{1,2,3\}$.
Following the approach adopted in the monochromatic case and considering  $\bm{\overline{u}}_{1,l} = \bm{\overline{u}}_{2,l} = \bm{\overline{u}}_{3,l} = \bm{\overline{x}}_{l}$ for $1 \leq l \leq L$, the tri-linear counter-part of the inverse problem~\eqref{eq:hyper_meas} becomes:
\begin{equation}\label{eq:hyper_non_linear}
\bm{y}_l = \big[ (\bm{\mathsf{T}}_{1,l} \bm{\overline{u}}_{1,l})\cdot (\bm{\mathsf{T}}_{2,l} \bm{\overline{u}}_{2,l})\cdot (\bm{\mathsf{T}}_{3,l} \bm{\overline{u}}_{3,l}) \big] + \bm{\eta}_l  .
\end{equation} 
Then, concatenating all the spectral channels, we define the ill-posed hyperspectral inverse problem as: 
\begin{equation} \label{eq:hyper_inv}
\bm{\mathsf{Y}} = \big[ \bm{\mathsf{T}}_1 (\bm{\overline{\bm{\mathsf{U}}}}_1)\cdot \bm{\mathsf{T}}_2 (\bm{\overline{\bm{\mathsf{U}}}}_2) \cdot \bm{\mathsf{T}}_3 (\bm{\overline{\bm{\mathsf{U}}}}_3) \big] + \bm{\mathsf{H}} , 
\end{equation}
where $\bm{\mathsf{Y}} = [\bm{y}_1, \ldots, \bm{y}_L] \in \mathbb{C}^{M\times L}$ is the measurement matrix, 
for every $p \in \{1,2,3\}$, $\bm{\overline{\bm{\mathsf{U}}}}_p = [\bm{\overline{u}}_{p,1}, \ldots, \bm{\overline{u}}_{p,L}] \in \mathbb{R}_+^{N \times L}$ is the image matrix, and $\bm{\mathsf{H}} = [\bm{\eta}_1, \ldots, \bm{\eta}_L] \in \mathbb{C}^{M\times L}$ is the noise matrix, and $\bm{\mathsf{T}}_1$, $\bm{\mathsf{T}}_2$, $\bm{\mathsf{T}}_3$ are the concatenated measurement operators such that, for $p \in \{1,2,3\}$, $\bm{\mathsf{T}}_p(\bm{\overline{\bm{\mathsf{U}}}}_p) = (\bm{\mathsf{T}}_{p,l} \bm{\overline{u}}_{p,l})_{1 \leq l \leq L}$. 
More precisely, column $l \in \{1,\ldots,L\}$ of $\bm{\overline{\bm{\mathsf{U}}}}_p$ represents the intensity image at wavelength $\lambda_l$, whereas row $n \in \{1,\ldots,N\}$ represents the variation of pixel values along the spectral channels.

In analogy with the monochromatic case and the minimization problem described in~\eqref{eq:overall_min} by symmetrizing the data fidelity term, we propose to define the estimate of $(\bm{\overline{\bm{\mathsf{U}}}}_1, \bm{\overline{\bm{\mathsf{U}}}}_2, \bm{\overline{\bm{\mathsf{U}}}}_3)$  as a solution to 
\begin{equation} \label{eq:overall_min_hyper}
\underset{(\bm{\mathsf{U}}_1,\bm{\mathsf{U}}_2,\bm{\mathsf{U}}_3)\in (\mathbb{R}^{N\times L})^3}{\text{minimize}} \,
\Big\{ h(\bm{\mathsf{U}}_1, \bm{\mathsf{U}}_2, \bm{\mathsf{U}}_3)  = \widetilde{f}(\bm{\mathsf{U}}_1, \bm{\mathsf{U}}_2, \bm{\mathsf{U}}_3) 
+ \sum_{p=1}^3 r(\bm{\mathsf{U}}_p) \Big\} ,
\end{equation}
where the same regularization term 
\begin{equation} \label{eq:hyper_reg}
(\forall \bm{\mathsf{X}} \in \mathbb{R}^{N\times L})\quad
r(\bm{\mathsf{X}}) = {\iota_{\mathbb{R}_+^{N\times L}}}(\bm{\mathsf{X}}) + \mu g(\bm{\mathsf{X}}) ,
\end{equation}
is chosen for $\bm{\mathsf{U}}_1,\bm{\mathsf{U}}_2,$ and $\bm{\mathsf{U}}_3$, and $\widetilde{f}$ is the symmetrized data fidelity term given by
\begin{align}
\widetilde{f}(\bm{\mathsf{U}}_1,\bm{\mathsf{U}}_2,\bm{\mathsf{U}}_3) 
&	= \frac{1}{2}\|{\widetilde{\bm{\mathsf{Y}}} - \widetilde{\bm{\mathsf{T}}}_1 (\bm{\mathsf{U}}_1)\cdot \widetilde{\bm{\mathsf{T}}}_2 (\bm{\mathsf{U}}_2) \cdot \widetilde{\bm{\mathsf{T}}}_3 (\bm{\mathsf{U}}_3)} \|_2^{2}	\nonumber  	\\ 
& = \sum_{l = 1}^L  \check{f}_l( \bm{u}_{1,l},  \bm{u}_{2,l}, \bm{u}_{3,l} ),  \label{eq:data_fid_hyper}
%&	= \sum_{l = 1}^L \frac12 \| \bm{y}_l - (\bm{\mathsf{T}}_{1,l} \bm{u}_{1,l}) \cdot (\bm{\mathsf{T}}_{2,l} \bm{u}_{2,l}) \cdot (\bm{\mathsf{T}}_{3,l} \bm{u}_{3,l}) \|_2^2,  \label{eq:data_fid_hyper}
\end{align}
with 
\begin{multline}  \label{eq:data_fid_hyper_l}
\check{f}_l( \bm{u}_{1,l},  \bm{u}_{2,l}, \bm{u}_{3,l} ) \\
= \frac12 \| \widetilde{\bm{y}}_l - (\widetilde{\bm{\mathsf{T}}}_{1,l} \bm{u}_{1,l}) \cdot (\widetilde{\bm{\mathsf{T}}}_{2,l} \bm{u}_{2,l}) \cdot (\widetilde{\bm{\mathsf{T}}}_{3,l} \bm{u}_{3,l}) \|_2^2. 
\end{multline}
$\widetilde{\bm{\mathsf{Y}}} = [\widetilde{\bm{y}}_1, \ldots, \widetilde{\bm{y}}_L] \in \mathbb{C}^{6 M\times L}$, and $\widetilde{\bm{\mathsf{T}}}_p(\bm{\mathsf{U}}_p) = (\widetilde{\bm{\mathsf{T}}}_{p,l} \bm{u}_{p,l})_{1 \leq l \leq L}$ are the symmetrized versions of the measurements matrix and the linear operators for $p \in \{1,2,3\}$, respectively in accordance with Section~\ref{ssec:Alter_min}.

As discussed in the earlier sections, given the voids in the $u-v$ coverage, ensuring data consistency is not sufficient to 
obtain a good estimation from the measurements, and imposing  \emph{a priori} information is essential. 
In the monochromatic case, we have considered promoting sparsity prior with a, possibly weighted, $\ell_1$ regularization term (Section~\ref{ssec:choose_reg}). 
In the context of hyperspectral imaging, joint sparsity gives an additional degree of possible regularization, in the spectral dimension, that should be leveraged to improve the overall image reconstruction quality compared to reconstructing each channel separately \citep{Soulez2011,Thiebaut2013,Abdulaziz2016}. Mathematically, joint sparsity is defined for a set of sparse signals such that the non-zero entries of each signal are located at the same spatial position. From physical point of view, if a source is absent, i.e., the corresponding pixel has a zero value in a spectral channel, then the pixels at the same spatial positions along all the spectral channels will be zero. Thus, the joint sparsity prior enforces spatial sparsity while imposing spectral continuity.
We propose to promote the joint sparsity prior using an $\ell_{2,1}$ norm \citep{Fornasier2008,Thiebaut2013} for the regularization term, defined as follows:
\begin{equation}\label{eq:l21}
g(\bm{\mathsf{X}}) 
= \sum_{j=1}^{J} \bigg(\sum_{l=1}^L \big| [\bm{\Psi}^\dagger \bm{x}_l]_j \big|^2 \bigg)^{1/2},
\end{equation}
where $\bm{\Psi}$ can either be identity matrix, or a given dictionary belonging to $\mathbb{R}^{J \times N}$. 
The $\ell_{2,1}$ norm is characterized by taking $\ell_{2}$ norm along the columns and then $\ell_1$ norm of the resultant vector. 

In order to solve the minimization problem (\ref{eq:overall_min_hyper}), we propose to adopt the same methodology as developed for monochromatic case.

\vspace*{-0.5cm}
\begin{figure}
\begin{tabular}{@{}c@{}}
\hspace*{-0.4cm} \includegraphics[width=1.05\columnwidth]{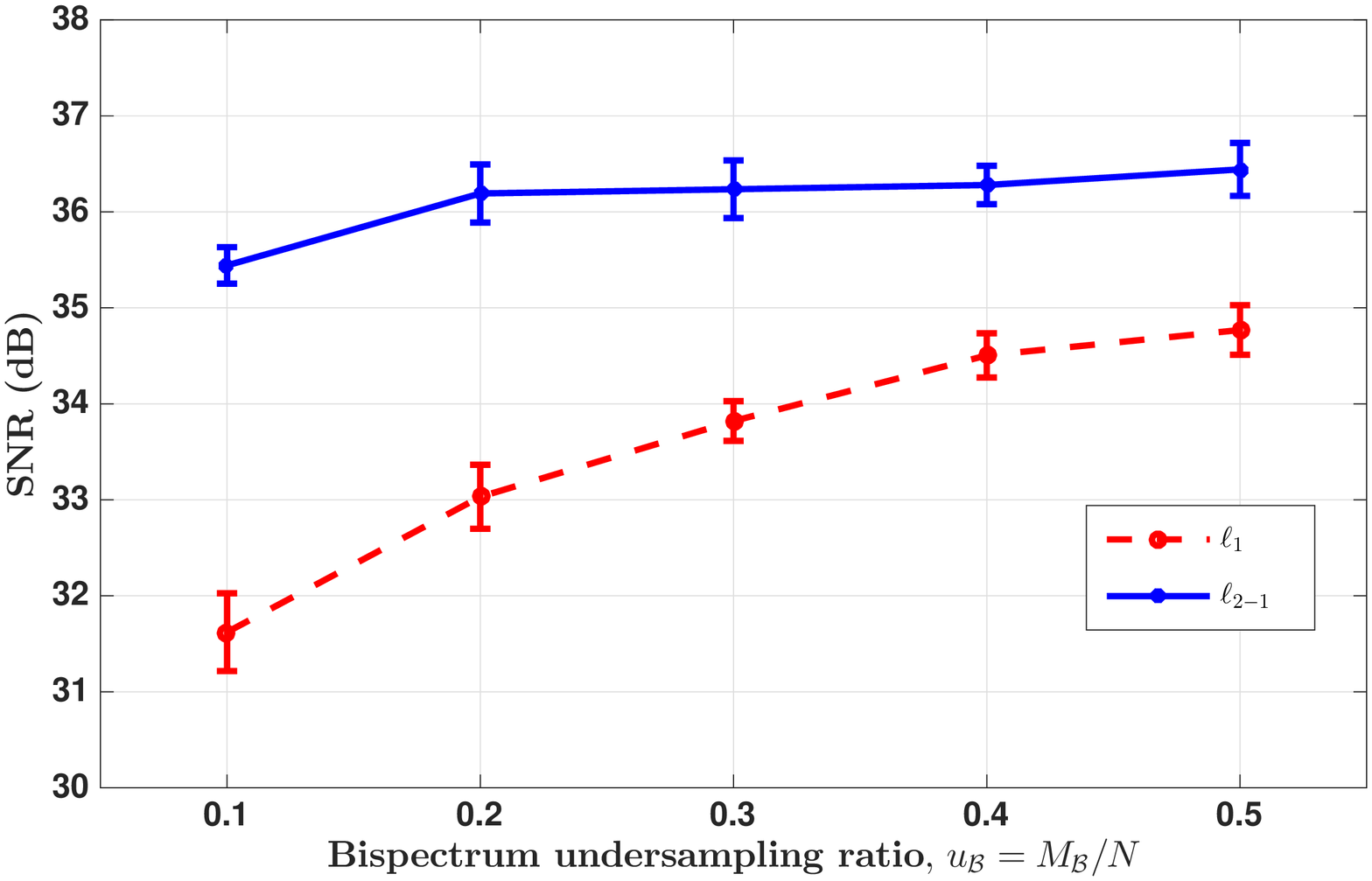} 
\vspace*{0.05cm} \\
\textbf{(a)} \\
%\vspace*{0.2cm}
\hspace*{-0.2cm} \includegraphics[width=1.065\columnwidth]{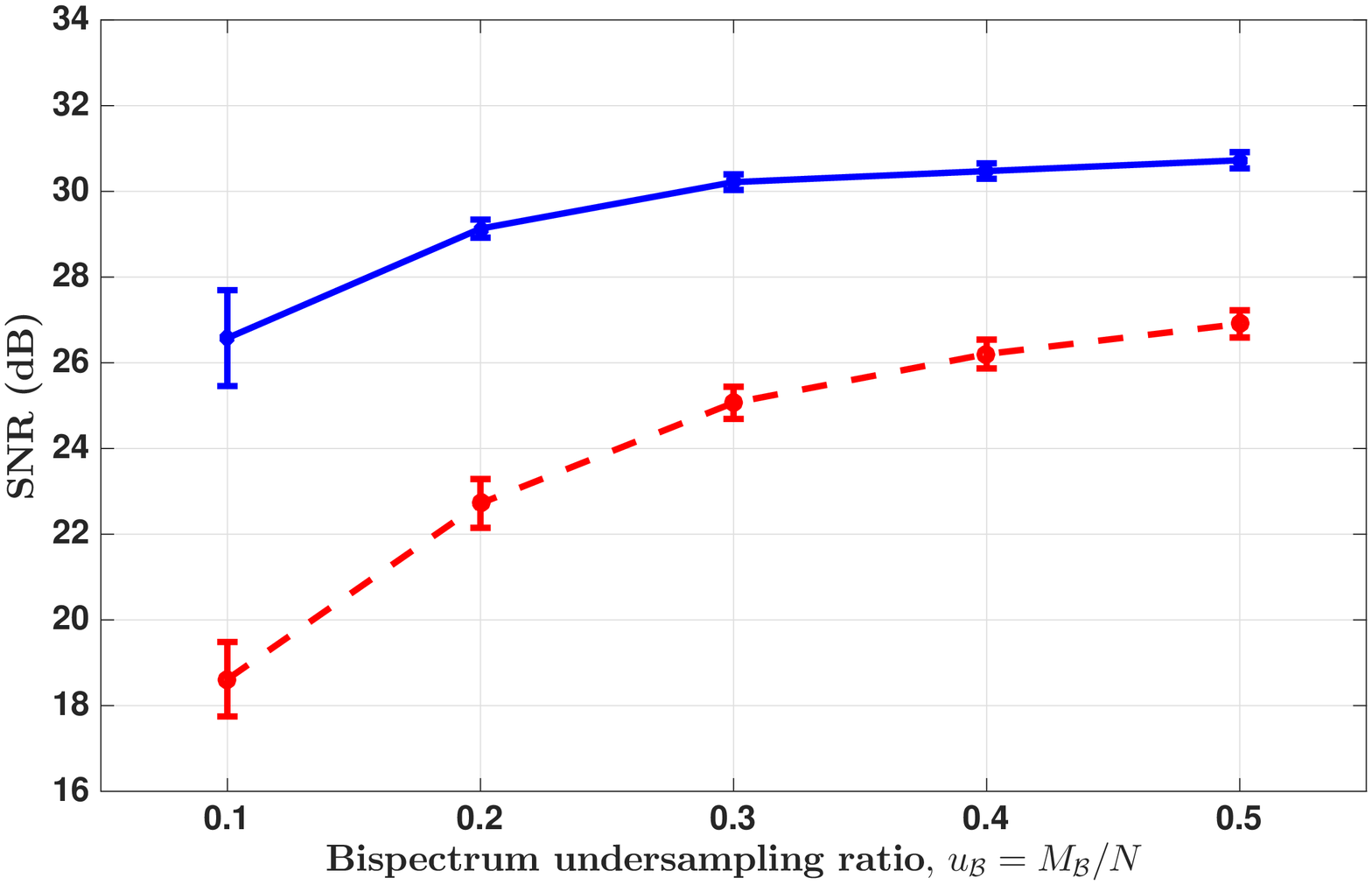} 
\vspace*{0.05cm} \\
\textbf{(b)}
\end{tabular}
\caption{SNR graphs obtained for the reconstruction of two different hyperspectral image cubes with the realistic u-v coverage, considering iSNR = 30 dB for each spectral channel, varying $u_{\bis}$. For the two graphs, the ground truth images at first spectral channel are given by: \textbf{(a)} \texttt{LkH$\alpha$} (top left image in Figure~\ref{Fig:hyper1}), and \textbf{(b)} \texttt{synthetic image} (top left image in Figure~\ref{Fig:hyper2}). Each graph depicts the comparison of the average SNR values (over 10 simulations) and corresponding 1-standard-deviation error bars, between single-channel reconstruction with $\ell_1$ regularization \eqref{eq:reg_l1_hyper} (red dashed) and reconstruction by considering joint sparsity with $\ell_{2,1}$ regularization \eqref{eq:l21} (blue solid).}
\label{Fig:hyper_SNR}
\end{figure}

\subsection{Implementation details}

The implementation of Algorithm~\ref{algo_blockfb} to solve \eqref{eq:overall_min_hyper} requires replacing the variables and the operators with the corresponding variables and operators for hyper-spectral case, as defined in Section~\ref{Ssec:Pb_hyper}. 

Firstly, according to \eqref{eq:data_fid_hyper}-\eqref{eq:data_fid_hyper_l}, for every $l \in \{1, \ldots, L\}$, partial gradients of $\check{f}_l$ are independent. Thus, the gradient descent step~\ref{algo:grad_step} of Algorithm~\ref{algo_blockfb} can be computed in parallel for each spectral channel. 
Secondly, the proximity operator of the non-smooth function $r$ defined by \eqref{eq:hyper_reg}, with $g$ given by \eqref{eq:l21}, does not have a closed form solution. In order to compute this, we propose to resort once more to Algorithm~\ref{algo_dualfb}.
In this case, step~\ref{algoDFB:proj} in Algorithm~\ref{algo_dualfb} requires performing the proximity operator of \eqref{eq:l21}, defined, for every $\bm{\mathsf{B}} \in \mathbb{R}^{J \times L}$ and $\nu >0 $, as
\begin{equation}
\text{prox}_{\nu \|.\|_{2,1}} (\bm{\mathsf{B}}) = 
\left[\begin{matrix}
\bm{p}_1	\\
\vdots	\\
\bm{p}_J
\end{matrix}\right] ,
\end{equation}
where, for every $j \in \{1,...,J\}$, $\bm{p}_j$ is a line vector given by
\begin{equation}
\bm{p}_j = 
\begin{cases}
    \bm{b}_{j} \frac{\|\bm{b}_j\|_2 - \nu}{\|\bm{b}_j\|_2} & \text{if }  \|\bm{b}_j\|_2 \geq \nu, \\
    \bm{0}              & \text{otherwise},
\end{cases}
\end{equation}
$\bm{b}_j$ denoting the $j$-th row of $\bm{\mathsf{B}}$. 
Thus, the proximity operator of the $\ell_{2,1}$ norm corresponds to a soft-thresholding operation row-wise.

\begin{figure}	
\centering
\begin{tabular}{@{}c@{}c@{}}
\includegraphics[height=3.5cm]{xtrue_1_circ.eps} &
\includegraphics[height=3.5cm]{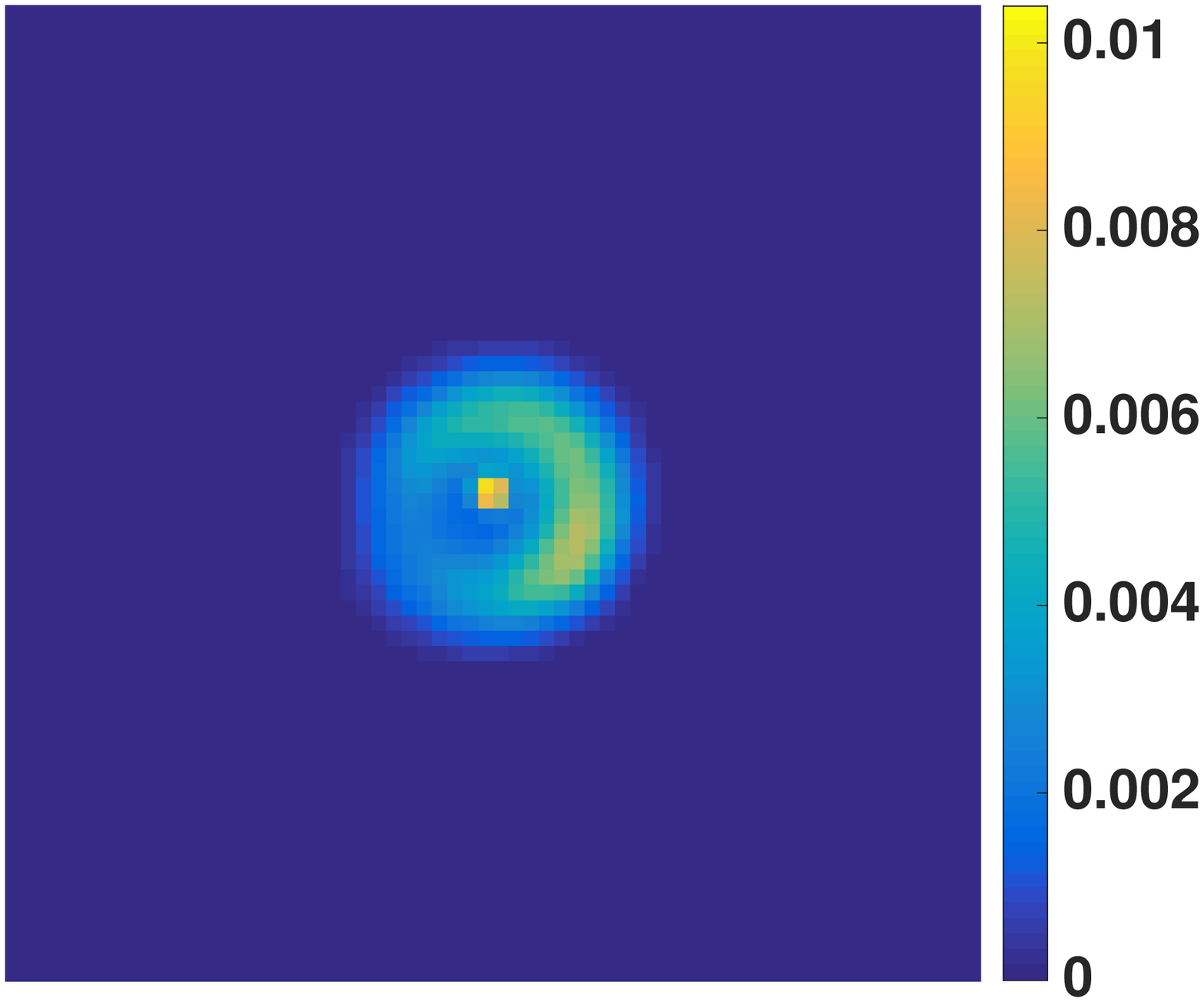} 
\vspace*{0.08cm}\\
\includegraphics[height=3.5cm]{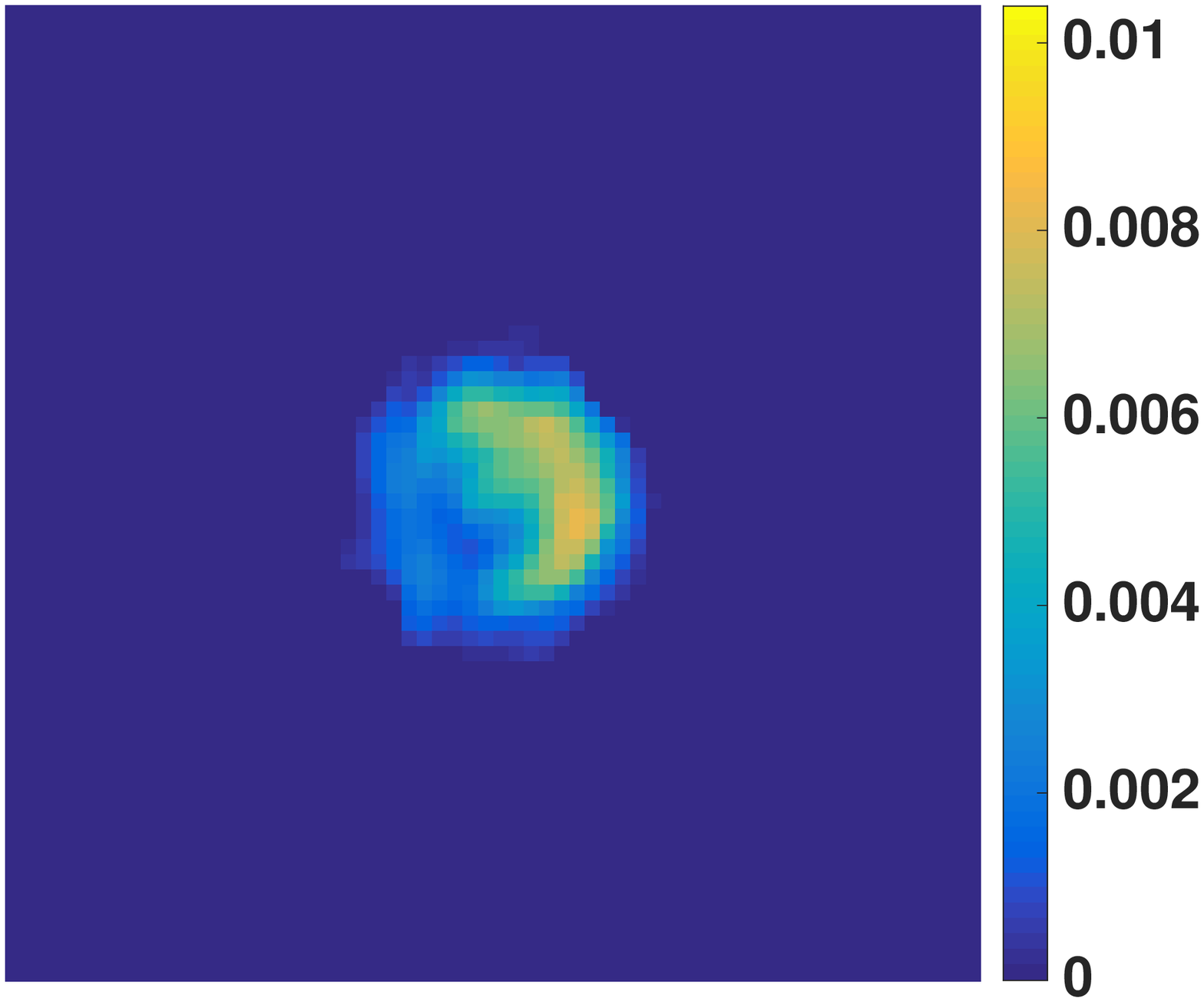} &
\includegraphics[height=3.5cm]{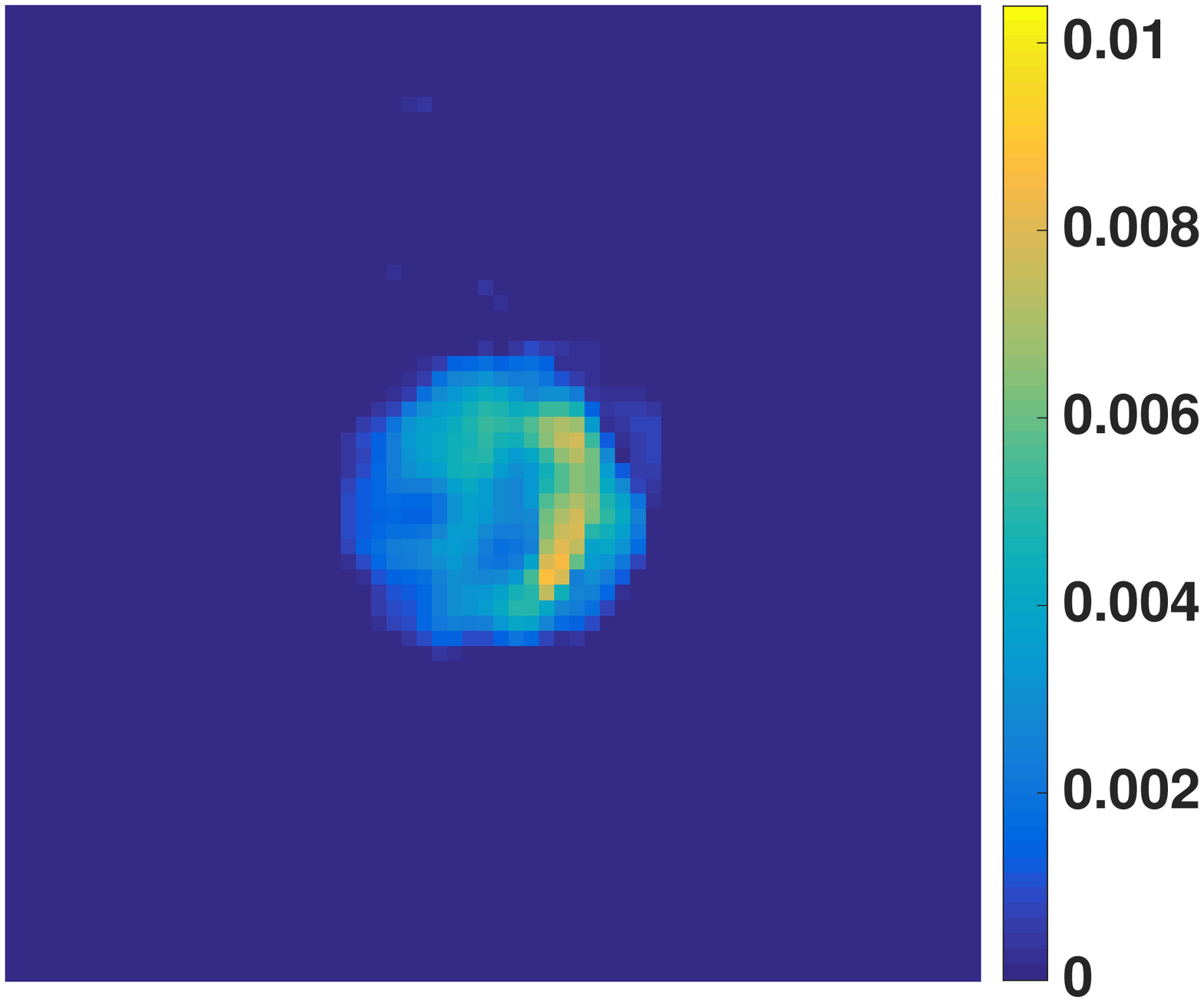} 
\vspace*{0.08cm}\\
\includegraphics[height=3.5cm]{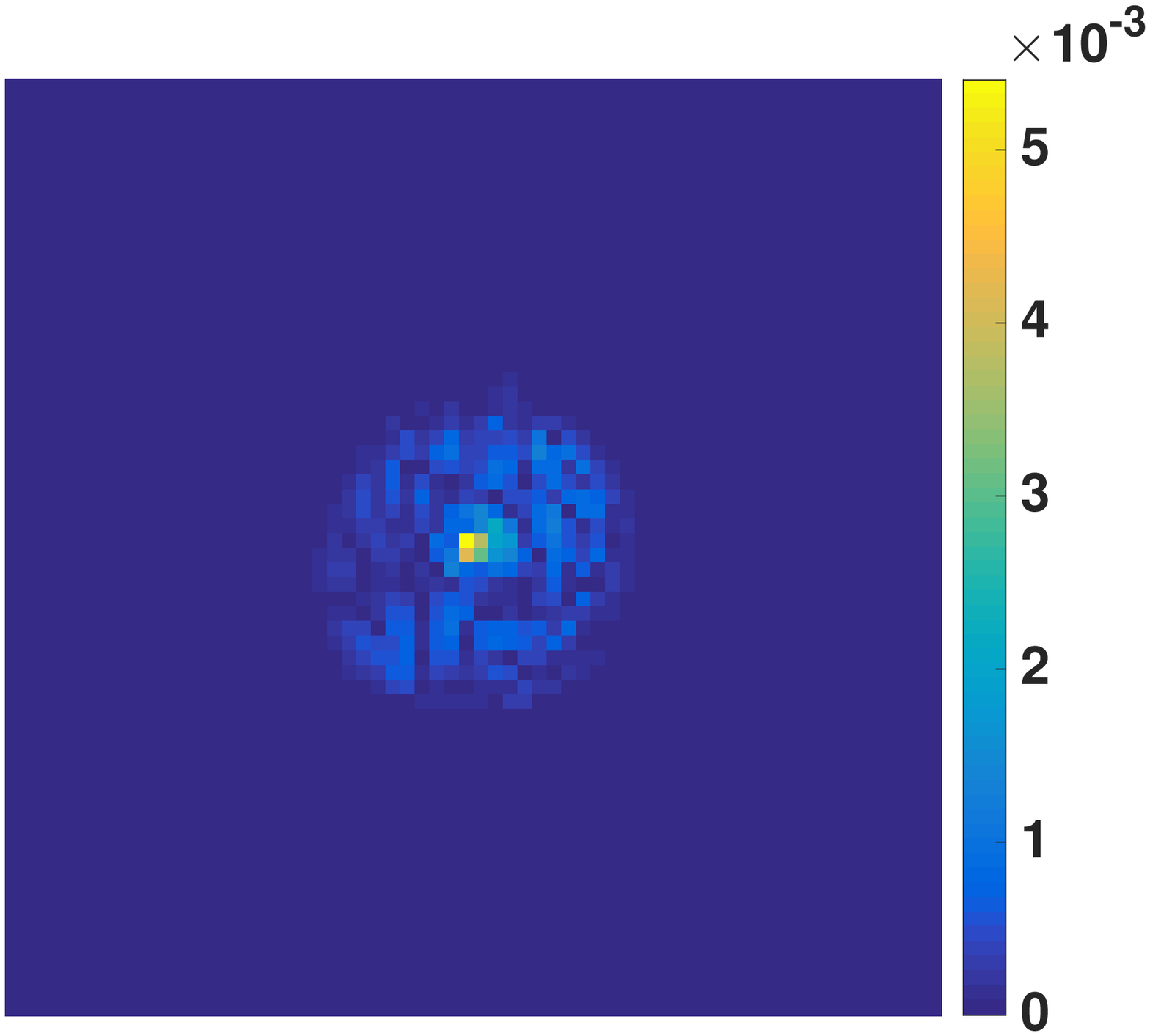} &
\includegraphics[height=3.5cm]{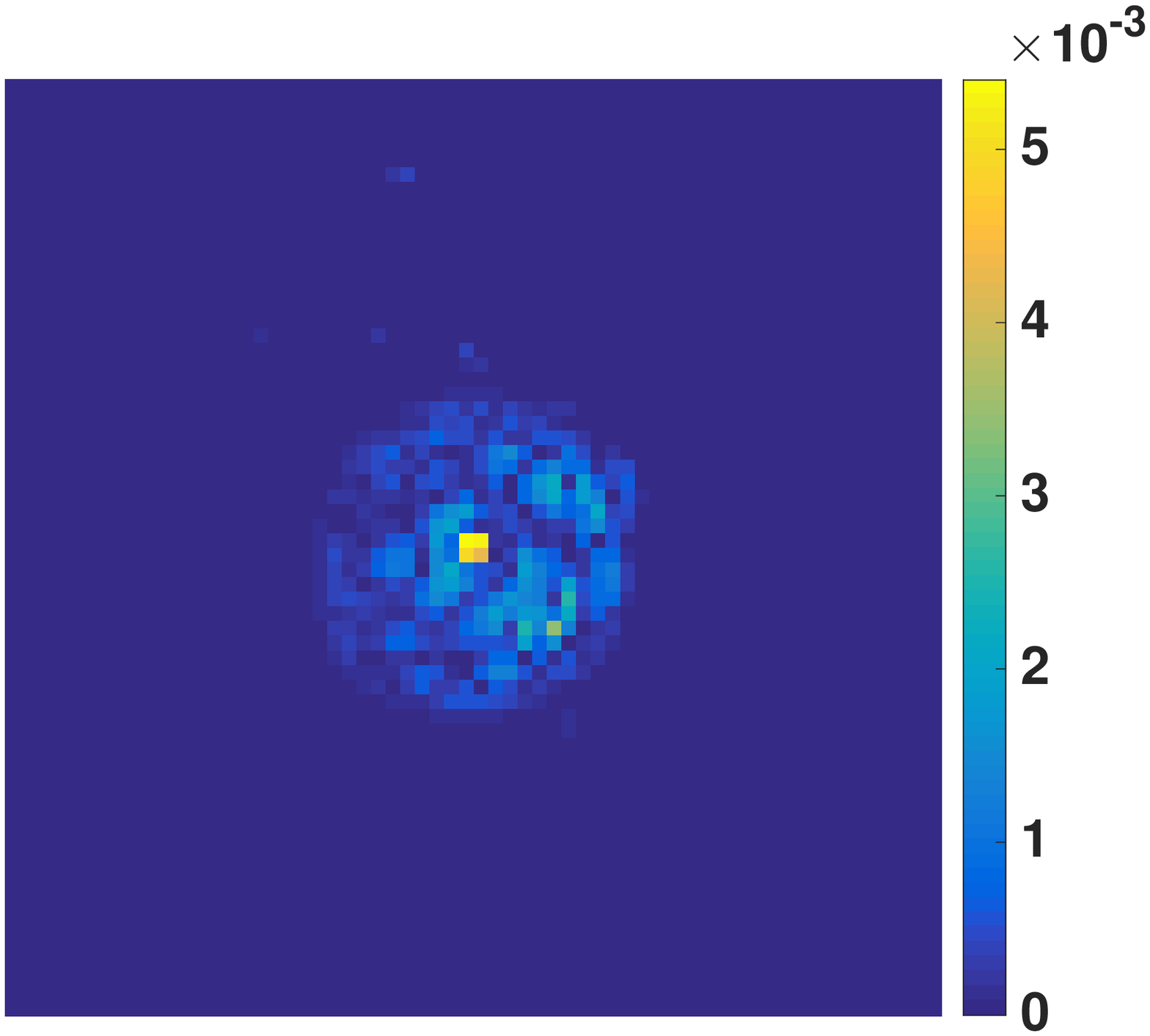} 
\vspace*{0.08cm}\\
\includegraphics[height=3.5cm]{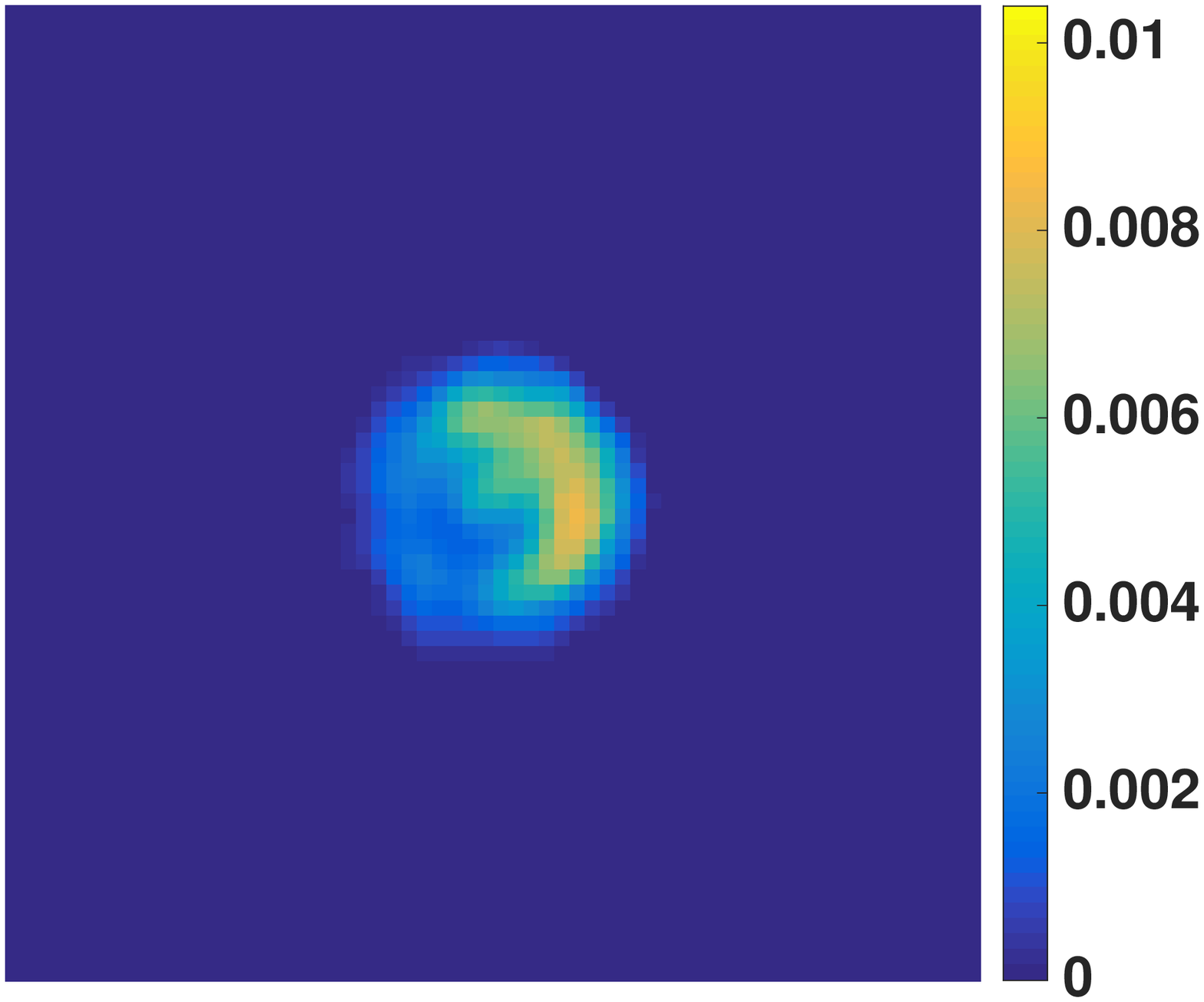} &
\includegraphics[height=3.5cm]{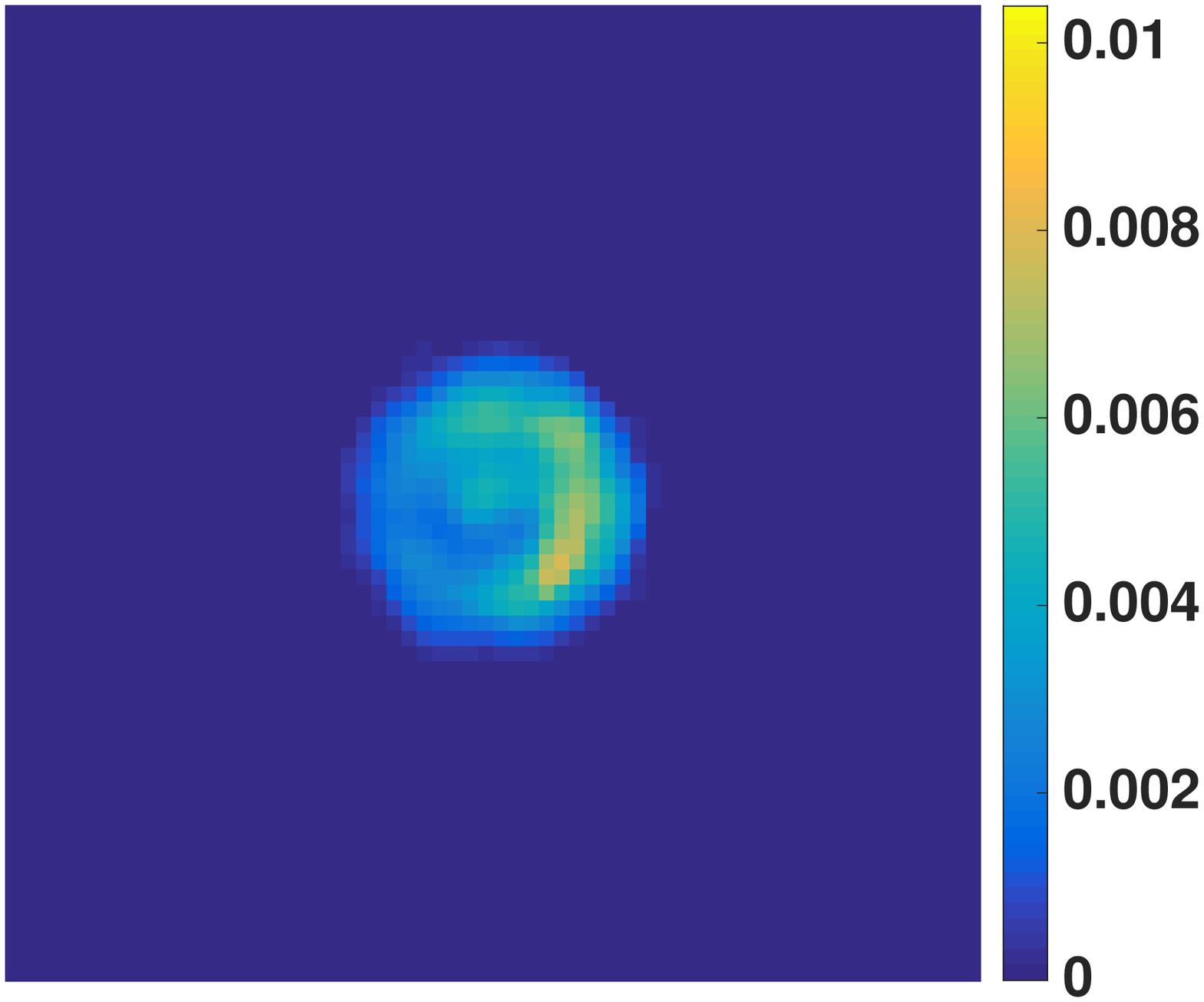}
\vspace*{0.08cm}\\
\includegraphics[height=3.5cm]{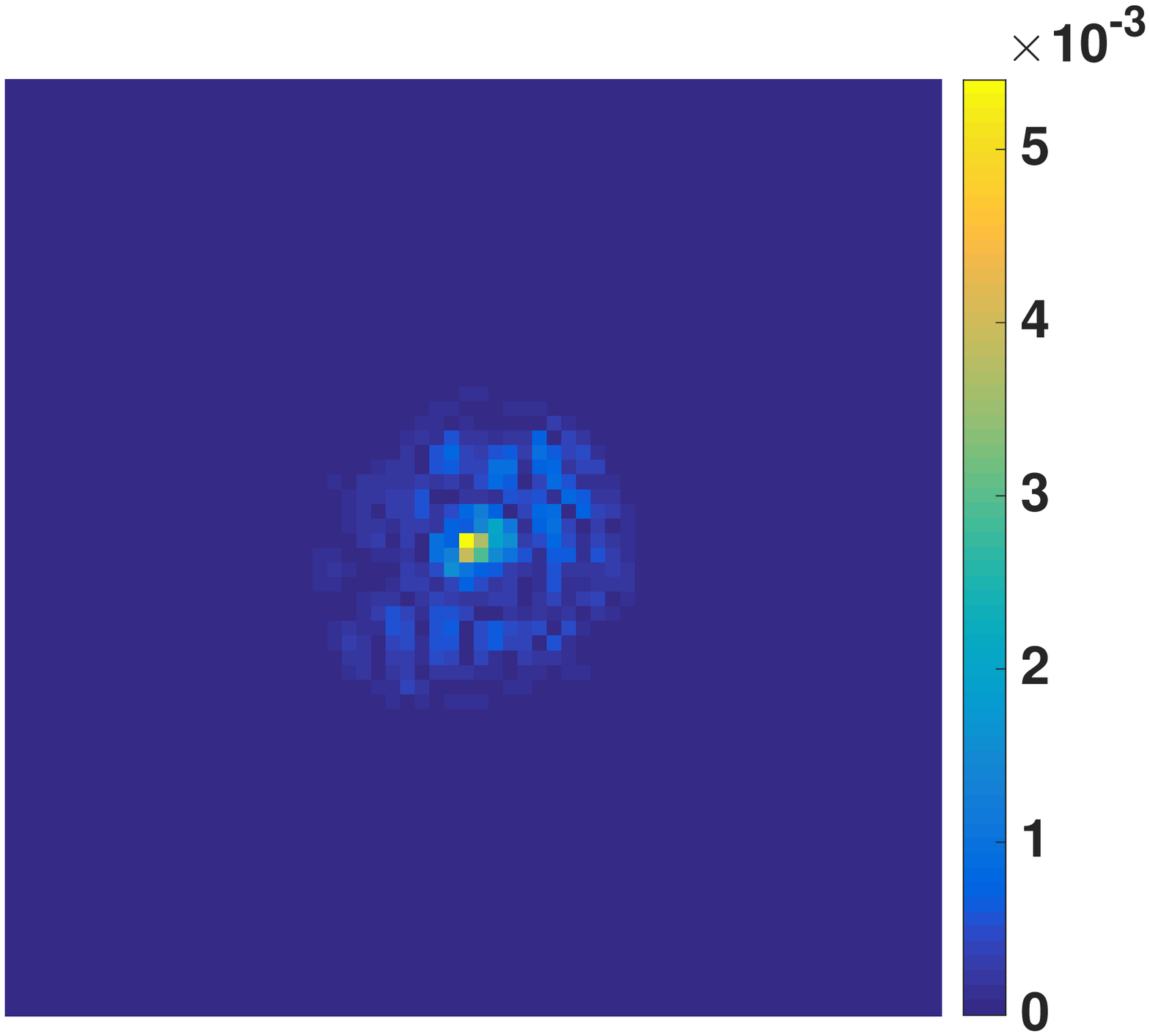} &
\includegraphics[height=3.5cm]{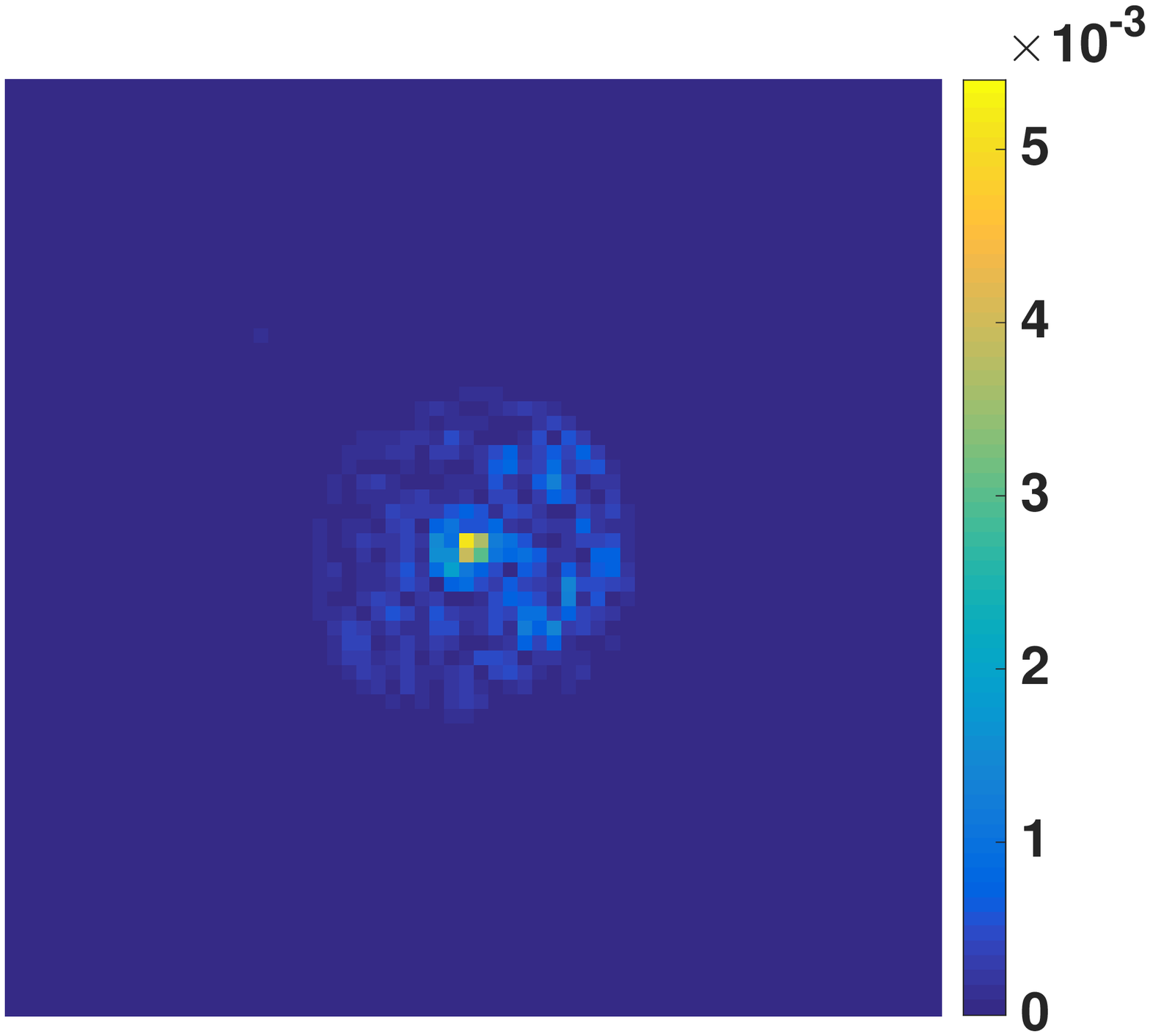} 
\end{tabular}
\caption{Results for hyperspectral imaging with realistic $u-v$ coverage for $L = 8$, $u_{\bis} = 0.1$ and \texttt{LkH$\alpha$} as the original image (top left). Left column: images corresponding to first spectral channel, $l = 1$; Right column: images corresponding to last spectral channel, $l = 8$. 
In each column, the images shown are- first row: original image (top), second row: reconstructed image with $\ell_1$ regularization \eqref{eq:reg_l1_hyper}, third row: error image with $\ell_1$ regularization, fourth row: reconstructed image with $\ell_{2,1}$ regularization \eqref{eq:l21}, and fifth row: error image with $\ell_{2,1}$ regularization \eqref{eq:l21}.
}
\label{Fig:hyper1}
\end{figure}

\begin{figure}	
\centering
\begin{tabular}{@{}c@{}c@{}}
%\vspace*{-0.05cm}
\includegraphics[height=3.5cm]{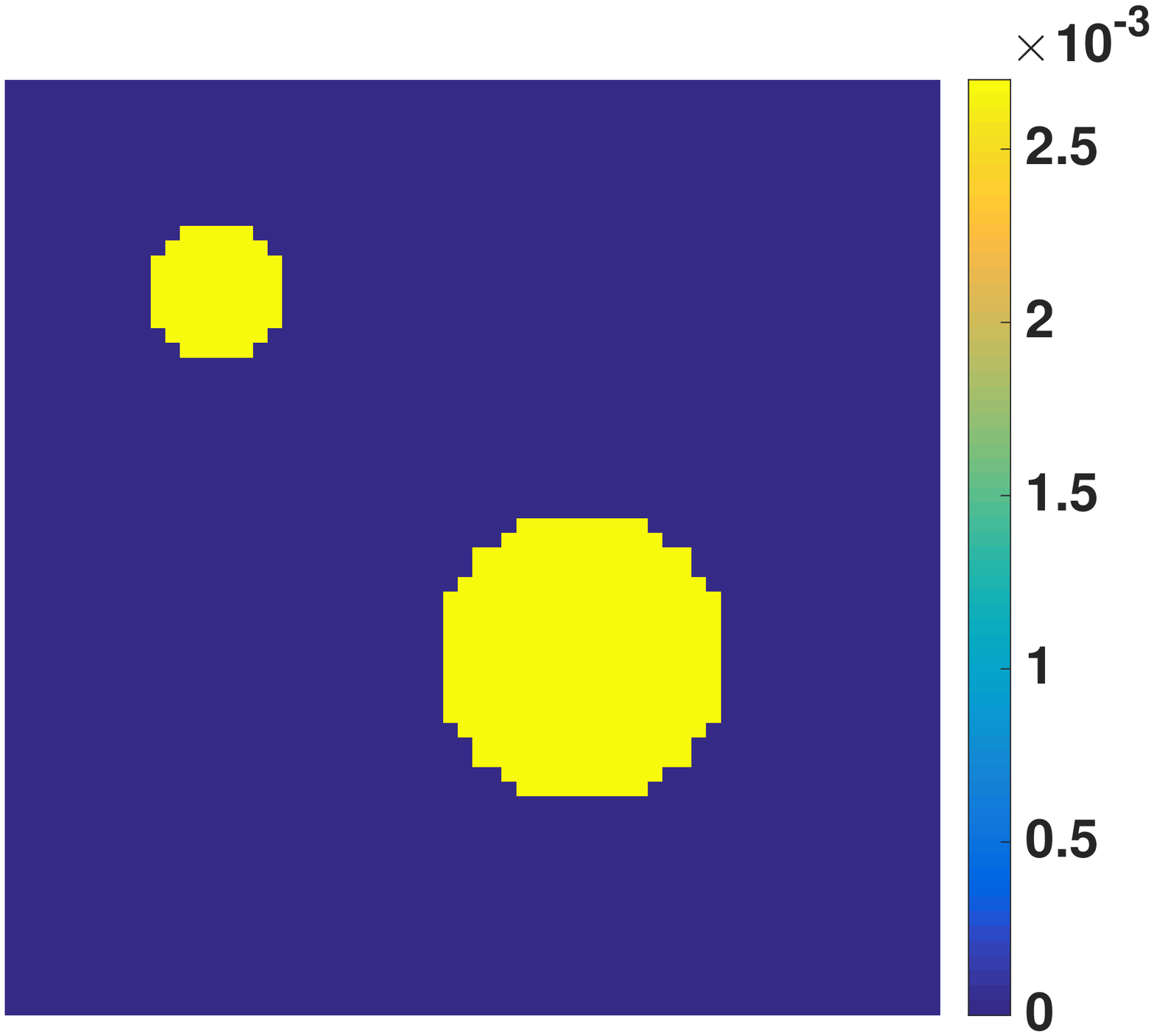} &
\includegraphics[height=3.5cm]{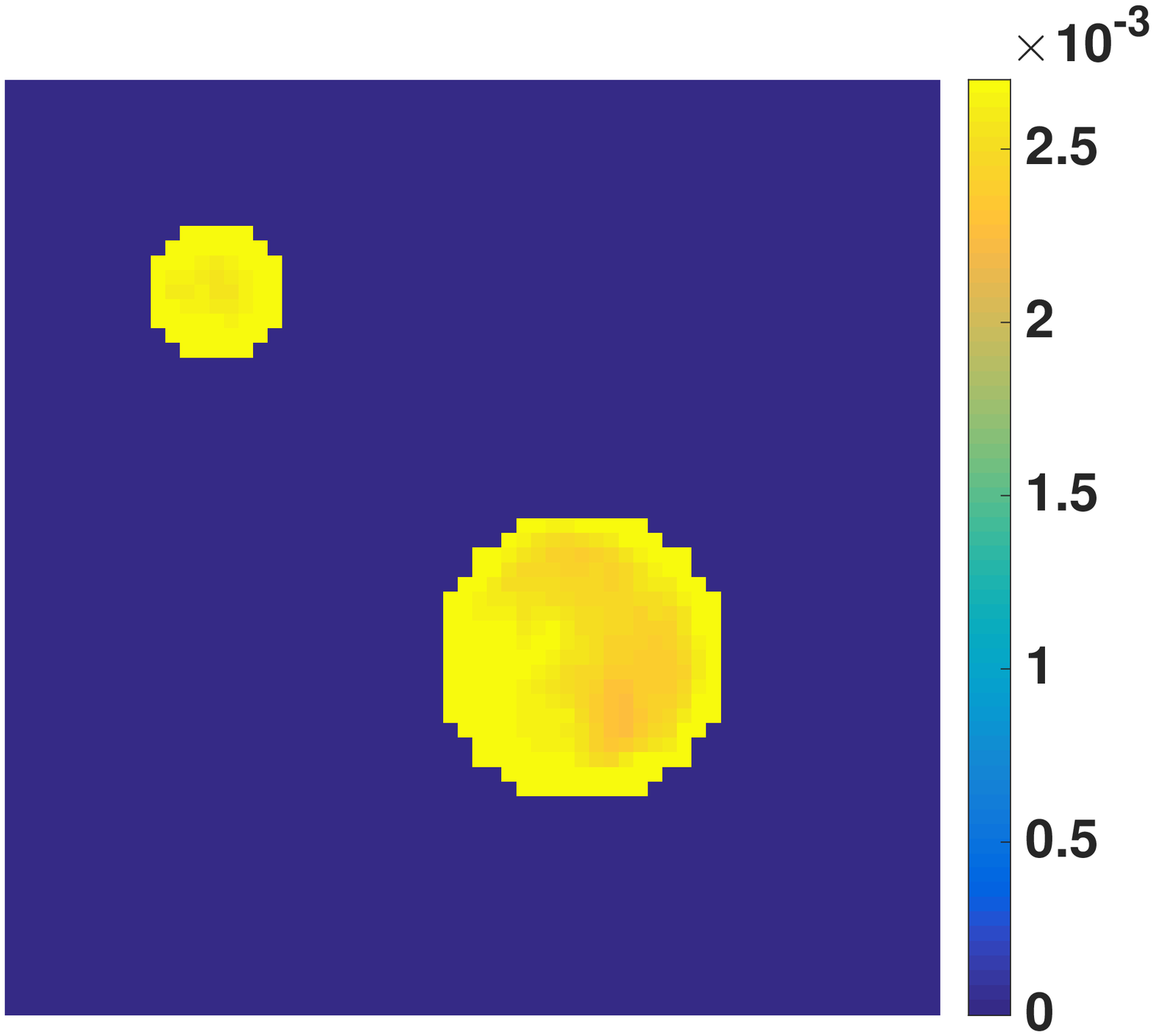} 
\vspace*{0.08cm}\\
\includegraphics[height=3.5cm]{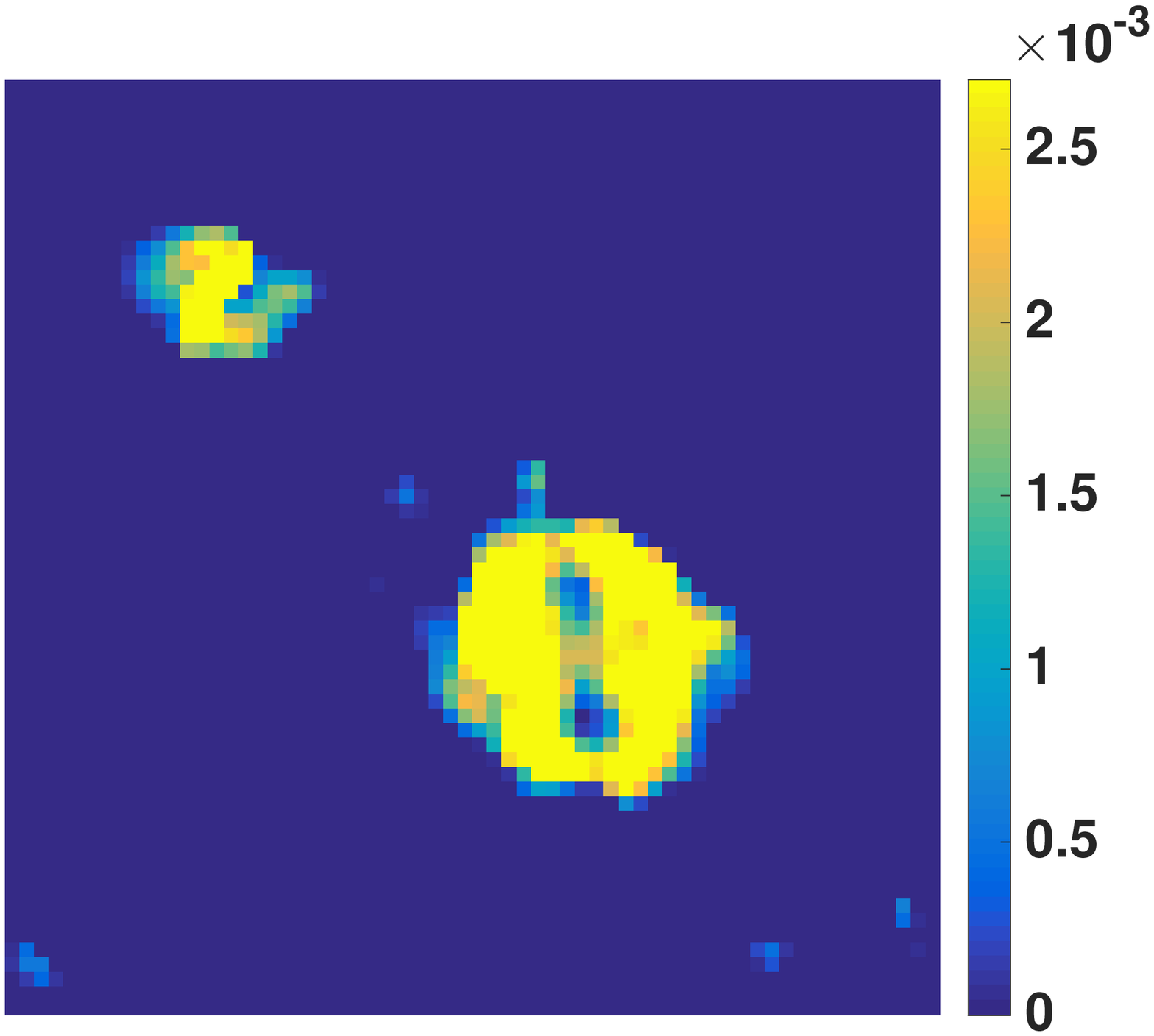} &
\includegraphics[height=3.5cm]{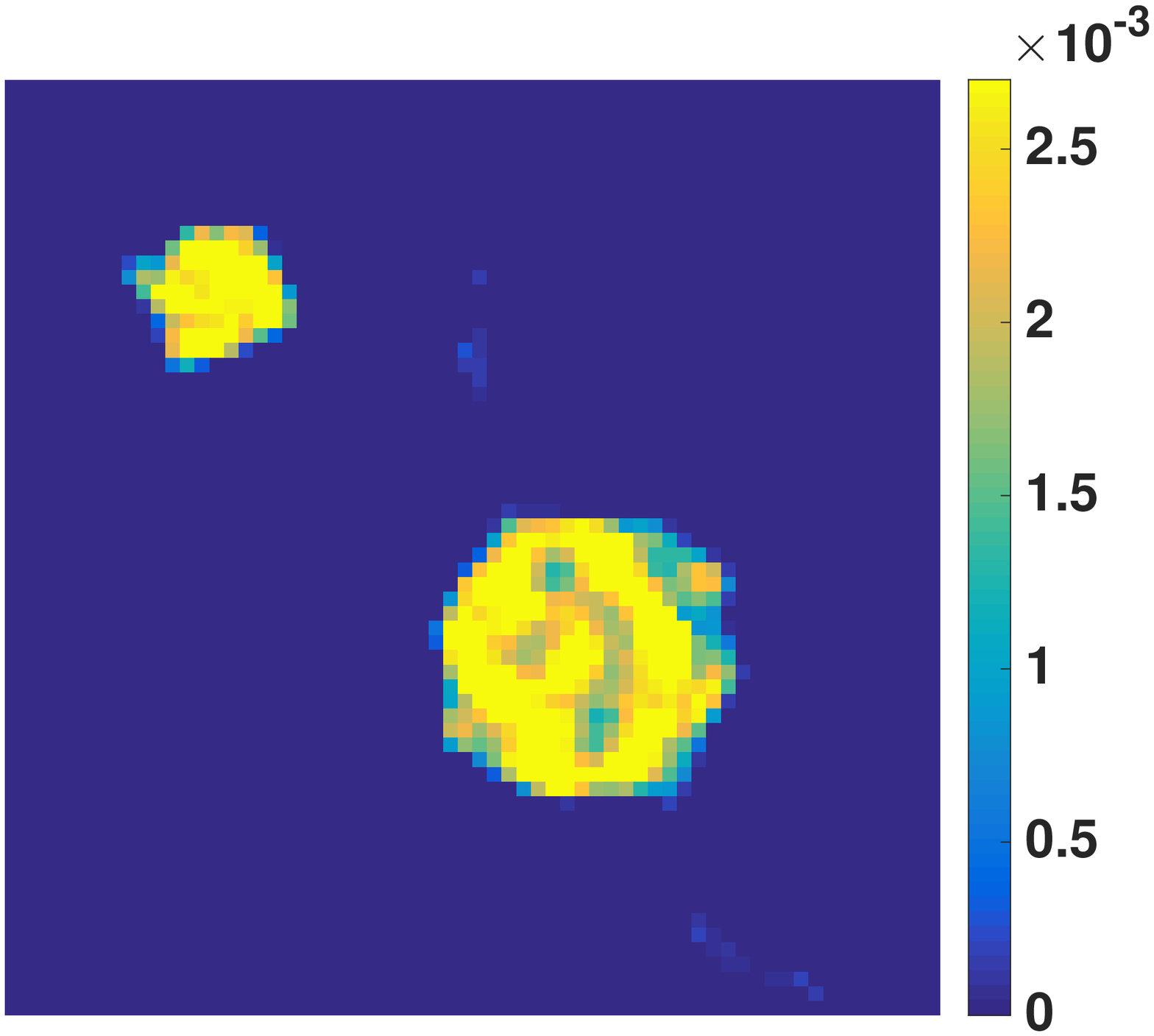} 
\vspace*{0.08cm}\\
\includegraphics[height=3.5cm]{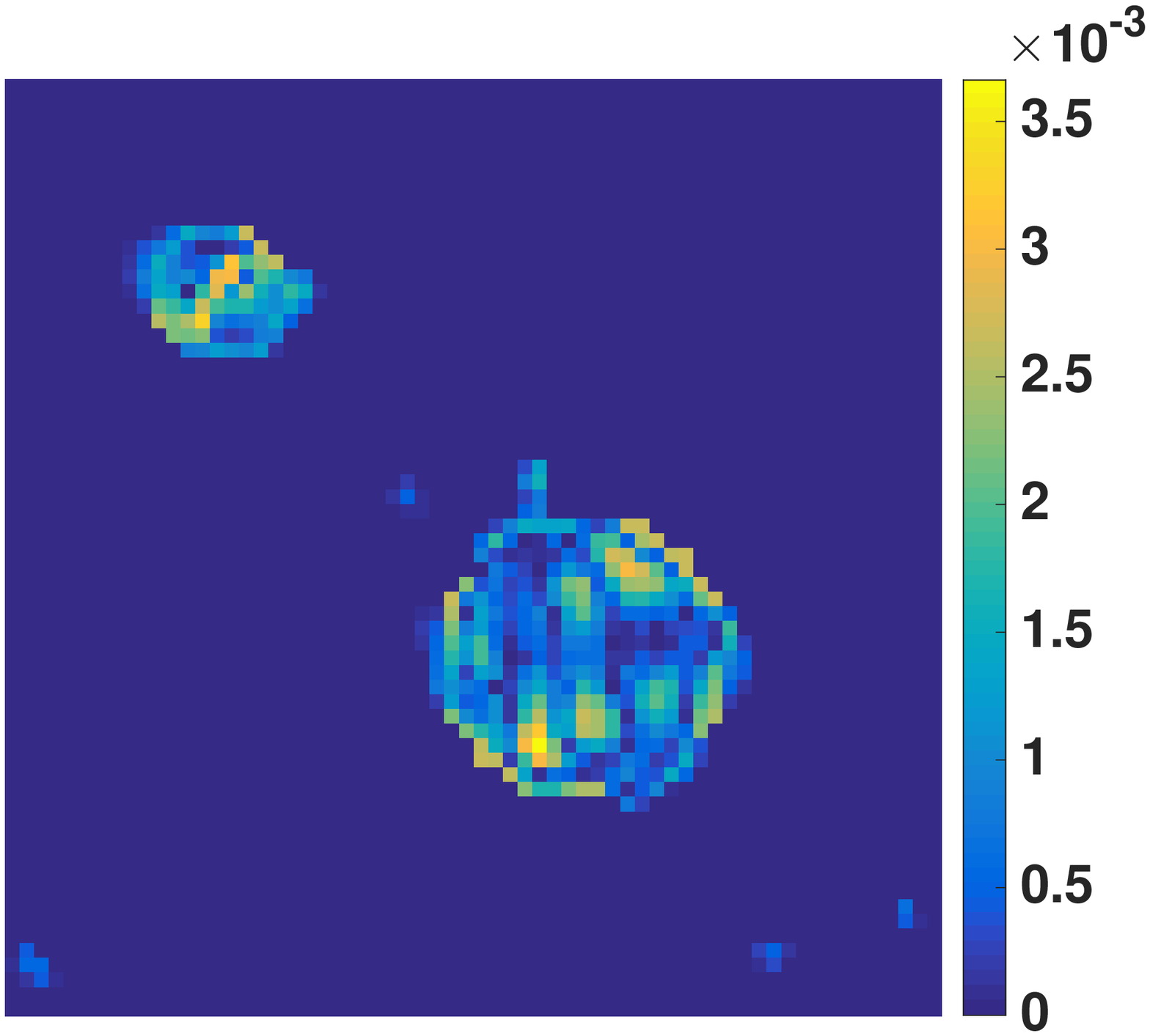} &
\includegraphics[height=3.5cm]{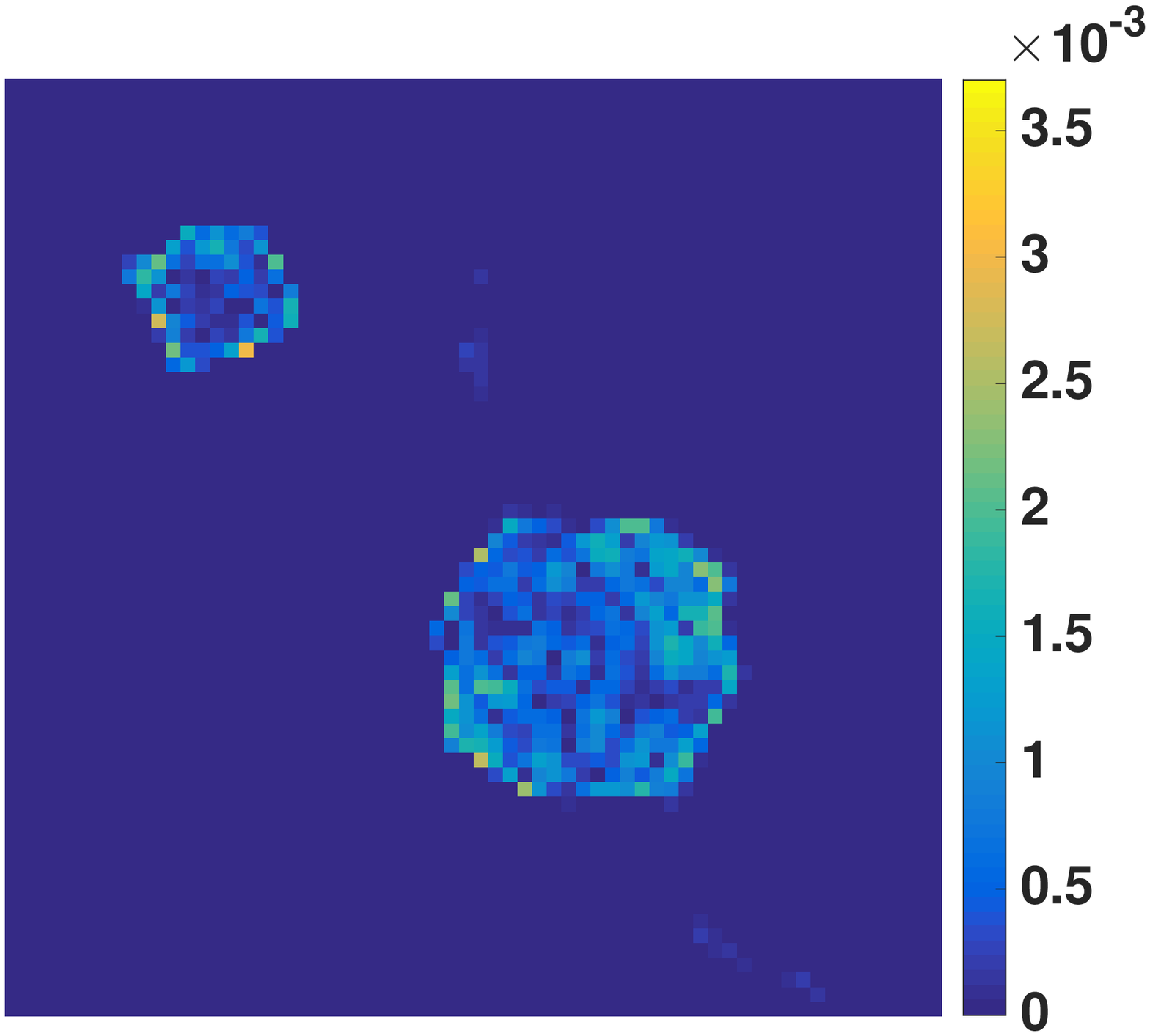} 
\vspace*{0.08cm}\\
\includegraphics[height=3.5cm]{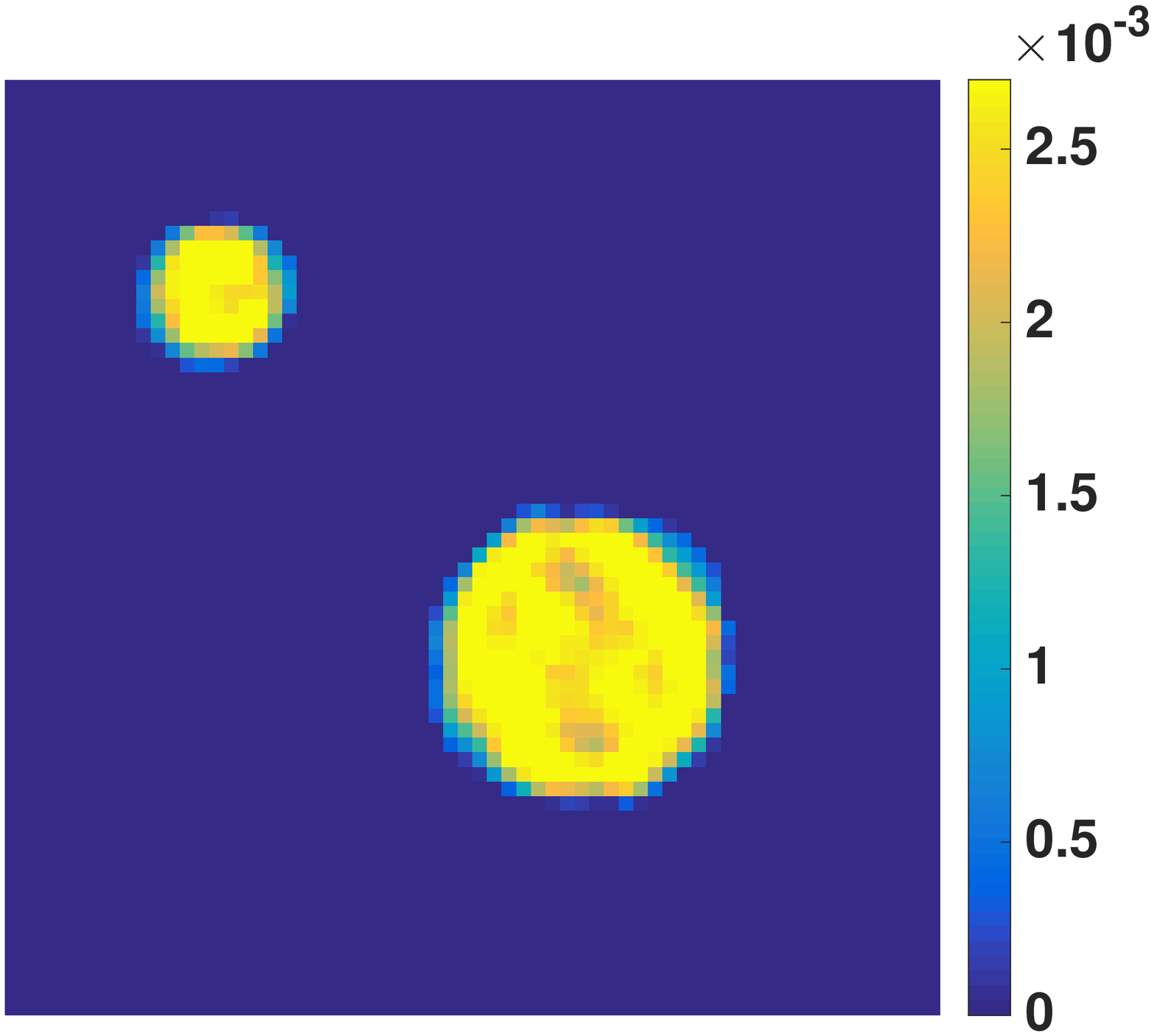} &
\includegraphics[height=3.5cm]{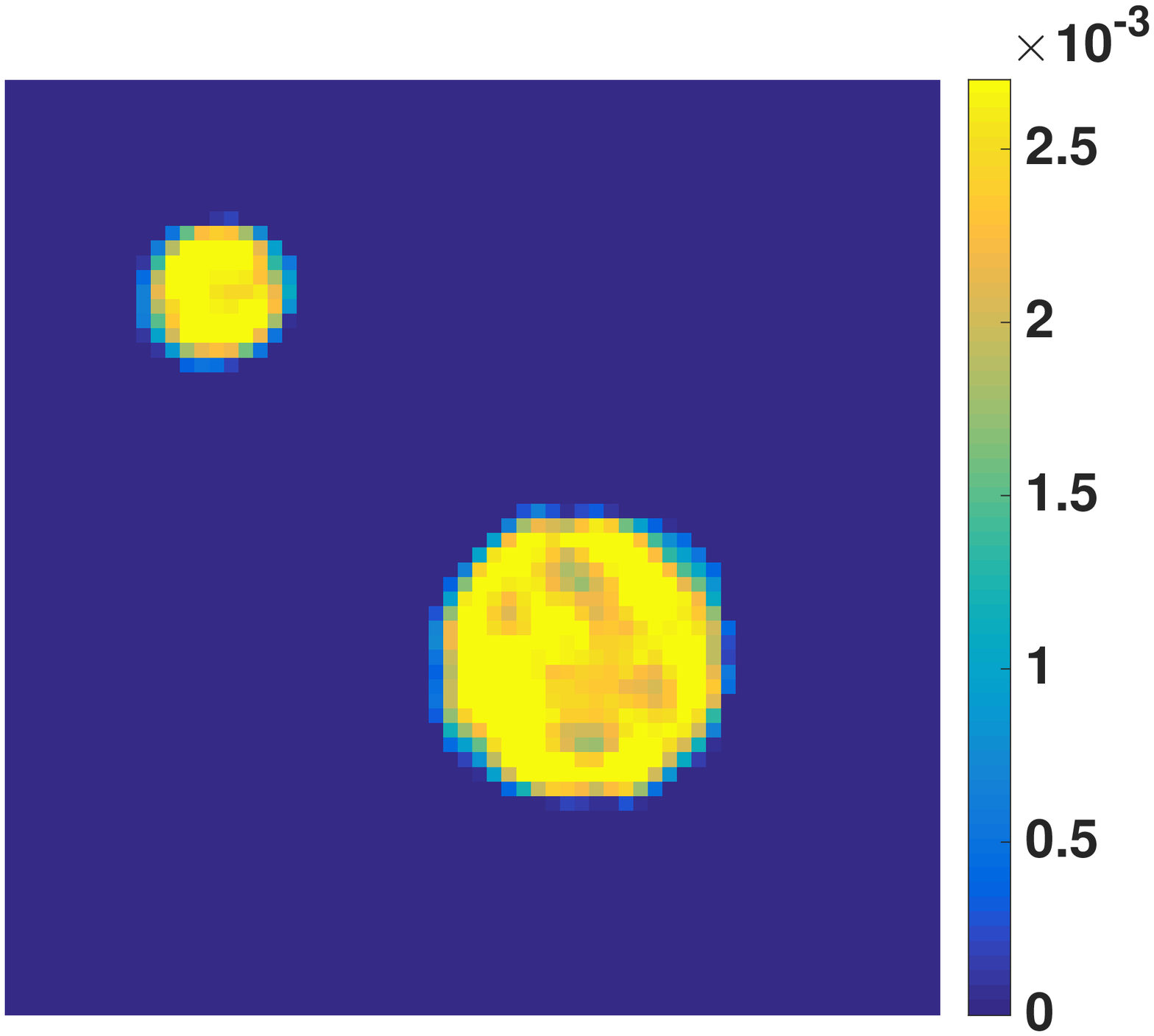}
\vspace*{0.08cm}\\
\includegraphics[height=3.5cm]{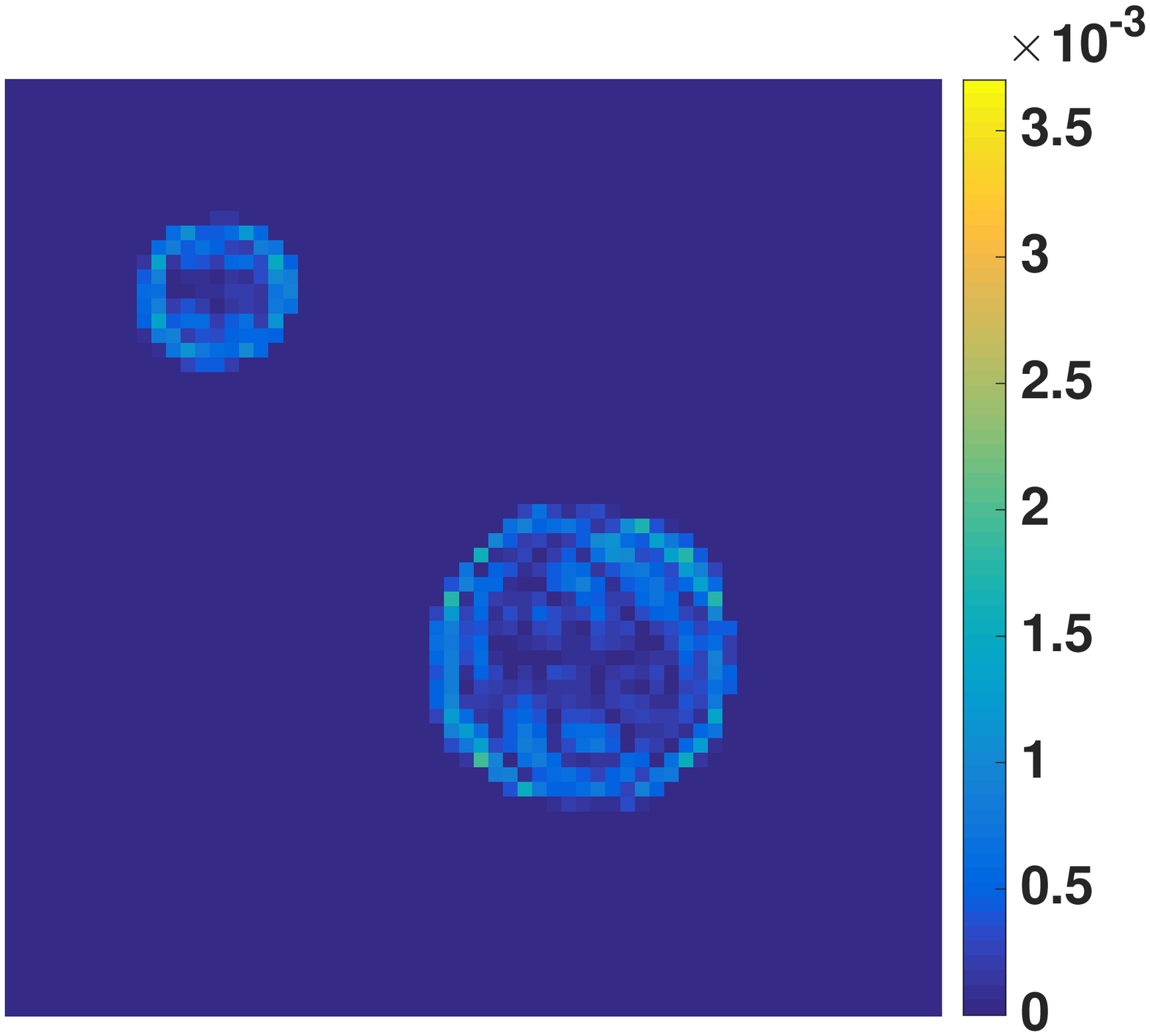} &
\includegraphics[height=3.5cm]{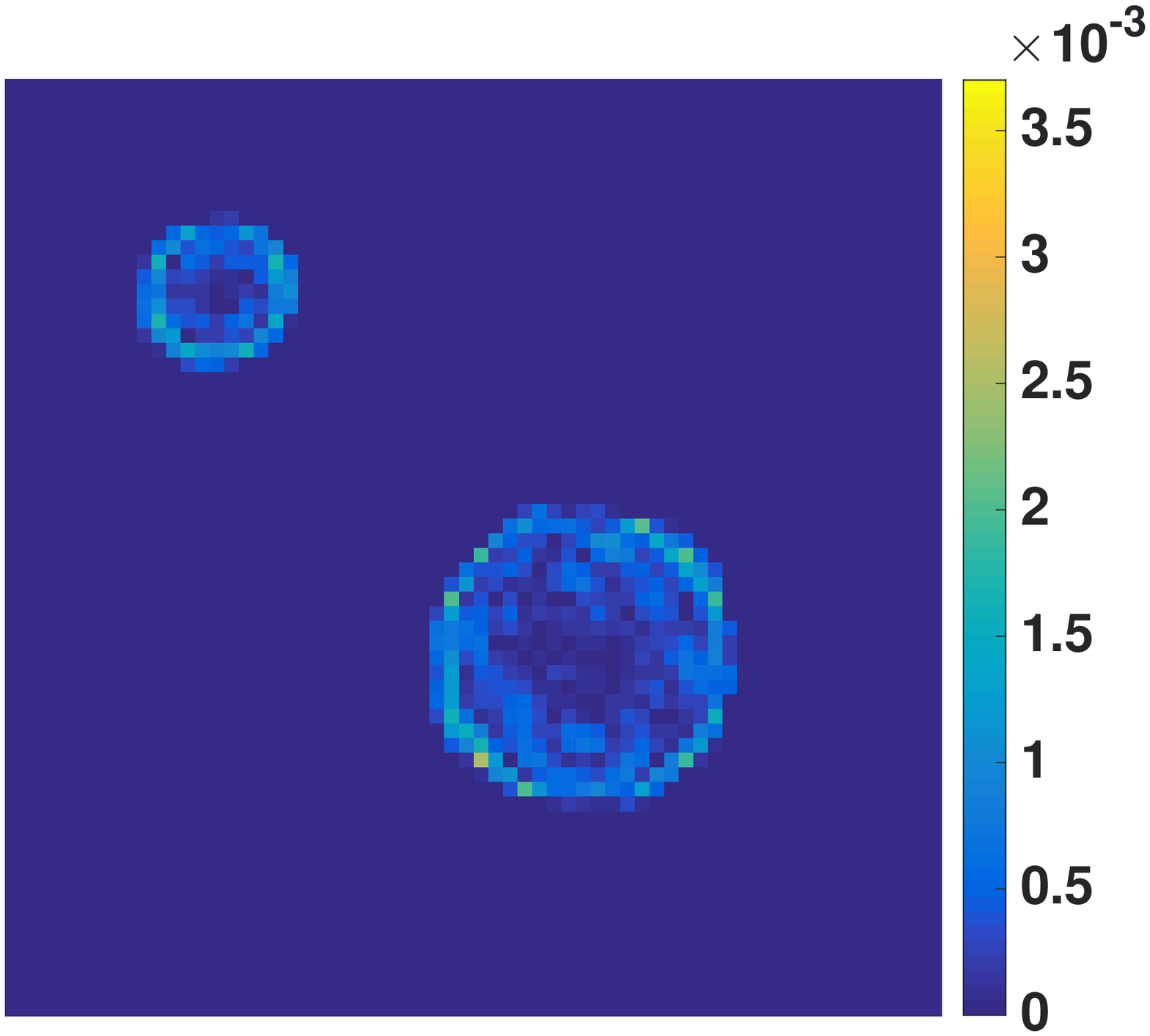} 
\end{tabular}
\caption{Results for hyperspectral imaging with realistic $u-v$ coverage for $L = 8$, $u_{\bis} = 0.1$ and \texttt{synthetic image} as the original image (top left). Left column: images corresponding to first spectral channel, $l = 1$; Right column: images corresponding to last spectral channel, $l = 8$. 
In each column, the images shown are- first row: original image (top), second row: reconstructed image with $\ell_1$ regularization \eqref{eq:reg_l1_hyper}, third row: error image with $\ell_1$ regularization, fourth row: reconstructed image with $\ell_{2,1}$ regularization \eqref{eq:l21}, and fifth row: error image with $\ell_{2,1}$ regularization \eqref{eq:l21}.}
\label{Fig:hyper2}
\end{figure}

\subsection{Simulations and results}

In this section, we will show the performance of the proposed algorithm~\ref{algo_blockfb} for hyperspectral imaging by solving \eqref{eq:overall_min_hyper}. 
Simulations are performed on two sets of images, with $N = 64^{2}$ for each image.
More precisely, two original images are considered: \texttt{LkH$\alpha$}, given in the top left of Figure~\ref{Fig:hyper1}, and an image consisting of two simulated uniform discs, which we refer to as \texttt{synthetic image}, shown as the top left image in Figure~\ref{Fig:hyper2}. These images correspond to the observed image at the first spectral channel $\overline{\bm{x}}_1$. 
Then, the images corresponding to other spectral channels $l \in \{2, \ldots,L\}$ are obtained by following power-law model. 
In this context, we have, for $\overline{\bm{x}}_{l} = (\overline{x}_{l,n})_{1 \leq n \leq N}$,
\begin{equation}
\overline{x}_{l,n} 
= \overline{x}_{1,n} \Big( \frac{\lambda_{1}}{\lambda_{l}} \Big)^{\alpha_n}, 
\end{equation}
where $\lambda_{l}$ denotes the wavelength at spectral channel $l$, and $\bm{\alpha} = (\alpha_n)_{1 \le n \le N}$ is the spectral indices' vector \citep{Rau2011}. 
Spatial correlation is ensured by taking $\bm{\alpha}$ to be a linear combination of a random Gaussian field and the reference image convolved with a Gaussian kernel of size $3 \times 3$ at FWHM \citep{Junklewitz2015}.

For both the images, $L=8$ spectral channels in the wavelength range 1.95-1.97 $\mu$m are considered. The corresponding $u-v$ coverage plan is given in Figure~\ref{fig:uv_cover}(b) for observation wavelength 1.95 $\mu$m. The generated ground-truth images for $l = 8$ are shown as top right images in Figures~\ref{Fig:hyper1} and \ref{Fig:hyper2}, respectively for \texttt{LkH$\alpha$} and \texttt{synthetic image}.

We compare the results obtained considering the $\ell_{2,1}$ norm with the case when each channel is treated separately, considering an $\ell_1$ norm on each image produced by each spectral channel: 
\begin{equation}	\label{eq:reg_l1_hyper}
(\forall \bm{\mathsf{X}}  \in \mathbb{R}^{N \times L}) \quad
g(\bm{\mathsf{X}}) = \sum_{l = 1}^L \| \bm{\bm{\Psi}}^\dagger \bm{x}_l \|_1.
\end{equation}
While the case considering $\ell_1$ norm is initialized with the solution of problem~\eqref{eq:overall_min_hyper} solved with only positivity and reality constrained case (i.e. $\mu = 0$ in \eqref{eq:hyper_reg}), the solution obtained for each channel by $\ell_1$ regularized case is in turn used to initialize $\ell_{2,1}$ regularized case. For both cases, the forward-backward iterations (steps~\ref{algo:fb_start}-\ref{algo:fb_end} in Algorithm~\ref{algo_blockfb}) are performed with $t_{\max} = 200$.
 
In the hyperspectral case, we observed that considering $\bm{\Psi}$ as the identity matrix gives better reconstruction results than using Daubechies wavelets. 
Moreover, the SNR of the reconstructed image matrix $\bm{\mathsf{X}}^{\star}$ is computed as the mean of the SNRs from the reconstructed images of each channel $(\bm{x}^{\star}_l)_{1 \le l \le L}$. The SNR comparisons between the regularizations \eqref{eq:l21} and \eqref{eq:reg_l1_hyper} are provided in Figure~\ref{Fig:hyper_SNR}. For both cases, average SNR curves with 1-standard-deviation error bars are presented (performed over 10 simulations, varying both the noise realization and the measured bispectrum). From these plots, we can observe that using $\ell_{2,1}$ norm as a regularization term leads to better reconstruction than considering only $\ell_1$ independently in each channel.

\modif{The reconstructed and the corresponding error images for the first and the last spectral channels, considering $\bm{\Psi}$ to be the identity matrix, are shown in Figures~\ref{Fig:hyper1} and \ref{Fig:hyper2}.}
For the two image examples, the figures demonstrate the superiority of solving globally for the hyperspectral channels over single-channel reconstruction, where no advantage of inter-channel information is taken.

%--------------------------------------------------------

%
\section{Conclusion} \label{sec:Conclusion}
We have presented a new method for image reconstruction in optical interferometry, based on the tri-linear data model proposed in \citet{Auria2013}. 
While only monochromatic imaging was considered in the previous work, we have extended this model to hyperspectral imaging. 
Furthermore, to improve the reconstruction quality, since in \citet{Auria2013} only positivity constraints were considered, 
we have proposed additionally to promote sparsity using either an $\ell_1$ or a weighted-$\ell_1$ regularization term in the monochromatic case, and an $\ell_{2,1}$ regularization term in the hyperspectral case. 
Moreover, in order to solve the resultant minimization problem, we have developed an alternated minimization algorithm, based on a block-coordinate forward backward algorithm.
This algorithm presents convergence guarantees, and benefits from the fact that it can be designed to work with smooth functions, using gradient steps, and with non-necessarily smooth functions thanks to proximity steps.
We have assessed the performance of the proposed method on several simulations both for synthetic and realistic $u-v$ coverages, in monochromatic and hyperspectral cases. 
On the one hand, for monochromatic imaging, adding a sparsity prior gives promising results.
On the other hand, for hyperspectral imaging, we have shown numerically that exploiting joint sparsity, using an $\ell_{2,1}$ norm, improves drastically the quality of reconstruction as compared to single-channel reconstruction. 
To summarise, we have proposed a method which presents a general framework, where the regularization term can be non-smooth and adapted either for the monochromatic case or for the hyperspectral case. 
Future work includes testing the proposed algorithm on realistic data sets and comparing our method with the state-of-the-art methods in optical interferometry.

\section*{Acknowledgements}

This work was supported by the UK Engineering and Physical Sciences Research Council (EPSRC, grant EP/M008843/1). We would like to thank Andr\'{e} Ferrari for insightful discussions.

%%%%%%%%%%%%%%%%%%%%%%%%%%%%%%%%%%%%%%%%%%%%%%%%%%

%%%%%%%%%%%%%%%%%%%% REFERENCES %%%%%%%%%%%%%%%%%%

\bibliographystyle{mnras}
%\bibliography{library_article} 

%%%%%%%%%%%%%%%%%%%%%%%%%%%%%%%%%%%%%%%%%%%%%%%%%%

%%%%%%%%%%%%%%%%% APPENDICES %%%%%%%%%%%%%%%%%%%%%

\appendix

%%%%%%%%%%%%%%%%%%%%%%%%%%%%%%%%%%%%%%%%%%%%%%%%%%

% Don't change these lines
\bsp	% typesetting comment
\label{lastpage}
\end{document}